\documentclass[prd,amsfonts,twocolumn,nofootinbib,showpacs]{revtex4-1}

\usepackage{graphicx}
\usepackage{dcolumn}
\usepackage{bm}
\usepackage{color}
\usepackage{epstopdf}
\usepackage{epsfig}
\usepackage{natbib}
\usepackage{amsmath}
\usepackage{amssymb}

\def\sigv{\langle\sigma v\rangle}

\def\he4{{\rm ^4He}}
\def\sigv{{\langle\sigma v \rangle}}

\pacs {26.35.+c, 98.80.Cq, 98.80.Ft}

\begin{document}

\title{Systematic Uncertainties In Constraining Dark Matter Annihilation From The Cosmic Microwave Background
}

\author{Silvia Galli$^{a,b}$, Tracy R. Slatyer$^{c}$, Marcos Valdes$^{d}$,  Fabio Iocco$^{e}$}
\affiliation{$^a$ UPMC Univ Paris 06, UMR7095, Institut d'Astrophysique de Paris, F-75014, Paris, France}
\affiliation{$^b$ CNRS, UMR7095, Institut d'Astrophysique de Paris, F-75014, Paris, France}
\affiliation{$^c$ School of Natural Sciences, Institute for Advanced Study, Princeton, NJ 08540, USA.} 
\affiliation{$^d$ Scuola Normale Superiore, Piazza dei Cavalieri 7, 56126 Pisa, Italy}
\affiliation{$^e$ The Oskar Klein Center for CosmoParticle Physics, Department of Physics, Stockholm University, Albanova, SE-10691 Stockholm, Sweden} 

\begin{abstract}
Anisotropies of the cosmic microwave background (CMB) have proven to be a very powerful tool to constrain dark matter annihilation at the epoch of recombination. However, CMB constraints are currently derived using a number of reasonable but yet un-tested assumptions that could potentially lead to a misestimation of the true bounds (or any reconstructed signal). In this paper we examine the potential impact of these systematic effects. In particular, we separately study the propagation of the secondary particles produced by annihilation in two energy regimes; first following the shower from the initial particle energy to the keV scale, and then tracking the resulting secondary particles from this scale to the absorption of their energy as heat, ionization, or excitation of the medium. We improve both the high and low energy parts of the calculation, in particular finding that our more accurate treatment of losses to sub-10.2 eV photons produced by scattering of high-energy electrons weakens the constraints on particular dark matter annihilation models by up to a factor of two. On the other hand, we find that the uncertainties we examine for the low energy propagation do not significantly affect the results for current and upcoming CMB data. We include the evaluation of  the precise amount of excitation energy, in the form of Lyman-$\alpha$ photons, produced by the propagation of the shower, and examine the effects of varying the Helium fraction and Helium ionization fraction. In the recent literature, simple approximations for the fraction of energy absorbed in different channels have often been used to derive CMB constraints: we assess the impact of using accurate versus approximate energy fractions. Finally we check that the choice of recombination code (between RECFAST v1.5 and COSMOREC), to calculate the evolution of the free electron fraction in the presence of dark matter annihilation, introduces negligible differences.
\end{abstract}

\keywords{dark matter; self--annihilating; CMB}

\maketitle

\section{Introduction}
\label{sect_introduction}
Within the framework of the currently favoured $\Lambda$ Cold Dark Matter
cosmological model,  only approximately 32\% of the current energy density of the Universe
is due to matter, of which approximately 85\% is of unknown nature, as 
particles belonging to the Standard Model (SM) of particle physics constitute only  
5\% of the total energy density of today's Universe \cite{PlanckCollaboration:2013}.
Several extensions of the SM exist that predict particles complying with the phenomenological requirements
of a {\it Dark} Matter (DM) candidate, see \cite{Taoso:2007qk} for a ten-point test.
Many such candidates are Majorana particles, bearing the remarkable property
of being their own antiparticle and thus ``self-annihilating'';
a popular and much-studied class of models is the Weakly Interacting Massive Particles
(WIMPs), where the natural size of the self-annihilation cross section gives rise to the observed DM density (see \cite{Bertone:2004pz} and \cite{Bergstrom:2012fi} for reviews).
A wide variety of present-day experimental searches seek to identify the nature of DM:
collider searches and direct detection experiments are complemented by indirect detection experiments that 
attempt to detect the astrophysical signatures left by DM,
which are expected to be particularly striking in the case of self-annihilating particles.
Clear positive indirect detections of DM annihilations taking place in the local Universe -- particularly in environments 
where DM over-densities are expected -- still elude us. However, in their absence, strong constraints on the 
parameters that control the annihilation energy release can be derived.
The most prominent approaches to indirect detection include searches for gamma-ray signals from the Galactic center
(e.g. \cite{Weniger:2012tx}) and dwarf galaxies \cite{Ackermann:2011wa}, charged particles in the Earth's neighbourhood \cite{FermiLAT:2011ab,Adriani:2009,Aguilar:2013}, contributions to the diffuse galactic gamma-ray background
\cite{Ackermann:2012rg} and associated multi-wavelength searches \cite{Fornengo:2011iq}.

However, DM annihilation is not limited to the local Universe, but 
-- if occurring at all -- should generically take place at all times;
DM annihilating during the recombination epoch and slightly after
is expected to affect the free electron fraction, and consequently the
anisotropy of the cosmic microwave background (CMB) \cite{Padmanabhan:2005es, Mapelli:2006es,zhang2006}.
Methods based on this effect can test the DM properties at a comparable sensitivity to techniques based on searching for signatures of DM annihilation in the present day \cite{Galli:2009zc, Galli:2011rz}. Furthermore, they have the significant qualitative advantage of depending only on the (well-measured) average DM density of the universe -- as favorable as the proximity of the targets is, all local Universe methods are
affected by the uncertainty in the DM distribution in the target, which
controls the overall annihilation rate and therefore the expected observable flux. In contrast, the annihilation rate producing the modification in the CMB depends
on the smooth, cosmological background of DM, before any significant structures form,
and therefore CMB constraints on DM annihilation are unaffected by this type of uncertainty (see e.g. \cite{Cirelli:2009bb, Hutsi:2011vx, Natarajan:2012ry,Lopez:2013}).
The DM-model dependence of such constraints -- i.e. the dependence on
the branching ratios for DM annihilation into different SM channels,
the absorption of the annihilation products by the cosmological plasma, and the uncertainties
on such estimates -- have been discussed in 
\cite{Galli:2011rz, Slatyer:2009yq, Finkbeiner:2011dx,Weniger:2013hja}. 

Constraints on the DM self-annihilation cross section and
mass obtained with the current CMB data  are 
already competitive with other existing searches; forecasts based on mock data 
show that a combination of \emph{Planck} temperature and polarization data\footnote{The \emph{Planck} collaboration's initial public data release in March 2013 included only temperature data.} could in principle allow the discovery of DM annihilation
at the level required by a ``thermal relic'' annihilation cross section, for an interesting 
range of DM masses (up to $\sim 50$ GeV/c$^2$, depending on the branching ratio). Given the forthcoming polarization data release by the \emph{Planck} collaboration, it is
most timely to assess all uncertainties affecting CMB constraints.
In this paper, we address some of the remaining uncertainties which can be considered
of systematic type, and affect the constraints from the CMB at comparable levels.

After illustrating the basic theory and equations in Section \ref{sec:theory}, we study the impact of assumptions previously adopted in the literature to calculate the fraction of dark matter annihilation energy that is absorbed by the medium. The study of the propagation of the secondary shower  is split into two phases: from the initial energy to the keV scale, and then from this scale to the final energy depositions into the IGM as heating, ionization, and excitation. We first study the uncertainties related to this second part of the calculation, assuming as a benchmark that the fraction of the initial annihilation energy transferred into keV-scale electrons (via decays and subsequent cooling of weak-scale annihilation products) is constant with redshift. The particular value of this constant does not affect our tests. We describe the code to calculate the fractions of energy going into different absorption channels in Section \ref{sec:fractions}, and study the impact of systematic uncertainties on the calculation of these energy fractions in Section \ref{sec:systematics}.

In Section \ref{sec:accuratevsSSCK} we determine the degree to which adopting a simplified parametrization for the energy fractions, e.g. the one proposed by Chen and Kamionkowski \cite{ChenKamion2004} and used afterward by a number of different authors \cite{Galli:2009zc,Galli:2011rz,Finkbeiner:2011dx,Hutsi:2011vx,Natarajan2009,Giesen2012,Lopez:2013}, can bias the constraints on dark matter annihilation, relative to the more accurate treatment of the fractions presented in this paper. In Section \ref{sec:lymanalpha} we emphasize the point that our code permits us to compute the precise energy fraction that goes into Lyman-$\alpha$ radiation. This is relevant as Lyman-$\alpha$ photons contribute to the ionization of the medium through an intermediate excited state, so the precise calculation of this fraction is important in improving the accuracy of the calculations. We study the differences in the calculated effects of dark matter annihilation due to the use of different recombination codes, such as RECFAST \cite{recfast} (v1.5) or the more recent and accurate COSMOREC \citep{cosmorec}, in Section \ref{sect:recfastvscosmorec}, and verify that the choice of recombination code has negligible impact on the derived constraints\footnote{Note that earlier versions of RECFAST (v1.4 and earlier) may have larger discrepancies with COSMOREC; v1.5, which we use throughout this work, includes correction functions that reproduce the behavior of COSMOREC.}.

In Section \ref{sec:higheng} we study the \emph{first} part of the cascade, from the initial energy at the GeV-TeV scale to the 3 keV threshold.  We describe the improvements to the calculation relative to previous works in the literature, including e.g.  a better understanding of the energy losses to sub-10.2 eV photons of inverse Compton scattered (ICS) secondary electrons. We then assess the impact of these improvements on the constraints. We find that for specific dark matter masses and annihilation channels, the improvements we introduce compared to previous calculations can significantly affect the recovered results, in some cases weakening the constraints by up to a factor of two.

Finally, we present our conclusions in Section \ref{sec:conclusions}.

\section{DM annihilation and the CMB}
\label{sec:theory}

Dark Matter annihilating after the thermal decoupling
-- and before the formation of sizable gravitationally bound structures --
injects energy into the Universe in the form of SM particles with a rate
per unit volume, $\frac{dE}{dt}$, given by

\begin{equation}
\label{enrateselfDM}
\frac{dE}{dt}(z)= \rho^2_c c^2 \Omega^2_\mathrm{DM} (1+z)^6 f(z)  \frac{\sigv}{m_\chi}
\end{equation}

where $n_{DM}(z)=\Omega_\mathrm{DM}\rho_c (1+z)^3$ is the relic DM abundance at a given redshift $z$,
$\sigv$ is the effective self-annihilation rate and $m_\chi$ the mass
of our dark matter particle, $\Omega_{DM}$ is the dark matter density
parameter and $\rho_c$ the critical density of the Universe today.
The parameter $f(z)$ indicates the fraction of annihilation energy that
-- at each redshift -- is absorbed by the plasma, via collisional heating, and atomic excitations and ionizations (energy may also escape in the form of neutrinos, or as photons which free-stream to the present day). The latter processes modify the evolution of the free electron fraction $x_e$, while heating mainly affects the temperature of baryons.
In the presence of annihilating particles, the evolution of the
ionization fraction $x_e$ satisfies:
\begin{equation}
\label{eq:dxe}
\frac{d x_e}{d z} = \frac{1}{(1+z)H(z)}
\left[ R_s(z) - I_s(z) - I_X(z)  \right],
\end{equation}
where $R_s$ is the standard recombination rate, $I_s$ the ionization
rate by standard sources, and $I_X$ the ionization rate due to
 particle annihilation.

Following the seminal papers \cite{recbase}, standard recombination is described by:

\begin{equation}
   \left[ R_s(z) - I_s(z) \right]=C \times{\big[x_{\rm e}^2 n_{\rm H} \alpha_{\rm B}
 - \beta_{\rm B} (1-x_{\rm e})
   {\rm e}^{-h_{\rm P}\nu_{2s}/k_{\rm B}T_{\rm b}}\big]}.
\end{equation}

Here $n_H$ is the number density of hydrogen nuclei, $\alpha_{\rm B}$ and $\beta_{\rm B}$ are
the effective recombination and photo-ionization rates for principal
quantum numbers $\ge 2$ in Case B recombination, $\nu_{2s}$ is the frequency of the $2s$ level from
the ground state, $T_b$ is the temperature of the baryonic gas, and the factor $C$ is given by:

\begin{eqnarray}
C= \frac{\big[1 + K \Lambda_{2s1s} n_{\rm H}(1-x_{\rm e})\big]}
 {\big[1+K \Lambda_{2s1s} n_{\rm H} (1-x_{\rm e})
 + K \beta_{\rm B} n_{\rm H}(1-x_{\rm e})\big]}.
\label{eq:standard_xe}
\end{eqnarray}
Here  $\Lambda_{1s2s}$ is
the decay rate of the metastable $2s$ level, $n_{\rm H}(1-x_{e})$ is
the number of neutral ground state $H$ atoms, and $K=\lambda_\alpha^{3}/ (8\pi H(z))$,
with $H(z)$ the Hubble expansion factor at redshift $z$ and $\lambda_{\alpha}$ the wavelength of the Lyman-$\alpha$
transition from the $2p$ level to the $1s$ level.

The $I_X$ term  of Eq. \ref{eq:dxe}
represents the contribution to the electron fraction evolution
from a ``non-standard'' source; in our case
it takes into account that during recombination DM
annihilations increase the ionization rate both by direct ionization
from the ground state, and by contributing additional Lyman-$\alpha$ photons.
The latter boosts the population at $n= 2$, increasing the rate of
photoionization by the CMB from these excited states. Therefore, the ionization rate due to particle annihilation is:
\begin{equation}
I_X(z) = I_{Xi}(z) + I_{X\alpha}(z) ,
\label{terms}
\end{equation}
where $I_{Xi}$ is the ionization rate due to ionizing photons, and $I_{X\alpha}$
the ionization rate due to additional Lyman-$\alpha$ photons.

Each of the terms in Eq. \ref{terms} is related to the rate of energy release as:

\begin{eqnarray}
I_{Xi}&=&\phantom{(1-}\;\, \phantom{)}\; \chi_i\frac{[dE/dt]}{n_H(z) E_i} \\
I_{X\alpha} &=&(1-C)\; \chi_\alpha\frac{[dE/dt]}{n_H(z) E_\alpha}
\end{eqnarray}
 where $E_i$ is the average ionization energy per baryon, $E_\alpha$ is the difference
in binding energy between the $1s$ and $2p$
energy levels of a hydrogen atom, $n_H$ is the number density of hydrogen nuclei and
$\chi_i,\chi_\alpha$ are the fractions of energy going to ionization and to
Lyman-$\alpha$ photons respectively.

Finally, a fraction of the energy released by annihilating particles goes into heating of the baryonic gas, adding 
an extra $K_h$ term in the standard evolution equation for the matter temperature $T_b$:

\begin{eqnarray}
(1+z)\frac{dT_b}{dz}&=&\frac{8\sigma_T a_R T_{CMB}^4}{3m_e
c H(z)}\frac{x_e}{1+f_{\rm He}+x_e} (T_b -T_{CMB})\nonumber \\
&& -\frac{2}{3
k_B H(z)} \frac{K_h}{1+f_{\rm He}+x_e} +2 T_b,
\end{eqnarray}
where the non standard term is given by:
\begin{equation}
K_h=\chi_h \frac{(dE/dt)}{n_H(z)}
\end{equation}
and $\chi_h$ is the fraction of energy going into heat.
In Section \ref{sec:fractions} we will describe how the $\chi_x$ fractions are calculated, and the uncertainties in their derivation that might impact the modeled effect of dark matter annihilation. 

The formalism presented here works under the assumption that it is possible to split the study of the cascade, from the DM mass scale down to the absorption channels, into two independent steps.
In the first step, the propagation of the shower is followed from the injection at the DM mass scale (GeV/TeV) down to the $\sim$keV scale (the exact energy threshold will be discussed in Section \ref{sec:sourcekev} and \ref{sec:higheng}); below this energy, all cooling processes are fast relative to a Hubble time, so once a particle's energy is degraded below this threshold it can no longer free-stream to the present day, or lose a significant fraction of its energy to redshifting before being absorbed (with the exception of photons below 10.2 eV). The fraction of the initial energy reprocessed into particles below this threshold is expressed through the $f(z)$ function in Eq. \ref{enrateselfDM}. In the second step, the copious particles degraded down to this atomic energy scale are followed, and the re-partition of their energy into heating, excitation and ionization channels (expressed by the $\chi_x$ functions) is studied in detail. We begin by examining the calculation of the $\chi_x$ functions in the following Section, while we will study the uncertainties of the $f(z)$ function in Section \ref{sec:higheng}.

\section{Calculation of the energy fractions}
\label{sec:fractions}
To assess how the ``deposited'' energy fraction $f(z)$ is divided between the different channels of heating, ionization and excitation, most of the literature today -- and for the past twenty years -- has adopted the values for $\chi_x$ taken from \cite{ShullVanSteen1985}. These results were obtained through Monte Carlo computations that followed in detail the secondary energy cascade generated by an electron of energy $E_\mathrm{in} \sim$~keV injected into a gas with $x_e \equiv n\mbox{(HII}\mbox{)}/n\mbox{(H)}\equiv n\mbox{(HeII}\mbox{)}/n\mbox{(He)}$. The code returned the fractions of the initial energy absorbed by the gas as heat, ionizations and excitations (i.e. the fractional energy depositions $\chi_h$, $\chi_i$ and $\chi_e$ respectively).

In recent years the problem of the energy depositions was tackled with considerable improvements over the seminal work by \cite{ShullVanSteen1985}: \cite{Valdes:2008cr, Furlanetto:2009uf} used updated cross sections and implemented new physical processes; \cite{Valdes:2009cq} extended the energy range of the primary electron up to $E_\mathrm{in} \sim$ 1 MeV $-$ 1 TeV; \cite{Evoli:2012zz} computed the cascade from the distributions of lepton and photons produced by the annihilation of selected DM candidates.

All the aforementioned works adopt a similar Monte Carlo scheme: for every particle, either primary or secondary, the code calculates the various interaction probabilities and samples among them via a random number generator. The considered physical processes when dealing with a primary electron of $E_\mathrm{in} \sim$~keV are the following: H, He, HeI ionization; H, He excitation; collisions with thermal electrons; free-free interactions with ionized atoms; recombinations. The Monte Carlo code degrades the energy of the particles down to $\sim$~eV energies and then provides the fractional energy depositions. Notice that in this energy range ($E<$10keV) it is a good approximation to consider that the whole cascade happens ``on the spot". The timescales of the two most relevant interactions, i.e. Coulomb scattering and collisional ionization, are in fact much smaller than a Hubble time for the considered values of $x_e$, therefore the primary electron energy is deposited locally in a short timescale (see \cite{Furlanetto:2009uf}). This approximation cannot be considered valid either for higher primary energies of order $\sim$~MeV or if the local value of the ionized fraction changes rapidly (e.g. around sources of ionizing radiation).

After a collisional ionization event the code has one extra secondary electron to follow, while collisional excitations off H or He atoms give line photons. If a He line photon has an energy higher than 13.6 eV it photo-ionizes an H atom in a short timescale. H excitations to levels $n > 2$ will generally cascade through $n=2$ and emit Ly$\alpha$ photons. Electron-electron collisions on the other hand transfer heat to the gas. When the energy of an electron is degraded to less than 10.2 eV then both collisional excitations and ionizations become impossible and the electron deposits all of its remaining energy into the gas as heat. When a photon has energy below 10.2 eV it will not interact further with the gas.

To compute the interaction probabilities for a given electron it is necessary to compute first the cross sections for the possible interactions. Then the mean free path is calculated by using the relative abundances of the target particles. The probability of interaction is then simply given by the inverse of the mean free path. The calculation is insensitive to H density, the important parameters are the relative abundance of He and the free electron fraction $x_e$: intuitively, an electron injected in a gas with a higher $x_e$ will have a higher probability of Coulomb interactions, and therefore the fractional energy depositions will present a higher $\chi_h$.

While the works cited above share this common Monte Carlo numerical structure, they present some crucial differences of implementation. In this work in particular we use a modified version of the code described in \cite{Valdes:2008cr} with updated interaction cross sections (see \cite{Valdes:2009cq, Evoli:2012zz}) and a separate treatment for hydrogen and helium which allows us to use as input different values of their ionized fractions and compute the fractional energy depositions for the processes relative to the two species. Another crucial improvement over \cite{ShullVanSteen1985} is that we are able to classify photons produced by collisional excitations into two main categories, Ly$\alpha$ or ``continuum'' if their energy is below 10.2 eV, as is the case, e.g. of photons produced by the two-photon forbidden transition $2s \rightarrow 1s$, previously neglected.

We perform a number of runs using realistic fractions for ionized hydrogen and helium ($x_{\rm HII}=n\mbox{(HII}\mbox{)}/n\mbox{(H)}$ and $x_{\rm HeII}=n\mbox{(HeII}\mbox{)}/n\mbox{(He)}$) rather than using the standard assumption that $x_e \equiv x_{\rm HII} \equiv x_{\rm HeII}$. In particular, we have calculated the fractions for different values of $x_{\rm HII}$ fixing the amount of ionized helium at different values, i.e. $x_{\rm HeII}=1\times10^{-10},1\times10^{-5},1\times10^{-4},1\times10^{-3}$. The choice of these values is motivated by the fact that in the redshift range of interest for CMB studies, $100 \lesssim z \lesssim 1500$, the  ionization of Helium in a standard recombination history is below $1\times 10^{-9}$ at redshift  $z\lesssim 1500$. However, in the presence of dark matter annihilation, ionization of helium might  be higher even at lower redshifts. We have calculated that even for an unrealistic level of dark matter annihilation, corresponding to a constant annihilation parameter of $p_\mathrm{ann}=1.78\times10^{-26}$cm$^3$/s/GeV\footnote{In the literature, $p_\mathrm{ann}$ has often been expressed in units of m$^3$/s/kg. In this work, we convert to cm$^3$/s/GeV units for convenience; $p_\mathrm{ann}=1.78\times10^{-26}$cm$^3$/s/GeV corresponds to $1\times10^{-5}$m$^3$/s/kg.} already strongly ruled out by current data (see e.g. \cite{Giesen2012,Galli:2011rz}), the level of ionized Helium does not exceed $5\times10^{-3}$. This justifies the range of values for $x_{\rm HeII}$ we choose to study (see Section \ref{sec:xhe}). 

Moreover, to study the possible deviations in the fractional energy depositions produced by a smaller/larger amount of Helium in the IGM, we perform three series of runs for different values of the helium fraction by mass, ${\rm Y_p} = 0.21, 0.24, 0.27$, where the median value is inferred from WMAP data (\cite{Pisanti:2007hk}) (see Section \ref{sec:helium}) .
For details on the cross sections we refer the reader to \cite{Valdes:2008cr,Valdes:2009cq, Evoli:2012zz} and references therein.

For each assumed cosmological evolution of $x_\mathrm{HII}$ and $x_\mathrm{HeII}$, and for each value of ${\rm Y_p}$ we performed 100 Monte Carlo realizations, a number sufficient to produce consistent results not biased by the random nature of the computation, as determined in the aforementioned studies. The numerical outputs for the fractional energy depositions can be found in tabulated form in Appendix \ref{app:engfracs}.

\section{Impact of the systematics on the CMB}
\label{sec:systematics}

In this Section we study how the variation of
{\it i)} the level of Helium ionization, and {\it ii)} the Helium abundance, affects the calculated energy fractions, and thus how misestimation of these quantities could modify the computed effect of DM annihilation on the recombination history and CMB power spectra. We also study {\it iii)} the impact of changing the energy of the original (keV-scale) electron, for the Monte Carlo code described in Section \ref{sec:fractions}.
Before showing these results however, we briefly describe the codes used to calculate the recombination histories, the CMB power spectra  and the constraints we will present in the following sections.

The evolution of the free electron fraction will be calculated using a modified version of the RECFAST code (v1.5) \cite{recfast}, while CMB spectra are calculated with CAMB\footnote{\url{http://camb.info/}, CAMB version $\rm Jan\_12$.}\cite{camb} (high\_accuracy\_default option always switched on). As a benchmark, we will assume dark matter annihilation with a constant annihilation parameter $p_\mathrm{ann}=1.78\times10^{-27}$cm$^3$/s/GeV. This choice is motivated by the fact that this value of $p_\mathrm{ann}$ is about the $2\sigma$
upper bound derived from current CMB data \cite{Giesen2012,Galli:2011rz}. It thus represents the strongest annihilation signal still allowed by current data that future CMB experiments might detect. As a fiducial model for the other $\Lambda$CDM parameters, we use the WMAP7 marginalized values \cite{wmap7}.

Whenever needed, we determine constraints on DM annihilation parameters using the publicly available Markov Chain Monte Carlo package \texttt{cosmomc} \cite{cosmomc}.
Together with $p_\mathrm{ann}$ we sample the following six-dimensional set of cosmological parameters,
adopting flat priors on them:
the physical baryon and CDM densities, $\omega_b=\Omega_bh^2$ and
$\omega_c=\Omega_ch^2$, the scalar
spectral index, $n_{s}$
the normalization, $\ln10^{10}A_s(k=0.002/Mpc)$,
the optical depth to reionization, $\tau$, and the Hubble constant $H_0$.

We  consider purely adiabatic initial conditions.
The MCMC convergence diagnostic tests are performed on $4$ chains using the
Gelman and Rubin ``variance of chain mean''$/$``mean of chain variances'' $R-1$
statistic for each parameter. Our  constraints are obtained
after marginalization over the remaining ``nuisance'' parameters, again using
the programs included in the \texttt{cosmomc} package.
We use a cosmic age top-hat prior of 10 Gyr $ \le t_0 \le$ 20 Gyr.
As a baseline, we assume
mock data for a \emph{Planck}-like experiment as described in \cite{planckbluebook},  with a fiducial model
given by the best fit WMAP7 model with standard recombination, and assuming either $p_\mathrm{ann}=0$ or $p_\mathrm{ann}=1.78\times 10^{-27}$ cm$^3$/s/GeV. In the latter case, we calculated the mock data with the energy fractions computed for $Y_p=0.24$ and $x_\mathrm{HeII}=1\times10^{-10}$. We model the  \emph{Planck}-like experimental noise as (see \cite{planckbluebook}):
\begin{equation}
N_{\ell} = \left(\frac{w^{-1/2}}{\mu{\rm K\mbox{-}rad}}\right)^2
\exp\left[\frac{\ell(\ell+1)(\theta_{\rm FWHM}/{\rm rad})^2}{8\ln 2}\right],
\end{equation}
with $w^{-1}=1.5\times 10^{-4} \mu K^2$ as the temperature noise level
(we consider a factor $\sqrt{2}$ larger for polarization
noise) and  $\theta_{\rm FWHM}=7.1'$ for the beam size.
We take $f_\mathrm{sky}=0.85$ as sky coverage. 

\subsection{Level of Helium ionization}
\label{sec:xhe}
\begin{figure}[tb!]
\begin{center}
\includegraphics[width=0.5\textwidth]{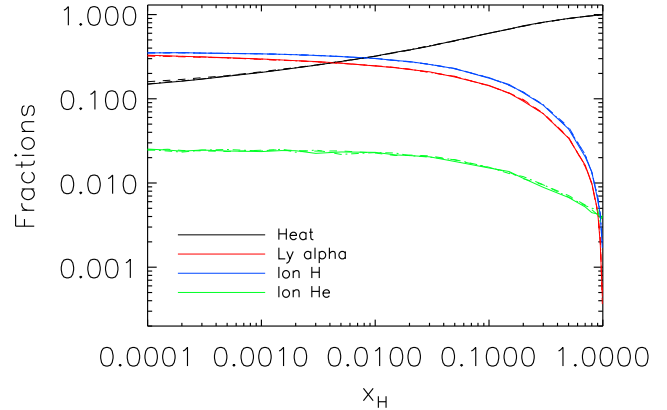}
\caption{Energy fractions for heat (black), Lyman-$\alpha$ (red), ionization of hydrogen (blue) and first ionization of helium (green), from top to bottom respectively. We plot the fractions as a function of $x_\mathrm{H}$ for different fixed values of the Helium ionization fraction $x_\mathrm{He}$. For clarity, we only plot the curves for the two extreme values of $x_\mathrm{He}$ described in the text, $x_\mathrm{He}=1\times10^{-10}$ (solid line) and $x_\mathrm{He}=1\times10^{-3}$ (dashed line). }
\label{plot_fractionsxhe}
\end{center}
\end{figure}

As anticipated and discussed in Section \ref{sec:fractions}, we have calculated the energy fractions for different values of the hydrogen ionization $x_{\rm HII}$ fixing the amount of ionized helium at different values, i.e. $x_{\rm HeII}=1\times10^{-10},1\times10^{-5},1\times10^{-4},1\times10^{-3}$.
Figure \ref{plot_fractionsxhe} shows the energy fractions as a function of $x_\mathrm{HII}$  for fixed values of the helium ionization $x_\mathrm{HeII}$. The different cases differ by a very small amount, at the level of a few percent for all the energy fractions, and are largely affected by numerical noise.

\begin{figure}[tb!]
\begin{center}
\includegraphics[width=0.5\textwidth]{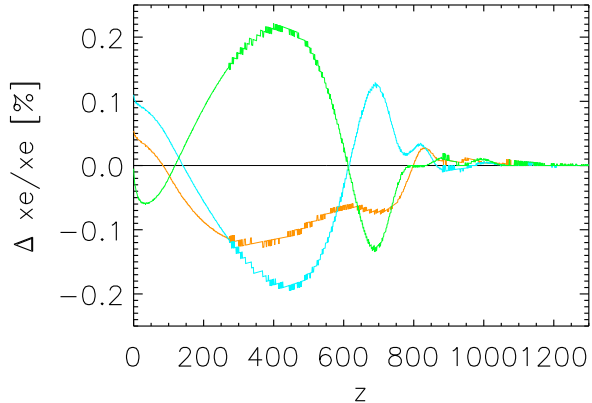}
\caption{Percentage difference, as a function of redshift, between (i) the evolution of the free electron fraction calculated with $p_\mathrm{ann}=1.78\times 10^{-27} \mathrm{cm}^3\mathrm{/s/GeV}$ and using energy fractions calculated assuming a fixed value of the Helium ionization fraction, $x_\mathrm{He} = 1\times10^{-5}$ (orange), $1\times10^{-4}$ (cyan), $1\times10^{-3}$ (green). 
and (ii) the evolution of the free electron fraction calculated assuming $p_\mathrm{ann}=1.78\times 10^{-27} \mathrm{cm}^3\mathrm{/s/GeV}$ and $x_\mathrm{He}=1\times10^{-10}$ (we show $(\rm xe_i-xe_{ii})/\mathrm{xe}_{ii}$).}
\label{plot_fractionsxhe_xe}
\end{center}
\end{figure}

We have then estimated the impact of using these different fractions to calculate the effect of dark matter annihilation on the recombination history. Figure \ref{plot_fractionsxhe_xe} shows  the percentage differences in the evolution of the free electron fraction when dark matter annihilation is present at the level of $p_\mathrm{ann}=1.78\times 10^{-27}\mathrm{cm}^3\mathrm{/s/GeV}$ and using i) $x_{\rm HeII}=1\times10^{-5},1\times10^{-4}$ or $1\times10^{-3}$, compared to using  fractions for ii) $x_{\rm HeII}=1\times10^{-10}$.

The differences between the resulting recombination histories are in all cases less than $0.2\%$. We have verified that the ensuing differences in the CMB power spectra are at the $\ll 0.1\%$ level and are dominated by numerical noise. These are negligible effects, as the current accuracy of CAMB is at the $\sim 0.1\%$ level.
We therefore conclude that the energy fractions to be used in a cosmological recombination context can be safely calculated by fixing the amount of helium ionization to a small value. In the following, we will assume  $x_{\rm HeII}=1\times10^{-10}$.

\subsection{Helium abundance $Y_\mathrm{He}$}
\label{sec:helium}
\begin{figure}[h!]
\begin{center}
\includegraphics[width=0.5\textwidth] {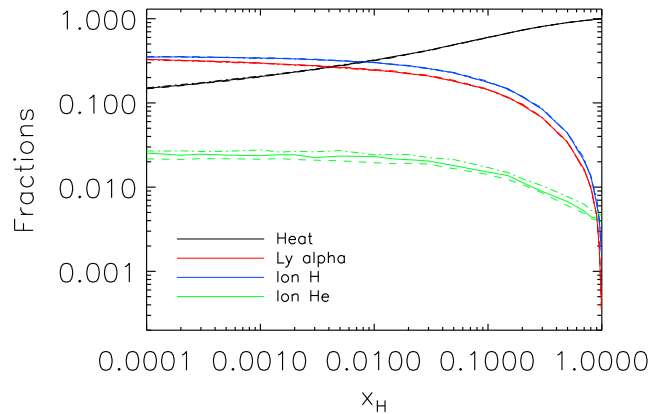}
\caption{Energy fractions for heat (black), Lyman-$\alpha$ (red), ionization of hydrogen (blue) and first ionization of Helium (green).We plot the fractions in function of $x_\mathrm{H}$ for different values of the Helium abundance, ${\rm Y_p}=0.21$ (dashed line), $0.24$ (solid line), $0.27$ (dot-dashed line). }
\label{plot_Yhe}
\end{center}
\end{figure}

The sensitivity of ongoing and upcoming CMB data from experiments such as \emph{Planck} \cite{planckgen}, ACTpol \cite{actpol} and SPTpol \cite{sptpol} will give tight constraints on the fraction of
$\he4$ present in the Universe at the time of recombination, providing a probe of primordial nucleosynthesis before any possible
enrichment by astrophysical sources.
The primordial $\he4$ fraction, usually
expressed through the proxy mass fraction
{\rm Y$_p$}$\equiv 4 X_\he4$ (with
$X_\he4$ being the particle fraction with respect
to hydrogen), is expected to be, from standard nucleosynthesis,
{\rm Y$_p$}=0.2467$\pm$0.0003 \cite{Pisanti:2007hk}, using WMAP7 values for
$\omega_b$. On the other hand, a combination of \emph{Planck}, ACT and SPT data currently provides an independent constraint of {\rm Y$_p$}=0.266$\pm$0.021 \cite{PlanckCollaboration:2013}.
s
To accommodate the possibility that future CMB experiments will measure a primordial Helium abundance different to that expected from BBN predictions, we study here the effect of different Helium abundances on the energy fractions. We have scanned a large range of parameters, from
{\rm $Y_p$}=0.21 up to  {\rm $Y_p$}=0.27 in order to probe even the most
unexpected of scenarios, and calculated the resulting energy fractions, fixing the ionization of Helium to $x_\mathrm{HeII}=1\times 10^{-10}$.
Results are presented in Figure \ref{plot_Yhe}.

\begin{figure}[h!]
\begin{center}
\includegraphics[width=0.5\textwidth] {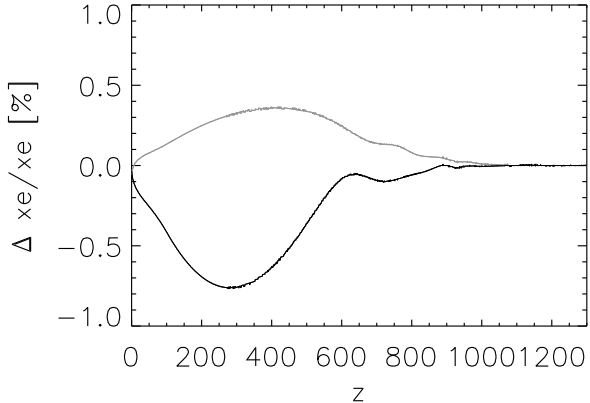}
\caption{Percentage difference in recombination histories for dark matter annihilation with $p_\mathrm{ann}=1.78\times 10^{-27}\mathrm{cm}^3\mathrm{/s/GeV}$, calculated using the energy fractions assuming Helium abundance equal to (i) ${\rm Y_p}=0.21$ (grey) or $0.27$ (black) compared to assuming (ii) ${\rm Y_p}=0.24$ (we show $(\mathrm{xe}_{i}-\mathrm{xe}_{ii})/\mathrm{xe}_{ii}$).}
\label{fig:xe_yhe}
\end{center}
\end{figure}

\begin{figure}[h!]

\begin{center}
\includegraphics[width=0.5\textwidth]{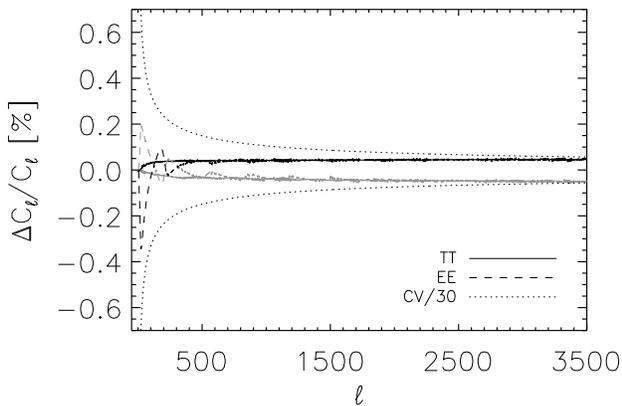}
\caption{Percentage difference in CMB angular power spectra for dark matter annihilation with $p_\mathrm{ann}=1.78\times 10^{-27}\mathrm{cm}^3\mathrm{/s/GeV}$, calculated using the energy fractions that assume Helium abundance equal to (i) ${\rm Y_p}=0.21$ (grey) or $0.27$ (black) compared to (ii) ${\rm Y_p}=0.24$ (we show $(C_l^{i}-C_l^{ii})/C_l^{ii}$). We plot both the cases for temperature (solid lines) and EE polarization (dashed lines). For comparison, the dotted black line shows cosmic variance, rescaled by a factor $30$ (we thus show $[\Delta C_l/C_l]^{CV}/30$).}
\label{fig:TTEE_yhe}
\end{center}
\end{figure}

Changing the helium abundance has a small impact on the energy fractions, with less than a $\lesssim 5\%$ effect on the heating and ionization fractions, and $\lesssim 20\%$ on the Helium ionization fraction. Again as in the previous section, such small variations of the energy fractions do not significantly modify the effects of dark matter annihilation on the recombination history, as shown in Figure \ref{fig:xe_yhe}. As a consequence, the effect on CMB power spectra is negligible, at least 30 times smaller than cosmic variance, as shows in Figure \ref{fig:TTEE_yhe}\footnote{In all these plots, the Helium abundance is fixed to $Y_p=0.24$, but we use energy fractions calculated for different values of $Y_p$.}.

This finding has the reassuring consequence that provided the abundance of helium inferred by future experiments is within the range suggested by the
cosmological concordance scenario, deviations from the central expected value will not affect the DM constraints (note that this agrees with \cite{Chluba:2009uv}, which studied the effect of DM annihilation on helium recombination and found it to be small).

\subsection{Energy of the original keV-scale electron}
\label{sec:sourcekev}

As described at the end of Section \ref{sec:theory}, in the formalism we adopt, the function $f(z)$ encodes the fraction of the original DM annihilation energy that is redistributed into particles below some threshold energy (typically $3$ keV). Below this threshold energy atomic processes become dominant and we switch over to the Monte Carlo code described in Section \ref{sec:fractions}. However, because the energy loss processes for relativistic electrons and photons are not purely continuous, the sub-threshold electrons produced by high-energy processes do not appear solely at the threshold energy, but in a continuum spectrum below it (this point will be elucidated further in Section \ref{sec:higheng}). Consequently, it is important to study the degree to which the energy fractions depend on the energy of the injected electron.

Since the original study of \cite{ShullVanSteen1985}, the ``low-energy band'' has been studied with a low-energy source electron of initial energy 3 keV. We are interested in the degree to which the derived energy fractions depend on this choice, and the resulting impact on the constraints. We have therefore computed the different fractions $\chi_x$
injecting ``primary'' electrons of energy $10$ keV, $3$  keV, $300$ eV, $100$ eV, $60$ eV, $30$ eV or $14$ eV (these electrons are ``primary'' for the study of energy deposition in the
atomic regime, ``secondary'' with respect to the original DM annihilation and subsequent degradation from the GeV/TeV regime).

\begin{figure}[h!]
\begin{center}
\includegraphics[width=0.5\textwidth] {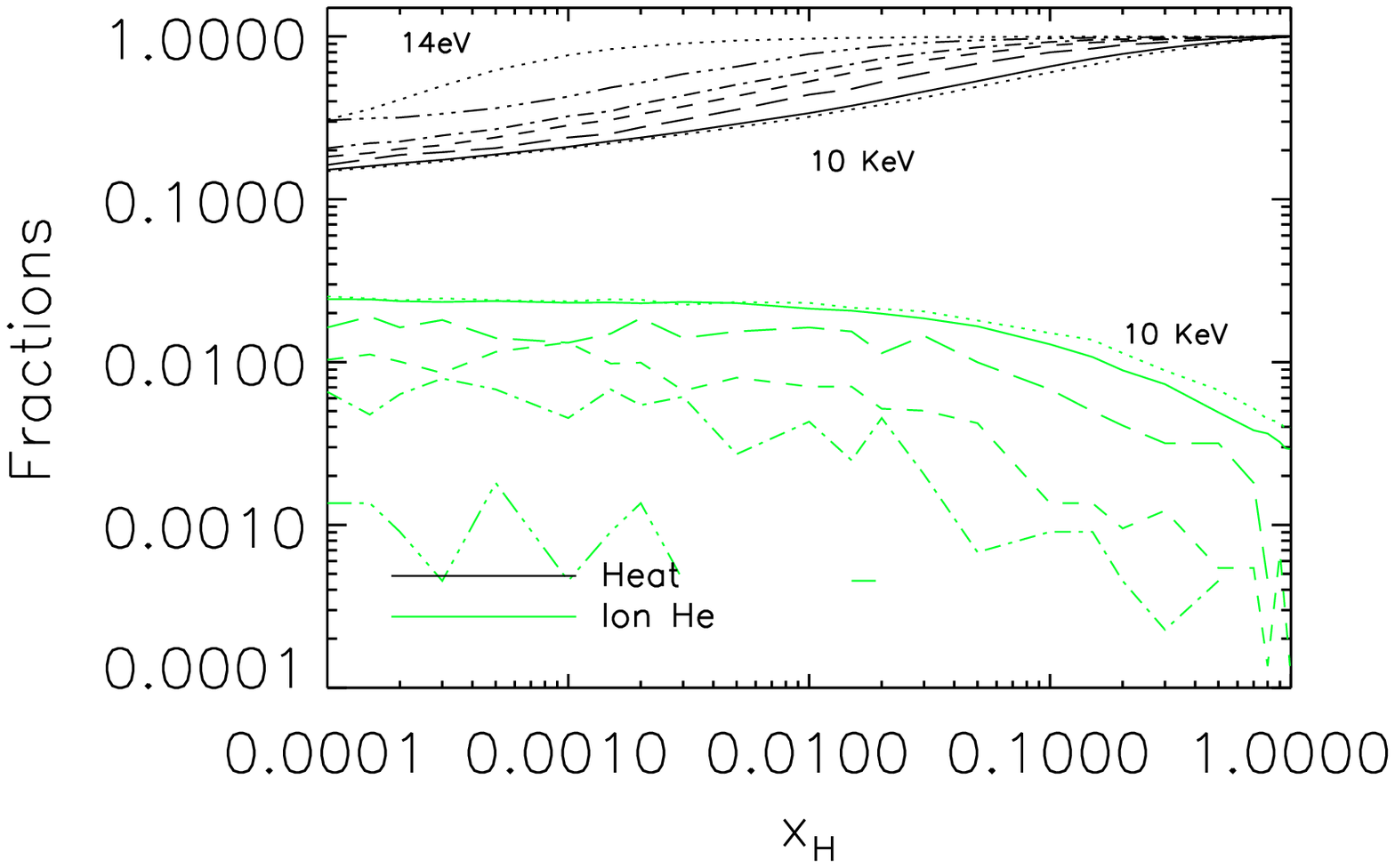}
\includegraphics[width=0.5\textwidth] {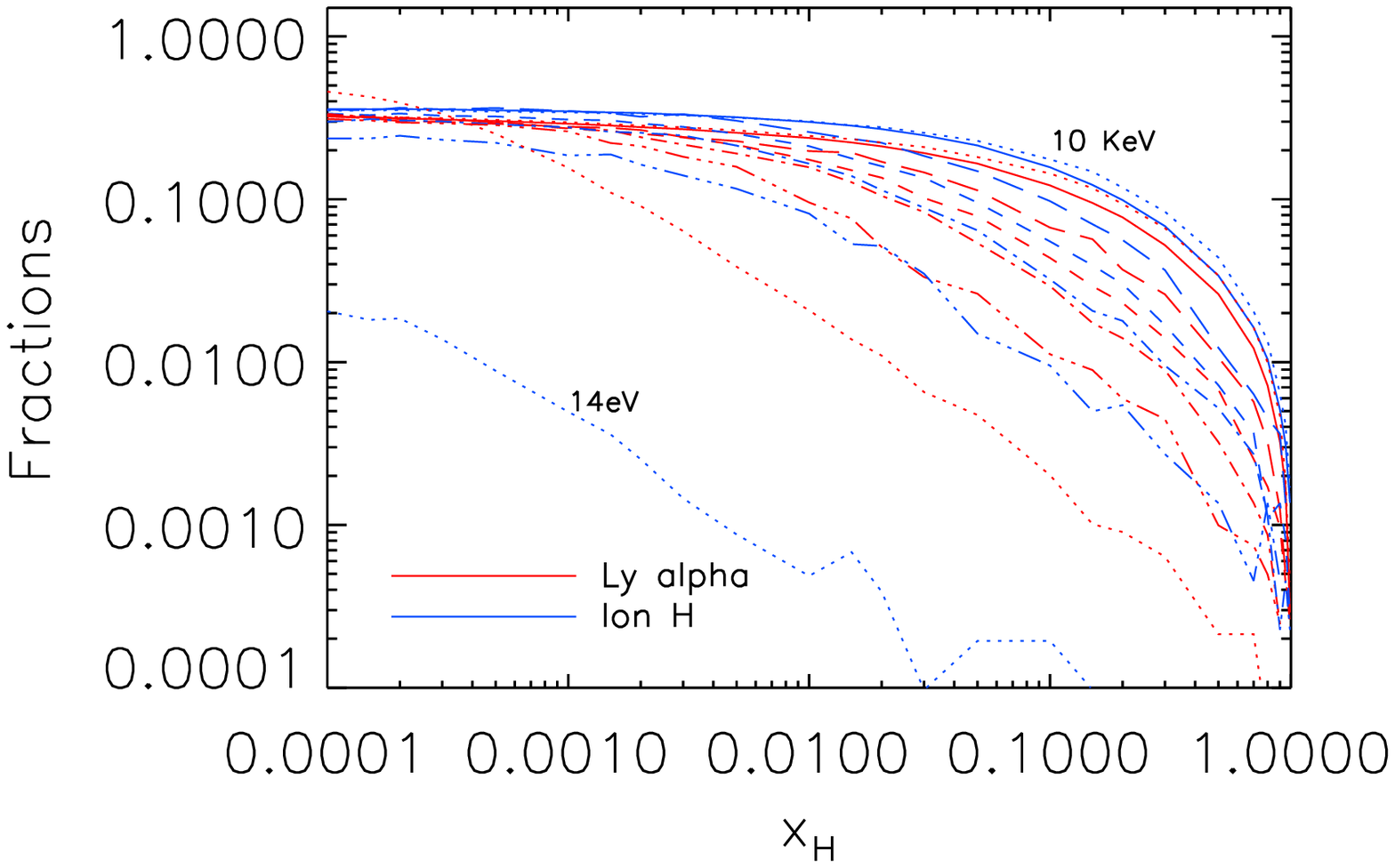}
\caption{Energy fractions for heat (black), Lyman-$\alpha$ (red), ionization of hydrogen (blue) and first ionization of Helium (green) as a function of $x_\mathrm{H}$ for different injected electron energy of $10$ keV (dotted line), $3$ keV (solid line), $300$ eV (long-dashed line),  $100$ eV (dashed line),  $60$ eV (  dot-dashed line), $30$ eV ( 3 dots-dashed line) and $14$ eV (dotted lines).}
\label{fig:injEn}
\end{center}
\end{figure}

\begin{figure}[h!]
\begin{center}
\includegraphics[width=0.45\textwidth]{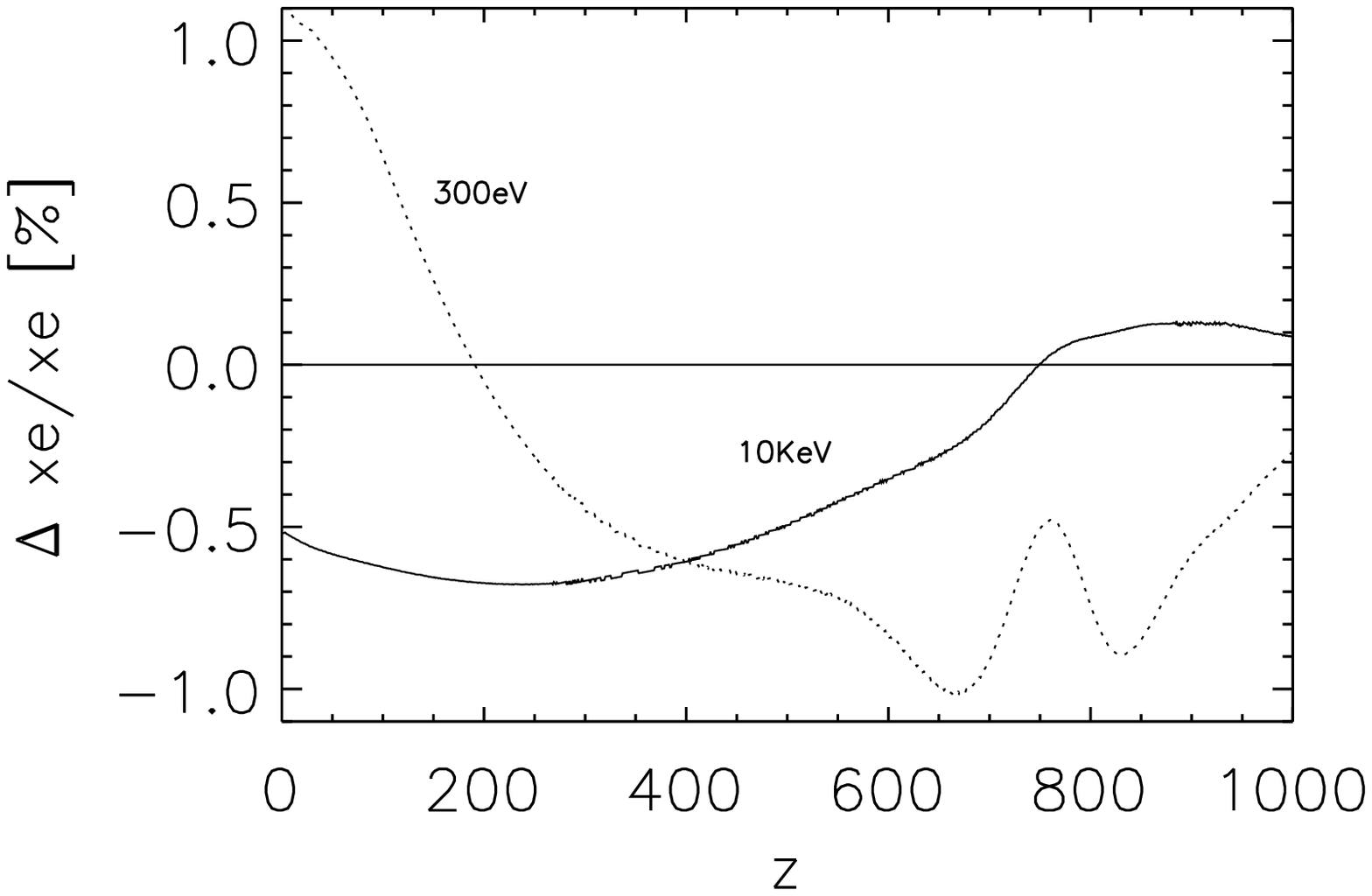}
\includegraphics[width=0.45\textwidth]{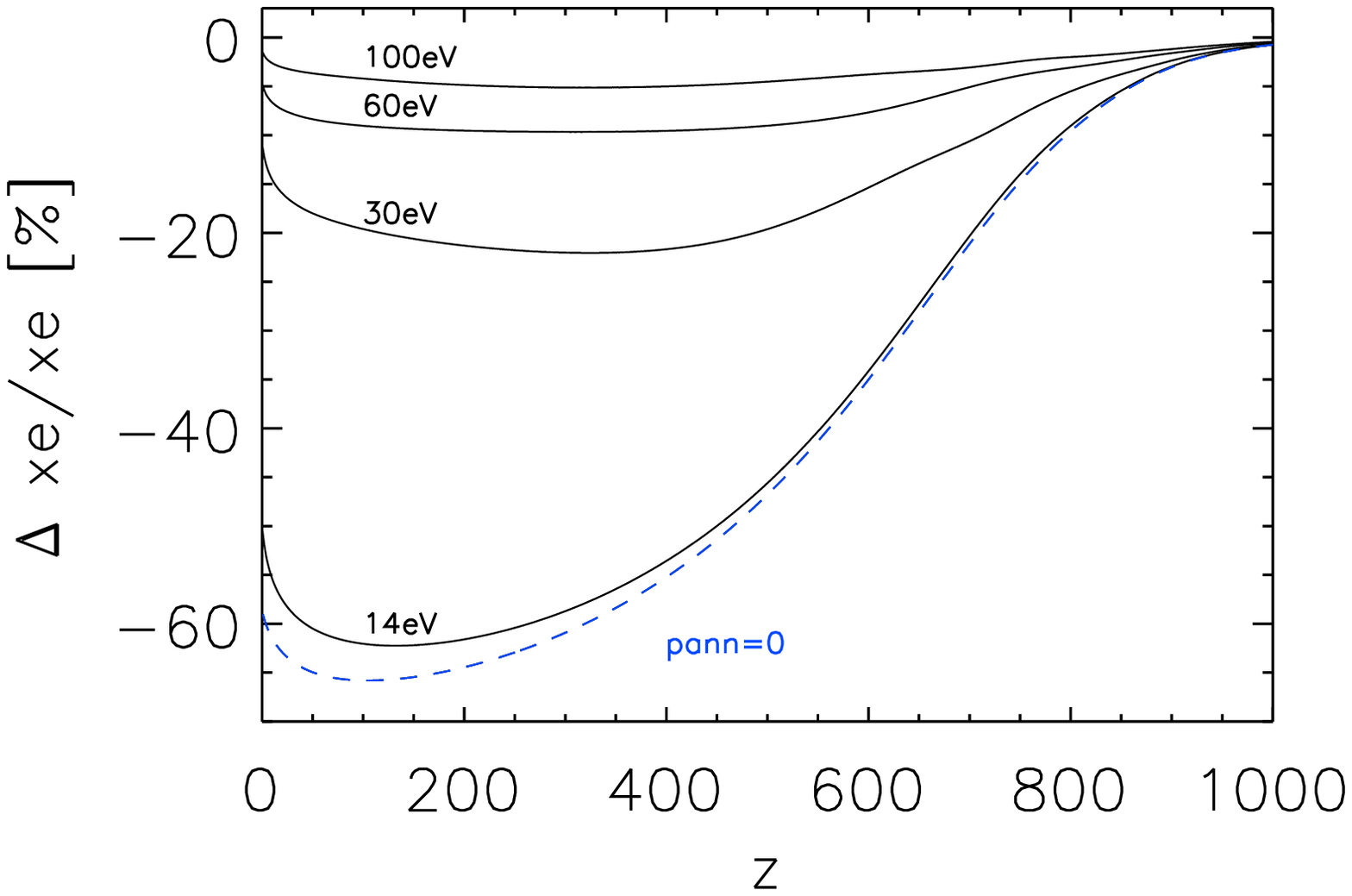}
\caption{Percentage difference in recombination histories for dark matter annihilation with $p_\mathrm{ann}=1.78\times 10^{-27}\mathrm{cm}^3\mathrm{/s/GeV}$, calculated using the energy fractions appropriate to initial electrons with energies of (i) $E_{(i)}$,compared to the case of an initial electron of (ii) $3$ keV (we show $(\mathrm{xe}_{i}-\mathrm{xe}_{ii})/\mathrm{xe}_{ii}$). In the top panel we show the cases for $E_{(i)}=$ $10$ keV or $300$ eV, while in the bottom panel we show the cases for  $E_{(i)}=$ $100$, $60$, $30$ or $14$ eV. For comparison, we also show the case of standard recombination with no dark matter annihilation ($p_\mathrm{ann}=0$) as a dashed blue line.}
\label{fig:injEnxe}
\end{center}
\end{figure}

Figure \ref{fig:injEn} shows the energy fractions for these different cases. Lowering the energy of the injected electron yields a higher heating fraction, at the expense of the ionization and Lyman-alpha fractions (in agreement with the findings of \cite{ShullVanSteen1985}). The effect on the recombination histories is shown in Figure \ref{fig:injEnxe}, where we plot the percentage difference between each of these scenarios and the case of an injected electron of energy $3$ keV, assuming the presence of dark matter annihilation corresponding to $p_\mathrm{ann}=1.78\times 10^{-27}\mathrm{cm}^3\mathrm{/s/GeV}$. 

\begin{figure}[h!]
\begin{center}
\includegraphics[width=0.5\textwidth]{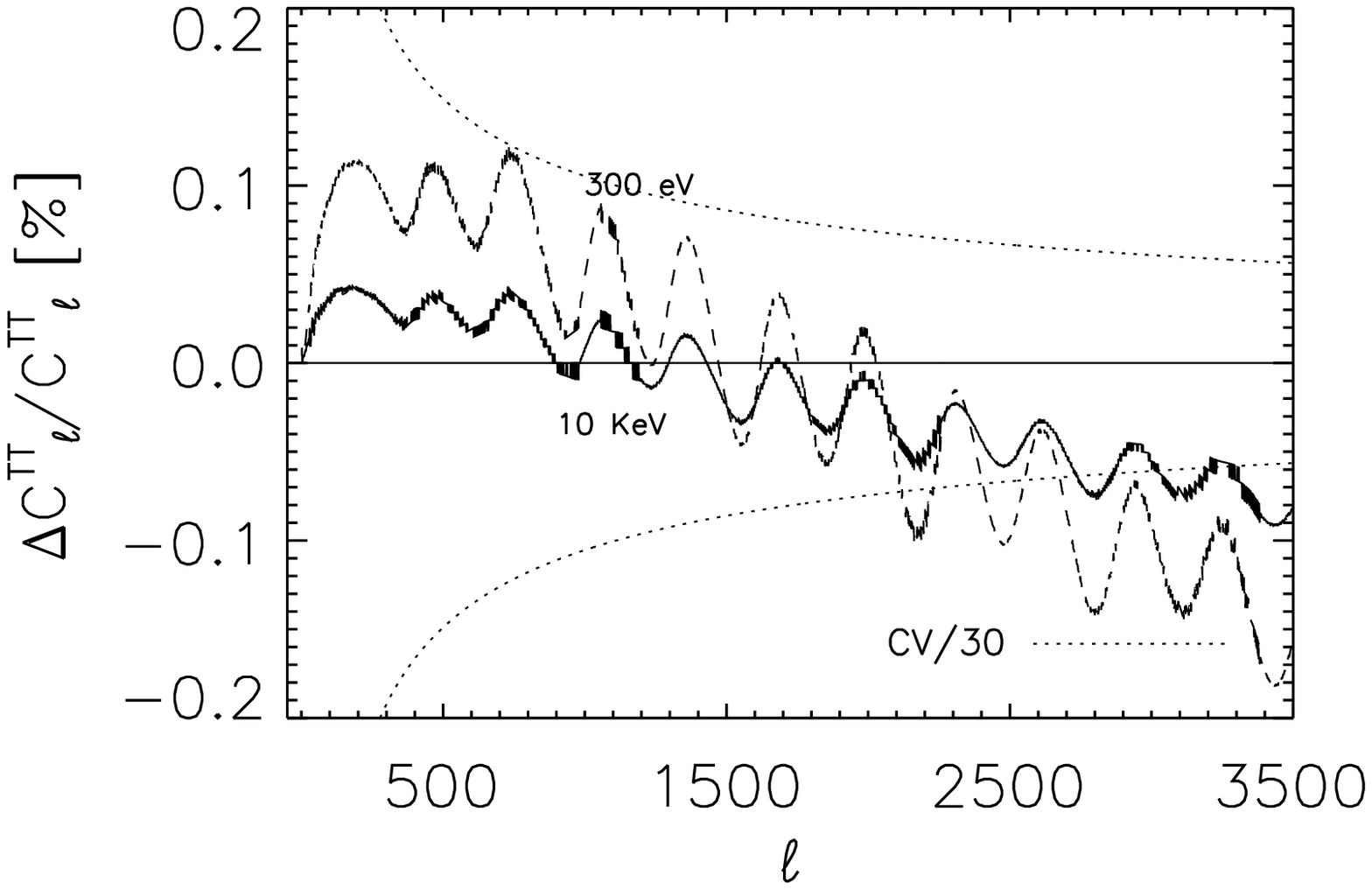}
\includegraphics[width=0.5\textwidth]{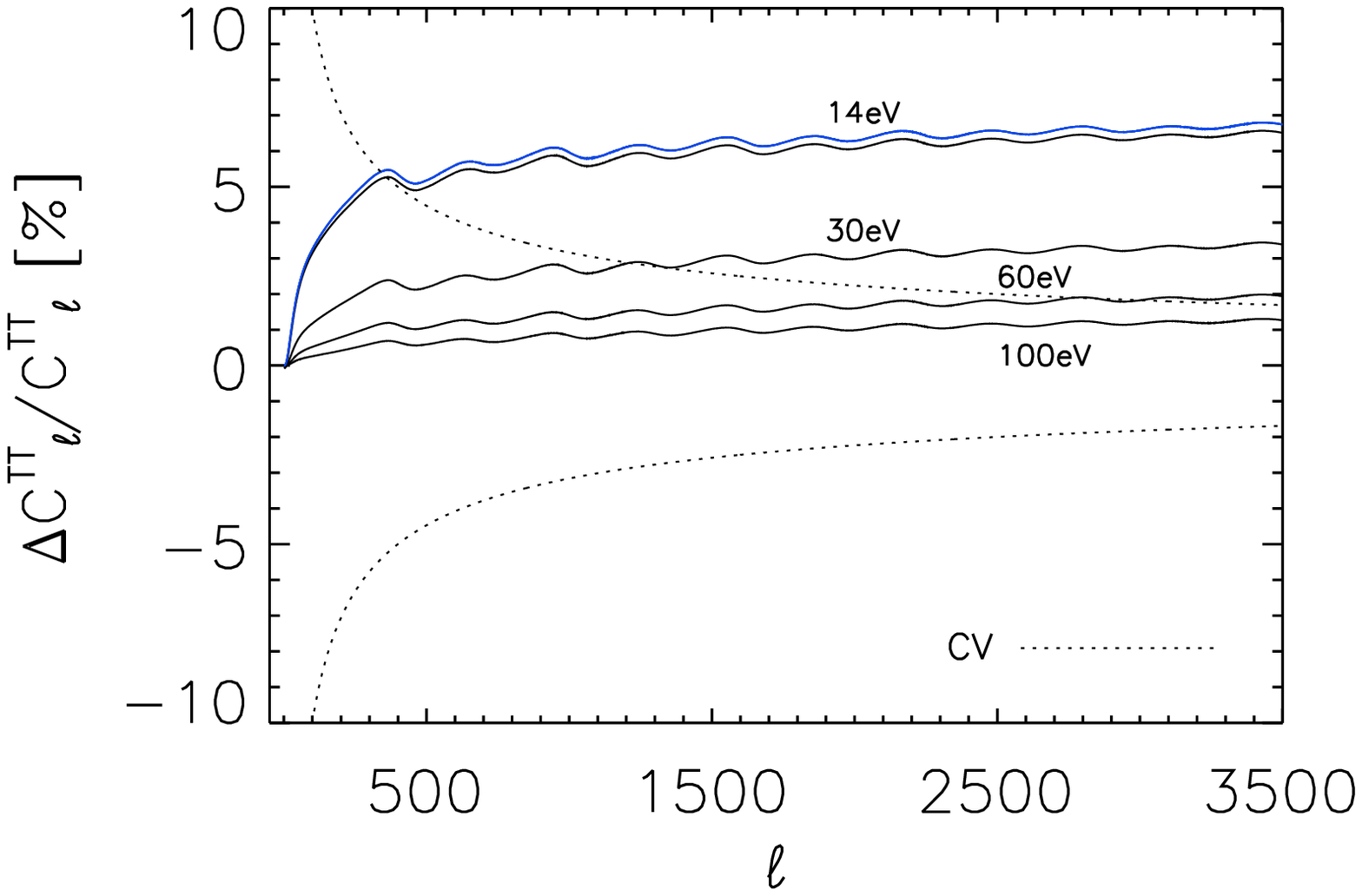}
\caption{Percentage difference in CMB TT angular power spectra for dark matter annihilation with $p_\mathrm{ann}=1.78\times 10^{-27}\mathrm{cm}^3\mathrm{/s/GeV}$, calculated using the energy fractions appropriate to initial electrons with energies of
(i) $E_{(i)}$, compared to the case of an initial electron of (ii) $3$ keV (we show $(C_l^{i}-C_l^{ii})/C_l^{ii}$). In the top panel we show the cases for $E_{(i)}=$ $10$ keV or $300$ eV, while in the bottom panel we show the cases for  $E_{(i)}=$ $100$, $60$, $30$ or $14$ eV. For comparison, we also show the case of standard recombination without dark matter annihilation ($p_\mathrm{ann}=0$) as a dashed blue line, as well as cosmic variance (dotted black line) rescaled by a factor $N=30$ $(1)$ in the top (bottom) panel (we thus show $[\Delta C_l/C_l]^{CV}/N)$.}
\label{fig:injEnTT}
\end{center}
\end{figure}

\begin{figure}[h!]
\begin{center}
\includegraphics[width=0.5\textwidth]{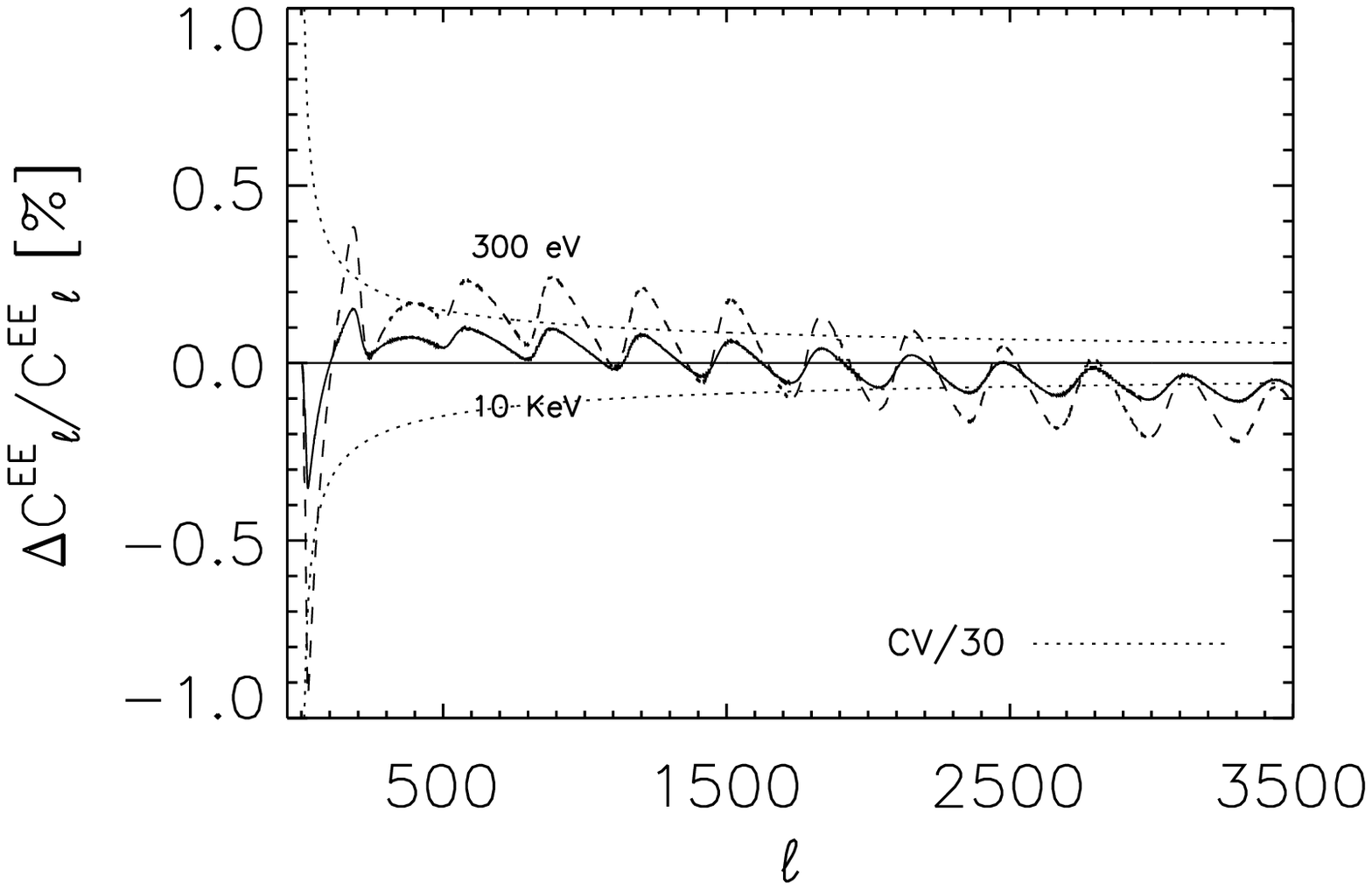}
\includegraphics[width=0.5\textwidth]{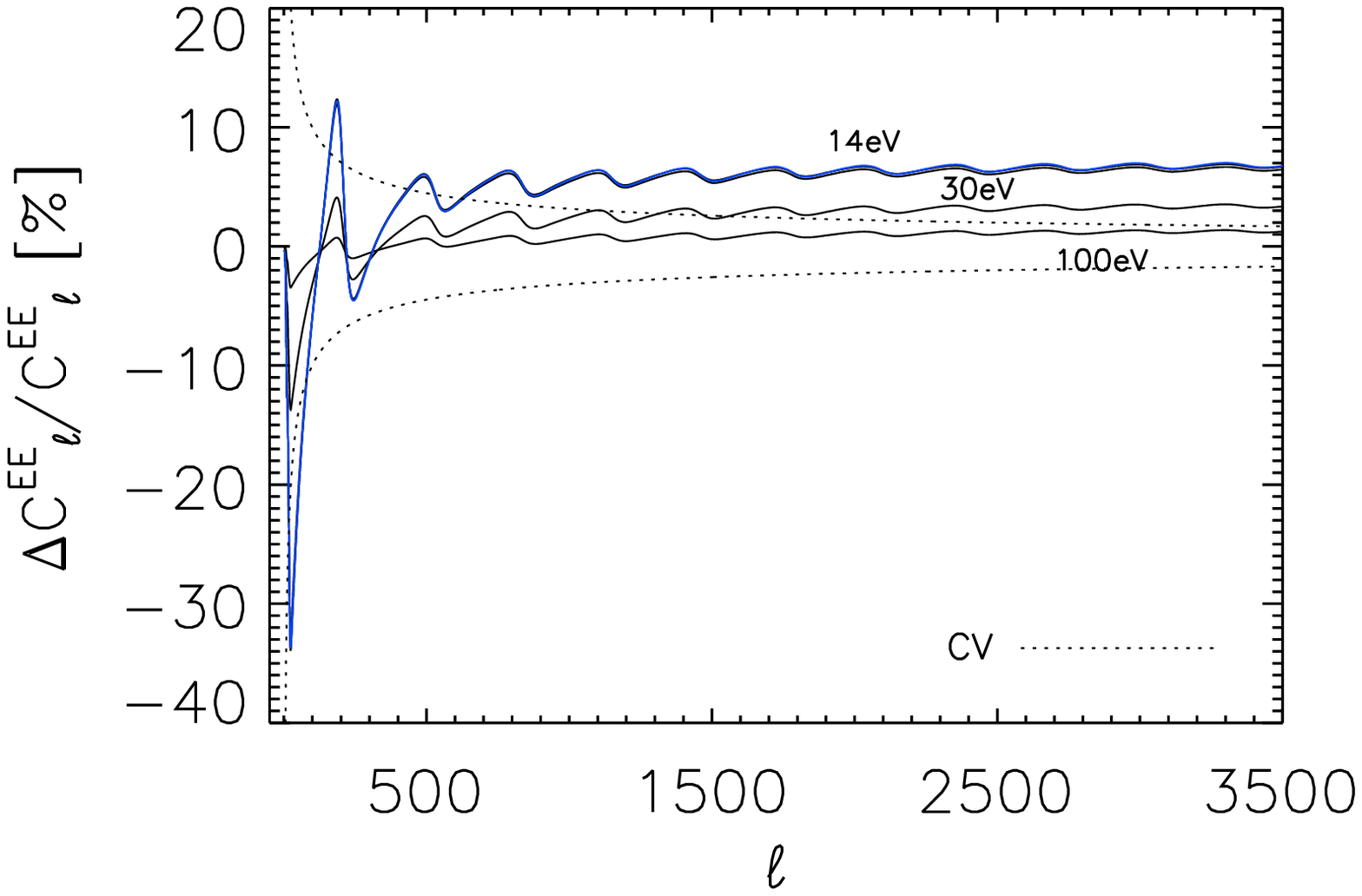}
\caption{Same as Figure \ref{fig:injEnTT}, but for EE polarization angular power spectra. For clarity, the $60$ eV case is not shown here as it is very close to the $100$ eV case.}
\label{fig:injEnEE}
\end{center}
\end{figure}

We find that the differences in recombination histories for the $10$ keV or the $300$ eV cases compared to the $3$ keV case are tiny, never exceeding 1.5\%, as shown in the top panel of Figure \ref{fig:injEnxe}. In the most important redshift range for the constraints (at $z \sim 600$, as identified from a PCA analysis,  \cite{Finkbeiner:2011dx}) the differences are less than 1\%. 
At the power spectrum level, this translates in differences smaller than $0.2\%$ for temperature (top panel of Fig. \ref{fig:injEnTT}) and $0.4\%$ for polarization, with a peak of $\sim 1\%$ at $l=24$ (top panel of Fig. \ref{fig:injEnEE}). These changes are about $20-30$ times smaller than cosmic variance, and we will show below that they do not impact constraints.

The situation is however different for electrons of lower energies. Differences between the $100$ eV  and the $3$ keV case are still small, but then rapidly grow as the energy of the injected   electron is lowered. In our most extreme case, that of an injected electron of $14$ eV, almost all the energy is absorbed as heat, and the ionization and excitation fractions are almost negligible. As a consequence, differences in the recombination history with the $3$ keV case are extremely large, comparable to the overall difference between a standard scenario with no dark matter annihilation and the $3$ keV case with $p_\mathrm{ann}=1.78\times 10^{-27}\mathrm{cm}^3\mathrm{/s/GeV}$, as also shown in the bottom panel of Fig. \ref{fig:injEnxe}. The resulting percentage differences in TT and EE power spectra are then shown in the bottom panels of Fig. \ref{fig:injEnTT} and \ref{fig:injEnEE}.

In order to assess whether and how much these differences impact the constraints on dark matter annihilation, we use the \texttt{cosmomc} code and the procedure described in Section \ref{sec:systematics}. We simulate mock-data for a \emph{Planck}-like experiment assuming standard recombination history with $p_{ann}=0$, and calculate the constraints using the energy fractions relative to different electron injection energies. Table \ref{tab:constraints} shows the $95\%$ c.l. upper bounds on $p_\mathrm{ann}$ for the different cases considered. As anticipated, differences in the constraints between the $10$ keV case and the $300$ eV case are negligible. On the other hand, using the $100$ eV case weakens the constraints by  $\sim 13$ \%. Finally, as the electron injection energy used to calculate the fractions becomes smaller, constraints become rapidly weaker.

\begin{table}[h!]
\begin{center}
\begin{tabular}{cc}
\multicolumn{1}{c}{Electron} &$95\%$ c.l. \\
\multicolumn{1}{c}{injection energy}&\\[1mm]
\hline \noalign{\vskip 1mm} 
$10$ keV &$p_\mathrm{ann}$ $< 2.7\times10^{-28}$\\
$3$ keV &$p_\mathrm{ann}$ $< 2.6\times10^{-28}$\\
$300$ eV &$p_\mathrm{ann}$ $< 2.7\times10^{-28}$\\
$100$ eV &$p_\mathrm{ann}$ $< 3.1\times10^{-28}$\\
$60$ eV &$p_\mathrm{ann}$ $< 3.5\times10^{-28}$\\
$30$ eV &$p_\mathrm{ann}$ $< 4.8\times10^{-28}$\\
$14$ eV &$p_\mathrm{ann}$ $< 9.9\times10^{-27}$\\
\hline
\end{tabular}
\caption{Constraints on $p_\mathrm{ann}$ from a \emph{Planck}-like experiment. We show results assuming no dark matter annihilation in the mock data (second column, upper limits at 95\% c.l.) using the fractions calculated assuming an electron injection energy as shown in column 1 . Constraints are in units of cm$^3$/s/GeV, that correspond to $1.78\times 10^{-28}$ $\rm cm^3$/s/GeV=$1\times 10^{-7}$ $\rm m^3$/s/kg.}
\label{tab:constraints}
\end{center}
\end{table}

We conclude that the differences in the calculated fractions for injected electrons with energies from $\sim 300$ eV -- $10$ keV can be safely neglected. This is an important point to consider when calculating constraints for specific dark matter models. As previously mentioned, in any dark matter scenario the  annihilation energy at GeV/TeV scale is degraded and redistributed to secondary particles by high-energy processes, until their energy falls below the threshold at the $\sim$keV scale. At this point, in order to calculate the effect of dark matter annihilation, we would need to determine the effect of these ``injected'' sub-threshold particles according to the appropriate fractions $\chi_x$ for their energies. 

We will show in Section \ref{sec:higheng} that for the models considered in this paper, the bulk of the energy is in particles above 300 eV, where our baseline fractions should be accurate. The component below 300 eV (and especially below 100 eV) must be treated separately, as we will discuss in more detail later, but this component is always subdominant.

Thus, for the tests in Sections \ref{sec:systematics}-\ref{sect:recfastvscosmorec} we use as a baseline fractions calculated for a $10$ keV injected electron. However, as we will further discuss in Section \ref{sec:higheng}, we find that the ideal energy threshold to split the high and low energy calculations is $3$ keV. The advantage of the high-energy code is that it properly accounts for inverse Compton scattering (ICS), which the low-energy code assumes to be negligible; on the other hand, the advantage of the low-energy code is that it treats the atomic cooling processes much more carefully. The optimal threshold is the highest energy at which atomic processes dominate over ICS; we show that this typically occurs at energies around $\sim 3$ keV (note that this is in conflict with some previous statements in the literature, a point which will be further discussed in Section \ref{sec:higheng}).

\section{Comparison with approximate fractions}
\label{sec:accuratevsSSCK}

In this section we address whether the use of approximate fractions -- in particular those proposed by Chen and Kamionkowski \cite{ChenKamion2004}, which have been widely used in the literature -- can affect the constraints on dark matter annihilation, when compared to the more accurate calculations of the energy fractions described in Section \ref{sec:fractions}.
Chen and Kamionkowski noted that the fractions derived by \cite{ShullVanSteen1985} can be approximated by a linear interpolation between two limiting cases: when the medium is completely ionized, all the energy of the electron goes into heat, whereas when the medium is neutral, the energy of the original electron is split about equally between heating, ionization and excitation \cite{ChenKamion2004}. This simplified approach has been adopted by several authors studying dark matter annihilation in the CMB \cite{zhang2006,Padmanabhan:2005es,Galli:2009zc,Galli:2011rz,Finkbeiner:2011dx,Hutsi:2011vx,Natarajan2009,Giesen2012,Lopez:2013}. Specifically, Chen and Kamionkowski proposed approximating the fractions calculated by Shull and Van Steenberg in the following way:
\begin{eqnarray}
 \chi_i&=&\chi_e=\frac{(1-x_\mathrm{H})}{3(1+f_\mathrm{He})} \nonumber\\ 
\chi_h&=&\frac{1+2x_\mathrm{H}+f_\mathrm{He}(1+2x_\mathrm{He})}{3(1+f_\mathrm{He})}\nonumber\\
 \chi^\mathrm{He}_i&=&\frac{f_\mathrm{He}(1-x_\mathrm{He})}{3(1+f_\mathrm{He})} 
\label{chi_iah}
\end{eqnarray}
where $\chi_i$ is the hydrogen ionization fraction, $\chi^\mathrm{He}_i$ is the Helium first ionization fraction, $\chi_h$ is the heating fraction, and $\chi_e$ is the hydrogen excitation fraction. Some studies have ignored this excitation term while others have implemented it as going entirely into additional Lyman-$\alpha$ photons, so $\chi_e = \chi_\alpha$. For the comparison here, we take the latter approach. $f_\mathrm{He}$ is equal to $f_\mathrm{He}=Y_p/(4(1-Y_p))$, with $Y_p$ the mass fraction of Helium, $x_\mathrm{H}$ is the ratio of ionized hydrogen to total hydrogen, and $x_\mathrm{He}$ is the ratio of ionized Helium to total Helium.

In the following we refer to the Chen and Kamionkowski fractions as the SSCK case. We  compare against the fractions calculated with our code (Section \ref{sec:fractions}) assuming $Y_\mathrm{He}=0.24$, Helium ionization $x_\mathrm{He}=1\times 10^{-10}$, and an initial electron energy of 10 keV.

\begin{figure}[tbh!]
\begin{center}
\includegraphics[width=0.5\textwidth]{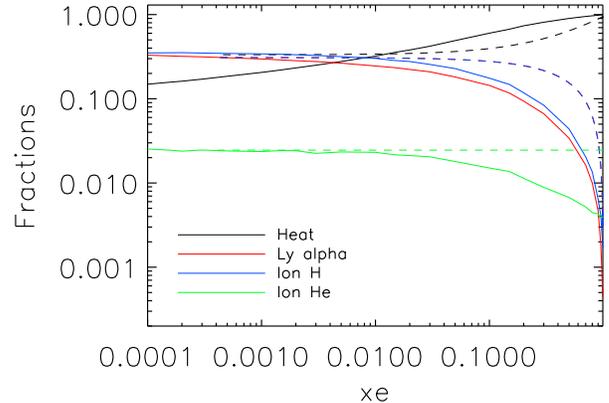}
\caption{Energy fractions calculated by our code (solid lines) and in the SSCK approximation (dashed lines). The fractions shown are for heat (black), Lyman-$\alpha$ (red), ionization of hydrogen (blue) and first ionization of Helium (green). For the SSCK case, the ionization and Lyman-$\alpha$ fractions are equal, so the lines are superimposed.}
\label{plotfractions}
\end{center}
\end{figure}

Figure \ref{plotfractions} shows the energy fractions obtained using our code and the SSCK case. For $x_e\gtrsim 0.01$, the ionization and excitation fractions in the SSCK case are higher and the heating fraction is lower compared to the fractions calculated by our code, while the situation is inverted for $x_e\lesssim 0.01$.

\begin{figure}[h!]
\begin{center}
\includegraphics[width=0.5\textwidth]{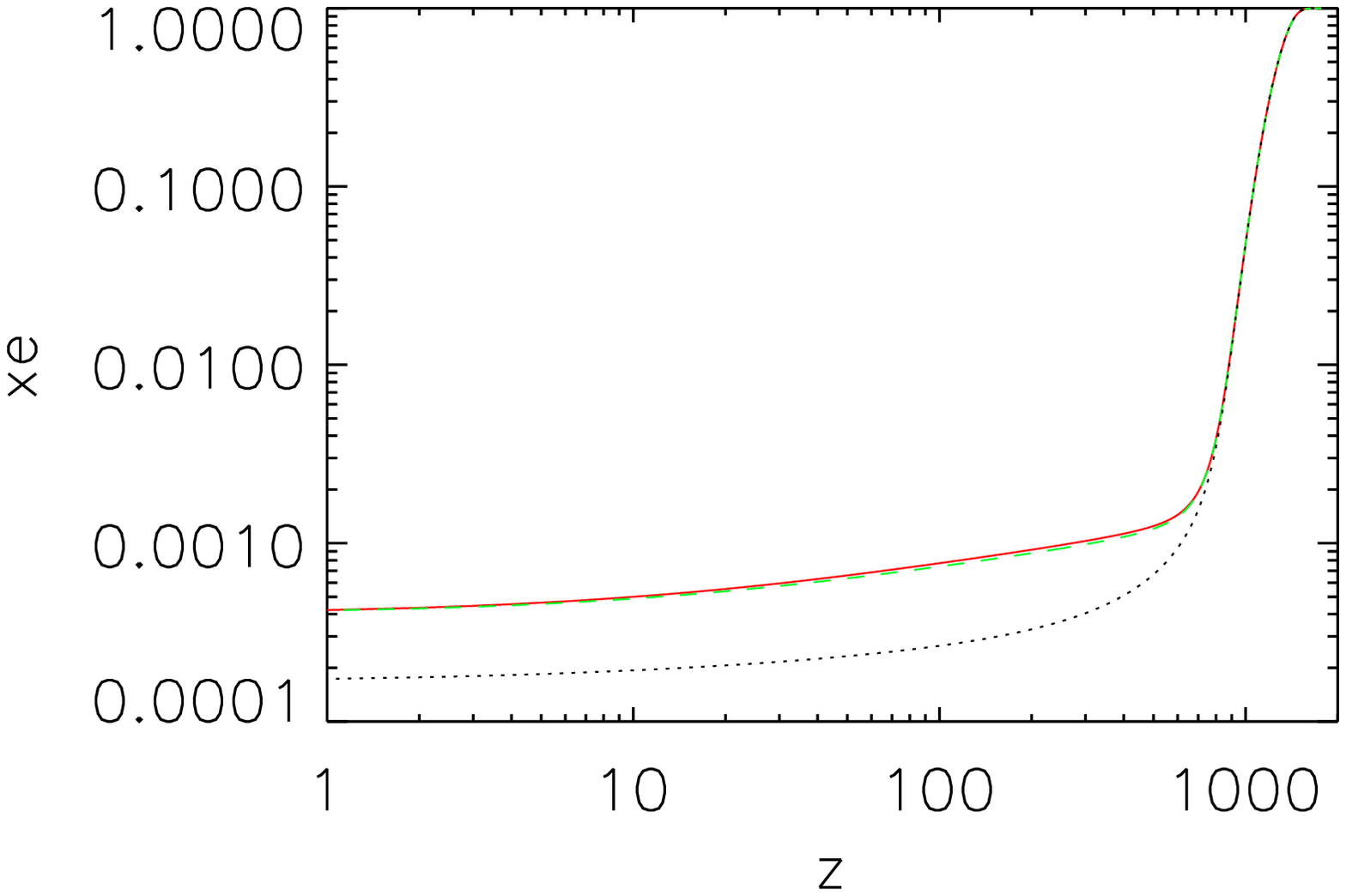}
\includegraphics[width=0.5\textwidth]{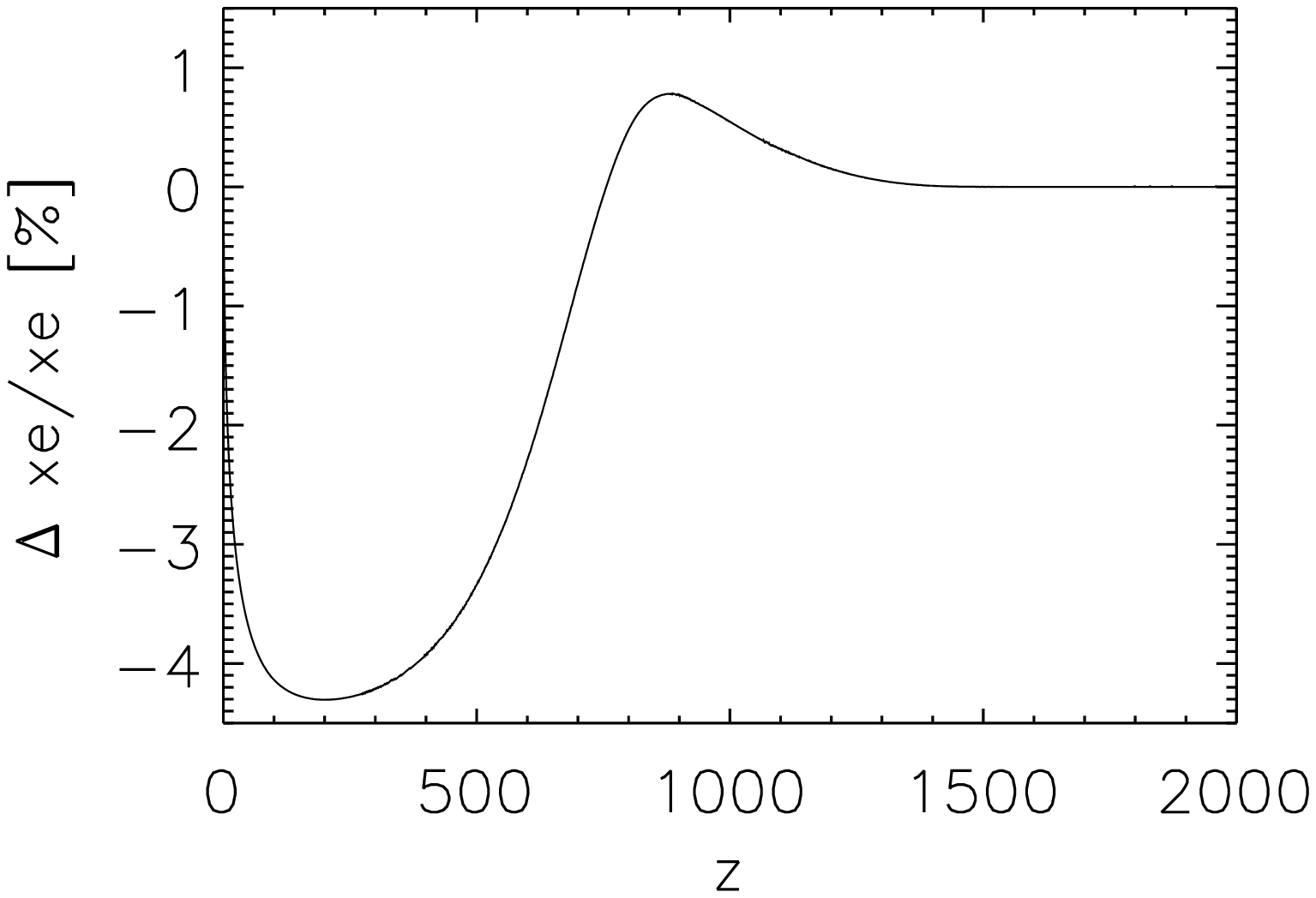}
\caption{Top: free electron fraction for dark matter annihilation with $p_\mathrm{ann}=1.78\times 10^{-27}$ cm$^3$/s/GeV, calculated using the energy fractions obtained by the code described in Section \ref{sec:fractions} (solid red), and the energy fractions in the SSCK approximation (dashed green). As a comparison, we also show the free electron fraction in the case of no dark matter annihilation (dotted black line). Bottom: the relative difference between the two recombination histories for $p_\mathrm{ann}=1.78\times 10^{-27}$cm$^3$/s/GeV, calculated in the SSCK approximation compared to the present work (we show $\rm  (xe_{\rm this work}-xe_{\rm SSCK})/\mathrm{xe}_{\rm this work}$).}
\label{plot_xe}
\end{center}
\end{figure}

The resulting differences in the recombination histories are shown in the top plot of Figure \ref{plot_xe}, where we show the evolution of the free electron fraction (in the presence of dark matter annihilation with $p_\mathrm{ann}=1.78\times 10^{-27} \mathrm{cm}^3\mathrm{/s/GeV}$) using the fractions obtained from our code, and the SSCK case. The SSCK case yields a slightly stronger effect for dark matter annihilation at $z\gtrsim 800$, while our code provides an effect stronger by $\sim 2\%$ at redshift $600$, increasing at lower redshifts to $4\%$, as shown in the lower plot of Figure \ref{plot_xe}.

\begin{figure}[h!]
\begin{center}
\includegraphics[width=0.5\textwidth]{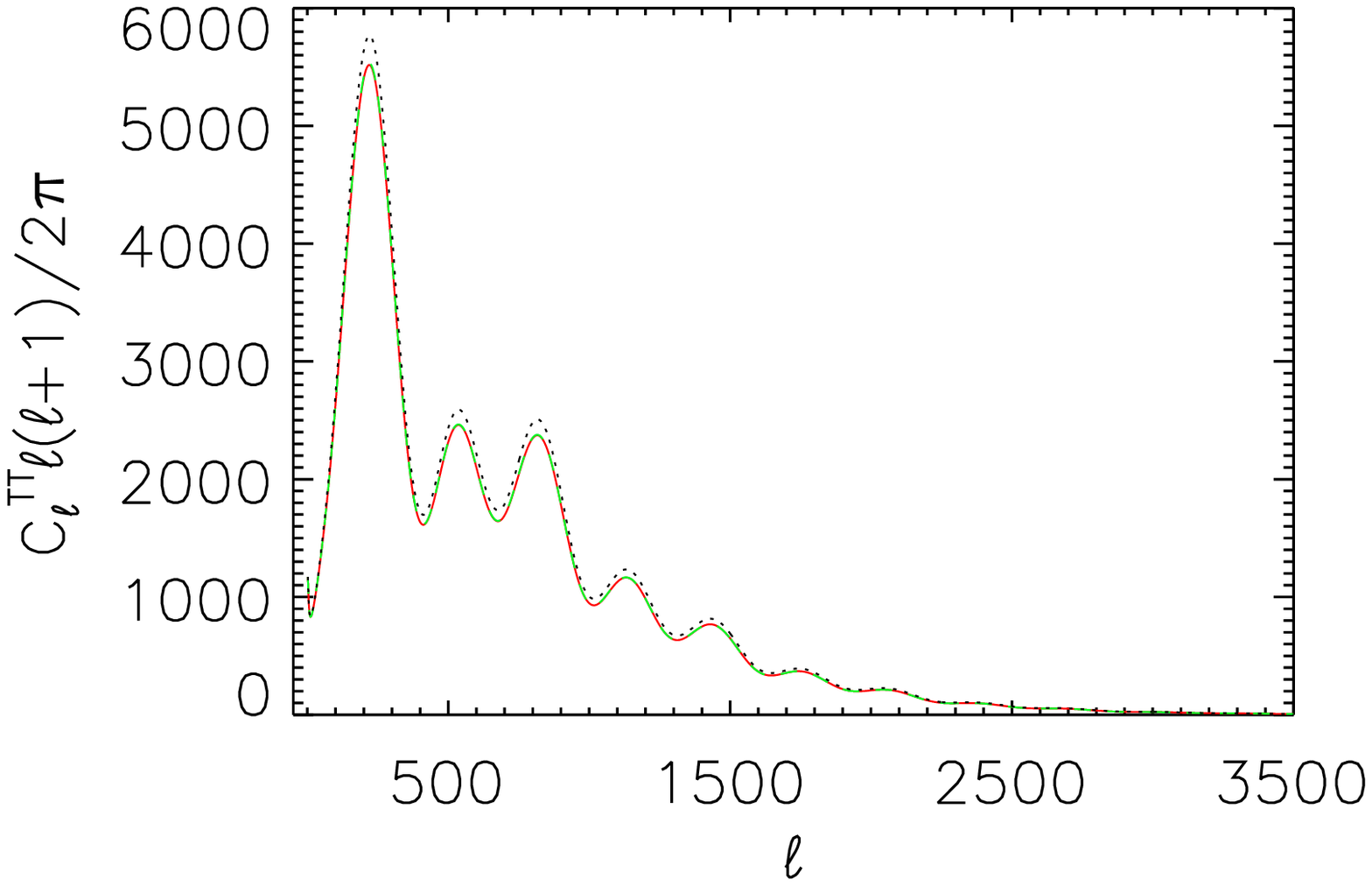}
\includegraphics[width=0.5\textwidth]{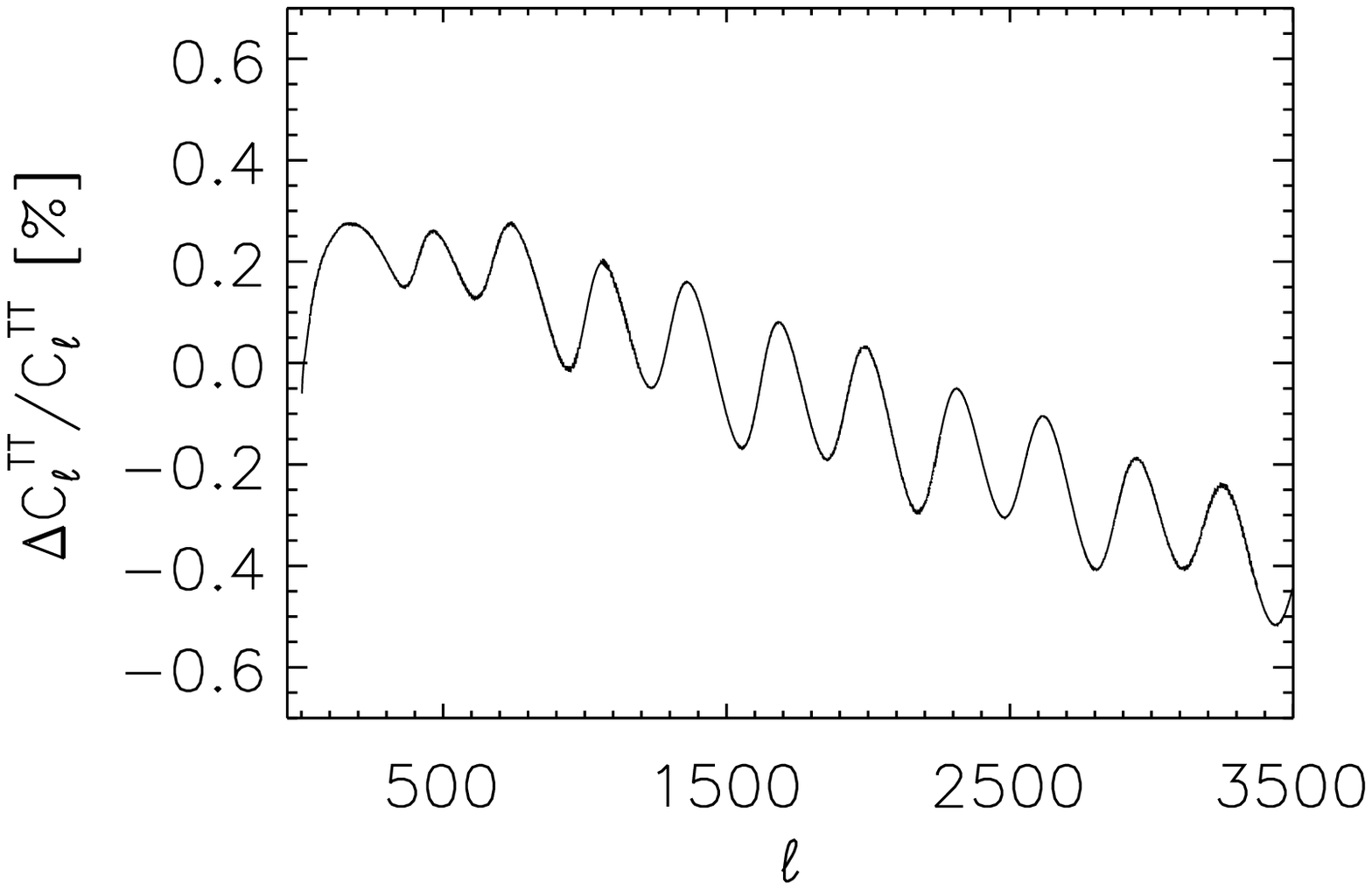}
\caption{Top: CMB TT power spectrum in the case of DM annihilation with $p_\mathrm{ann}=1.78\times 10^{-27}$ cm$^3$/s/GeV calculated with the energy fractions from the present work (solid red) and the energy fractions of SSCK (dashed green).  As a reference, we also show the TT spectrum calculated in a standard recombination scenario, without DM annihilation. Bottom: percentage differences between the TT power spectrum with energy fractions from the present work, and the one calculated with the SSCK energy fractions (we show $(C_l^{\rm SSCK}-C_l^{\rm this work})/C_l^{\rm this work}$).}
\label{plot_spectraTT} 
\end{center}
\end{figure}

\begin{figure}[h!]
\begin{center}
\includegraphics[width=0.5\textwidth]{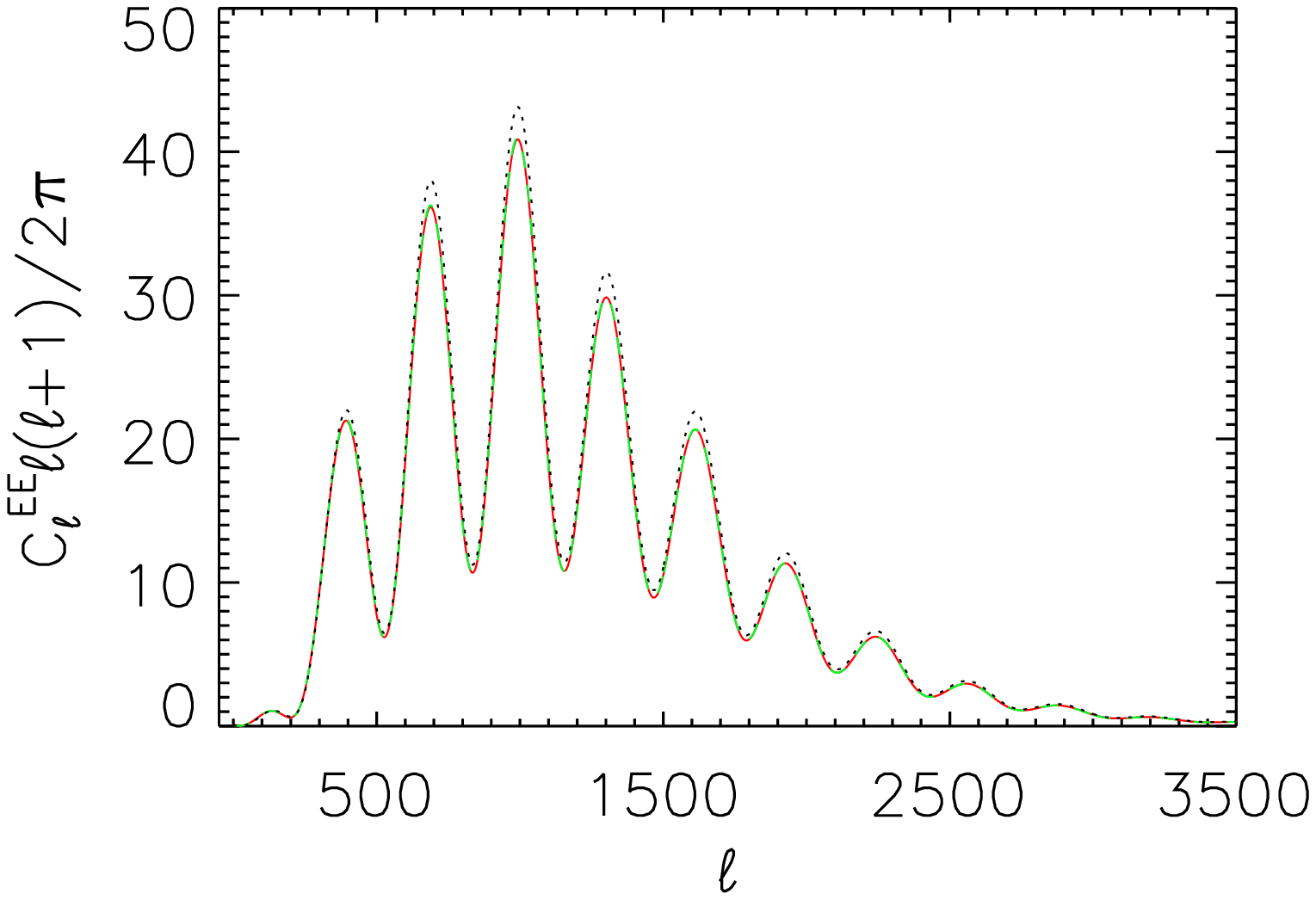}
\includegraphics[width=0.5\textwidth]{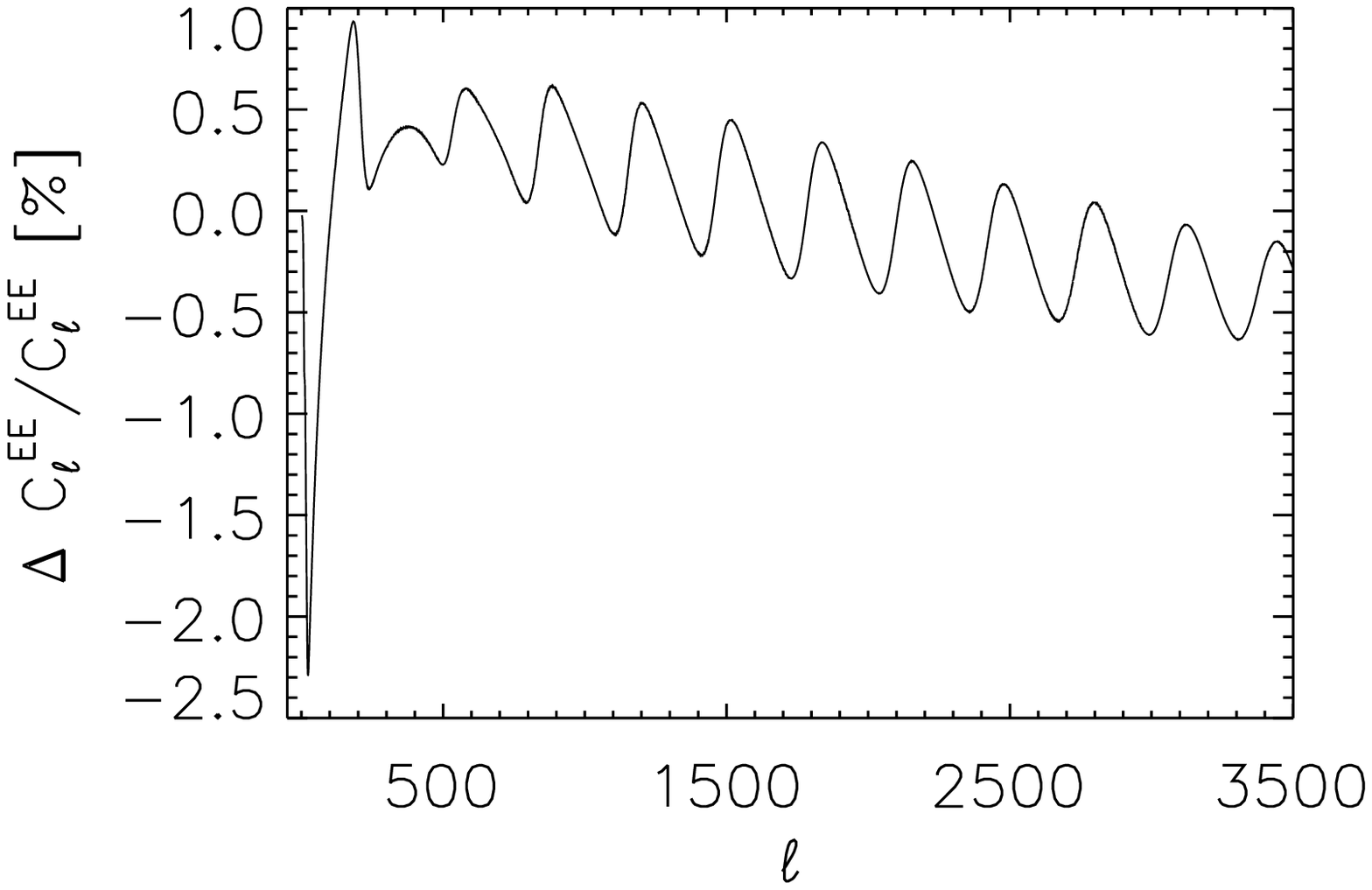}
\caption{Top: CMB EE power spectrum in the case of DM annihilation with $p_\mathrm{ann}=1.78\times 10^{-27}$ cm$^3$/s/GeV calculated with the energy fractions from the present work (solid red) and the energy fractions of SSCK (dashed green).   As a reference, we also show the EE spectrum calculated in a standard recombination scenario, without DM annihilation. Bottom: percentage differences between the EE power spectrum with energy fractions from our code, and the one calculated with the SSCK energy fractions (we show $(C_l^\mathrm{SSCK}-C_l^{\rm this work})/C_l^{\rm this work}$).}
\label{plot_spectraEE}
\end{center}
\end{figure}

The differences in the obtained  CMB anisotropy power spectra using the two sets of fractions are shown in Figure \ref{plot_spectraTT} for temperature and in Figure \ref{plot_spectraEE} for polarization. For temperature, differences between the two cases are at the sub-percent level. In polarization,  there is a slightly stronger effect on large scales, where using our fractions provides a difference with the SSCK fractions of at most $\sim 2\%$ at $l\sim 30$.

\subsection{Effect on the constraints from MCMC}
\label{sec:MCMC}
\label{sec:results1}
We now wish to determine how the differences in the CMB spectra, due to using our fractions as opposed to the SSCK fractions, impact the constraints on DM annihilation. To this end, we use the \texttt{cosmomc} code as already detailed in Section \ref{sec:systematics}. We remind the reader here that we simulate mock data for a \emph{Planck}-like experiment with a fiducial model
given by the best fit WMAP7 model with standard recombination, and assuming either $p_\mathrm{ann}=0$ or $p_\mathrm{ann}=1.78\times 10^{-27}$ cm$^3$/s/GeV. In the latter case, we calculated the mock data assuming our energy fractions calculated for $Y_p=0.24$ and $x_\mathrm{HeII}=1\times10^{-10}$. 

Table \ref{tab:exp} shows the constraints on $p_\mathrm{ann}$ obtained using the fractions from the current work or the SSCK fractions when analyzing the mock data.

When the mock data assume no DM annihilation, the 95\% c.l. upper limits are weaker in the SSCK case compared to the improved analysis, by about $\sim 10\%$. When the mock data assume DM annihilation, two effects can be noted in the results. The first is that using the ``wrong'' model, i.e. the SSCK case, to analyze the data leads to a bias on the recovered value of $p_\mathrm{ann}$ of about  $0.5 \sigma$. The second is again that using our fractions provides slightly stronger constraints, by about 10\%.
We thus conclude that the effect on the constraints of using accurate versus approximate fractions could hardly jeopardize the detection of dark matter annihilation in CMB data, and would in any case only slightly change the upper limits in case of null detection.

\begin{table*}[th]
\begin{center}
\begin{tabular}{lcc}

Experiment & $p_\mathrm{ann}=0$&$p_\mathrm{ann}=1.78\times 10^{-27}$ cm$^3$/s/GeV \\[1mm]
\hline \noalign{\vskip 1mm} 
\emph{Planck}: SSCK $p_\mathrm{ann}$& $< 3.0\times10^{-28}$&$ (1.90\pm 0.20)\times10^{-27}$\\
\emph{Planck}: Present work $p_\mathrm{ann}$& $< 2.7\times10^{-28}$&$ (1.78\pm 0.19)\times10^{-27}$\\[1mm]

\hline
\end{tabular}
\caption{Constraints on $p_\mathrm{ann}$ from a \emph{Planck}-like experiment. We show results both assuming no dark matter annihilation in the mock data (second column, upper limits at 95\% c.l.), and assuming dark matter annihilation with $p_\mathrm{ann}=1.78 \times 10^{-27}$ cm$^3$/s/GeV with the fiducial energy fractions  calculated by our code (third column, error bars at 68\% c.l.). The first row shows results using the SSCK approximation to analyze the data, while the second shows results using the fractions from the present work. Constraints are in units of cm$^3$/s/GeV.}
\label{tab:exp}
\end{center}
\end{table*}

\begin{figure}[th!]
\begin{center}
\includegraphics[width=0.5\textwidth]{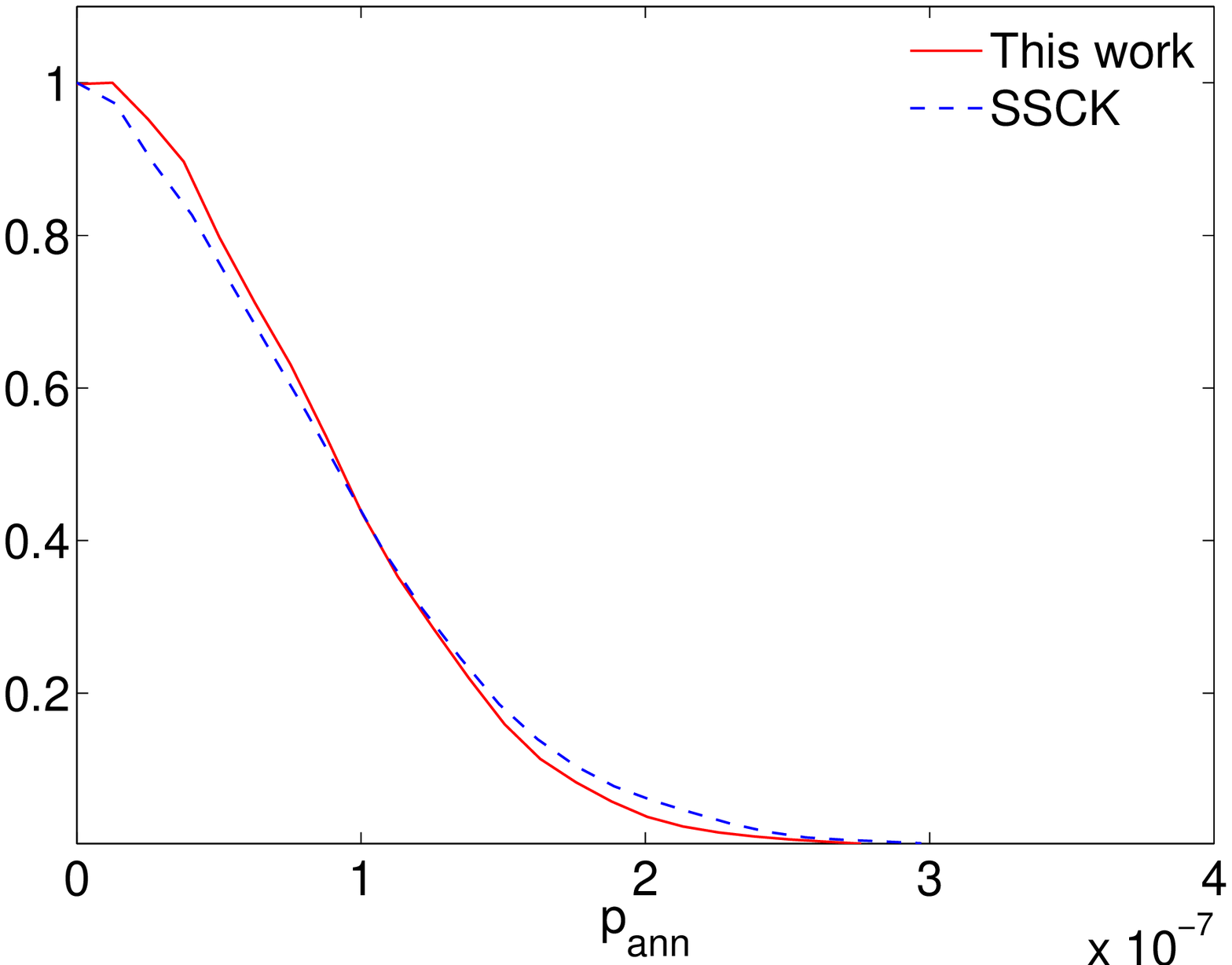}
\includegraphics[width=0.5\textwidth]{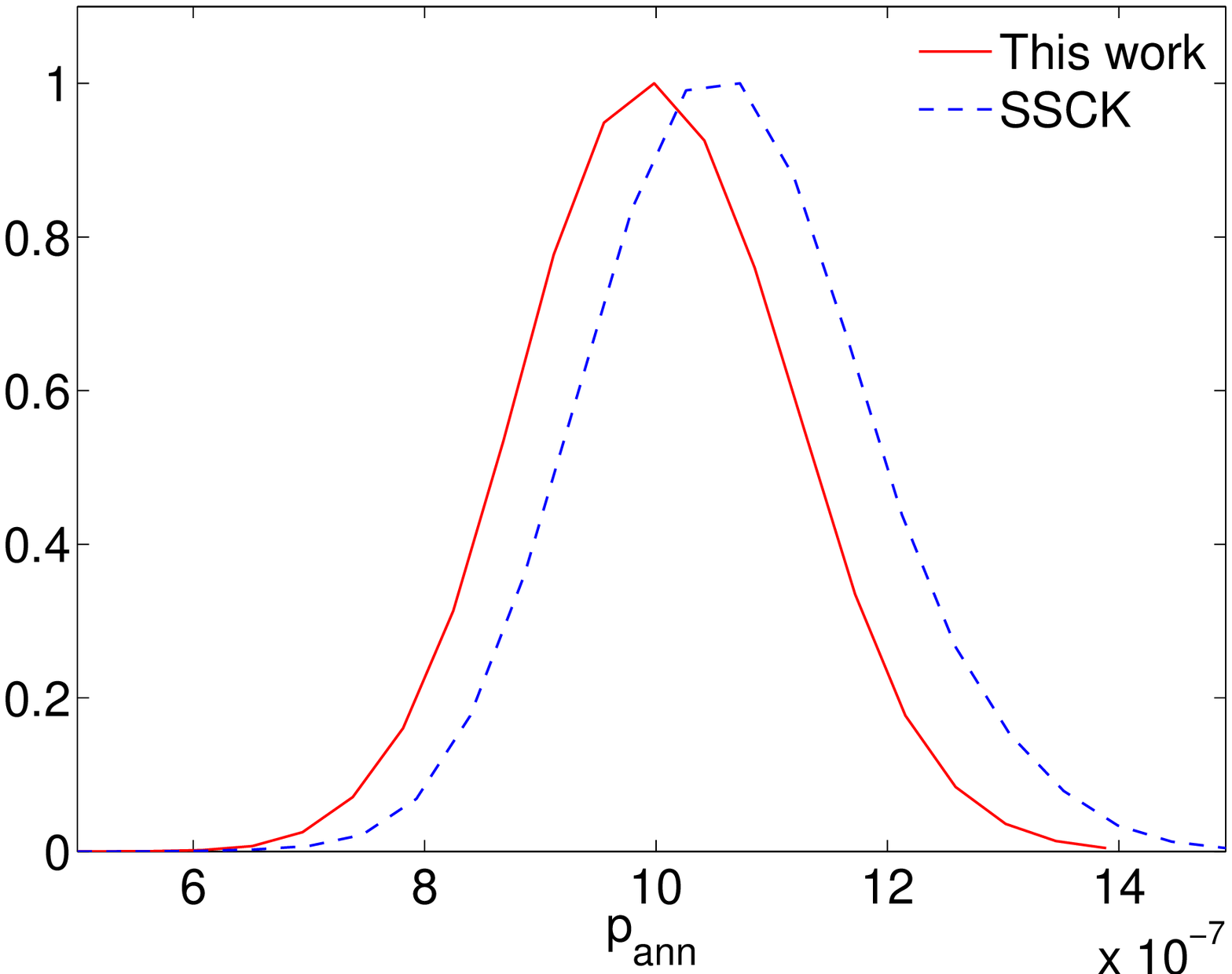}
\caption{Forecasts of marginalized one-dimensional likelihood distributions for $p_\mathrm{ann}$ from a \emph{Planck}-like experiment. We have simulated \emph{Planck}-like data assuming no dark matter annihilation (top) or dark matter annihilation (bottom) with $p_\mathrm{ann}=1.78\times 10^{-27}$ cm$^3$/s/GeV and with fiducial energy fractions calculated with our code. The red solid lines use our fractions when analyzing the data, while blue dashed lines use the SSCK case.}
\label{plot_cosmomcplanck_pann0_pann1eN6}
\end{center}
\end{figure}

\section{The effect of Lyman alpha}
\label{sec:lymanalpha}

\begin{table*}[ht]
\begin{center}
\begin{tabular}{lcc}
 & $p_\mathrm{ann}=0$&$p_\mathrm{ann}=1.78\times 10^{-27}$cm$^3$/s/GeV \\[1mm]
\hline \noalign{\vskip 1mm}
\emph{Planck}: No Lyman-$\alpha$ $p_\mathrm{ann}$& $< 2.8\times10^{-28}$&$ (1.91\pm 0.21)\times10^{-27}$\\[1mm]
\hline
\end{tabular}
\caption{Constraints on $p_\mathrm{ann}$ from a \emph{Planck}-like experiment. We show results both assuming no dark matter annihilation in the mock data (second column, upper limits at 95\% c.l.), and assuming a dark matter annihilation with $p_\mathrm{ann}=1.78\times 10^{-27}$ cm$^3$/s/GeV with the fiducial energy fractions  calculated by our code (third column, error bars at 68\% c.l.). These results assume no energy goes to Lyman-$\alpha$, when analyzing the data. Constraints are in units of cm$^3$/s/GeV.}
\label{tab:lyman}
\end{center}
\end{table*}

As we have already mentioned, in this work we  calculate the exact fraction of energy that is absorbed by the plasma. We assess here how much neglecting the Lyman-$\alpha$ contribution can change the constraints from current CMB data. This issue has already been addressed by \cite{Galli:2011rz,Hutsi:2011vx} using the SSCK case. In this scenario,  the fraction of energy that goes into excitations of the medium is predicted to be about equal to the one going into ionization. It is  however not specified how much of the exciting energy goes into Lyman-$\alpha$ photons, that can additionally contribute to ionizing the medium. The authors of \cite{Galli:2011rz,Hutsi:2011vx} assessed the possible variation in the constraints by considering two extreme cases, i.e. no energy goes into Lyman-$\alpha$ or all the energy going into excitations goes into Lyman-$\alpha$, and found that the size of the effect should be less than $\sim 15\%$.

Here we wish to verify this claim, as the energy fractions we present in this work include the precise calculation of the energy  fraction going into Lyman-$\alpha$. We follow a procedure similar to that described in Section \ref{sec:MCMC}. We generate a \emph{Planck} mock dataset with a fiducial model
given by the best fit WMAP7 model with standard recombination, assuming either a null $p_\mathrm{ann}$ or $p_\mathrm{ann}=1.78\times 10^{-27}$ cm$^3$/s/GeV. In the latter case, we calculate the mock data assuming the energy fractions computed in this work for $Y_p=0.24$ and $x_\mathrm{HeII}=1\times10^{-10}$ and with the correct Lyman-$\alpha$ fraction. We then analyze the mock data setting the Lyman-$\alpha$ fraction to the correct amount or to zero. (Note that the case where we use the SSCK fractions, and assume all the energy assigned to ``excitation'' goes into Lyman-$\alpha$ photons, has been studied already in Section \ref{sec:accuratevsSSCK}.)

We present results in Table \ref{tab:lyman}. We find that the effect of ignoring Lyman-$\alpha$ is rather small. In the case of null  $p_\mathrm{ann}$, it weakens the constraint at $2\sigma$ by $\sim 5\%$, while it biases the results by $0.5\sigma$ in the case with $p_\mathrm{ann}=1.78\times 10^{-27}$cm$^3$/s/GeV.

\section{The choice of recombination code}
\label{sect:recfastvscosmorec}
So far we have investigated how different systematics can impact  the effect of DM annihilation on recombination histories using the RECFAST code. However, we might wonder if the choice of recombination code can  represent an even larger source of uncertainty. To this end, we perform a very basic test, comparing the results obtained using the COSMOREC code \cite{cosmorec} and the RECFAST code in presence of DM annihilation. COSMOREC currently supports the study of dark matter annihilation using SSCK energy fractions and neglecting the effect of extra Lyman-$\alpha$ photons. We therefore use this same benchmark to compare COSMOREC to RECFAST (v1.5).

\begin{figure}[h!]
\begin{center}
\includegraphics[width=0.5\textwidth]{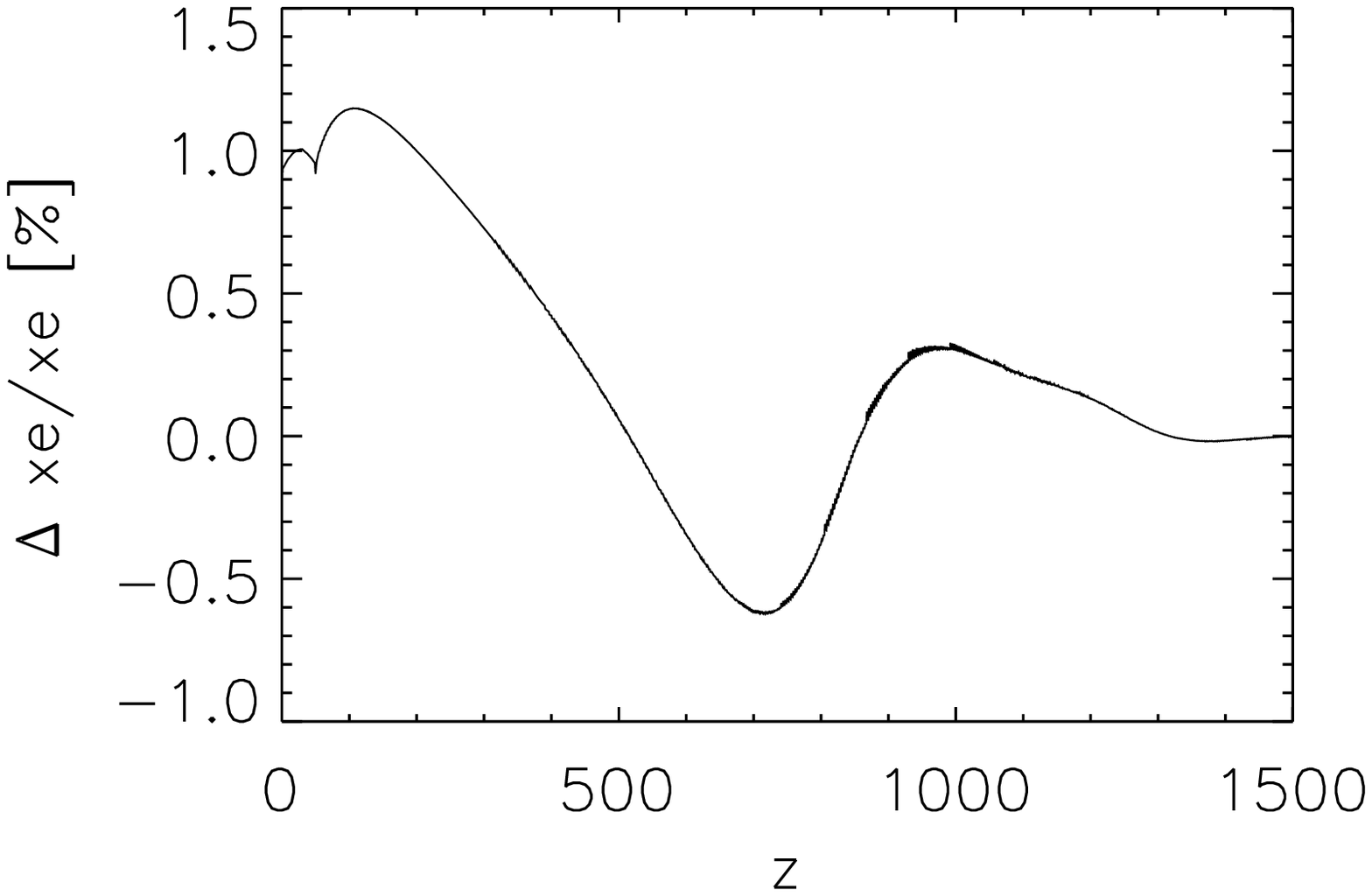}
\caption{Percentage difference in the free electron fraction with dark matter annihilation $p_\mathrm{ann}=1.78\times 10^{-27}\times$ cm$^3$/s/GeV calculated using RECFAST and COSMOREC. In both cases, to calculate the free electron fraction we used the energy fractions of the SSCK case, ignoring extra Lyman-$\alpha$ photons (we show $(\mathrm{xe}_{\rm COSMOREC}-\mathrm{xe}_{\rm RECFAST})/\mathrm{xe}_{\rm RECFAST}$).}
\label{plot_xecosmorec}
\end{center}
\end{figure}

Figure \ref{plot_xecosmorec} shows the percentage difference between the free electron fraction calculated with the two codes, in presence of DM annihilation with $p_\mathrm{ann}=1.78\times 10^{-27}$cm$^3$/s/GeV.  The differences between the two  are at  sub-percent level, with COSMOREC providing a stronger effect at redshifts higher than $z\sim800$ and a weaker one between $400\lesssim z\lesssim 800$. COSMOREC again provides a stronger effect at redshifts lower than $z\lesssim 400$, at the $\lesssim 1\%$ level.
We thus infer that the differences in the recombination codes can hardly change the constraints on dark matter annihilation\footnote{A similar conclusion was also found by \cite{Giesen2012} comparing recombination histories calculated with the HYREC code \cite{hyrec} or with the RECFAST code in presence of dark matter annihilation.}.

\section{Calculation of the high energy cascade}
\label{sec:higheng}
 
In the previous Sections we have described in detail the techniques used to compute the energy fractions that go into heating, excitations and ionizations $\chi_x$, and studied their uncertainties. We now turn to the study of the first part of the energy propagation, aiming to assess how much of the initial annihilation energy at the GeV-TeV scale reaches the keV scale in the form of electrons and photons, where it can be absorbed by the plasma in the previously studied channels.

We first review the previous treatments of $f(z)$ and the meaning of ``deposited'' or ``absorbed'' energy in that context, in Section \ref{sec:fz}. We have adapted the code described in \cite{Slatyer:2009yq} to interface more readily with the low-energy calculation described in Section \ref{sec:fractions}, by carefully following all electrons and photons down to keV energies; we describe these modifications in Section \ref{sec:highcalc}. One point that deserves particular attention is the treatment of inverse Compton scattering (ICS) by keV-MeV electrons, as energy lost through this channel contributes to sub-10.2 eV ``continuum'' photons (i.e. spectral distortion of the CMB), rather than heating, excitation or ionization of the gas. ICS is not included in the code of Section \ref{sec:fractions}, and we have found that prior treatments in the literature that claimed ICS was subdominant and could be ignored for electrons below several hundred keV relied on an incorrect extrapolation of the relativistic energy loss rate to non-relativistic energies. These points are discussed in Section \ref{sec:continuum}.

In Section \ref{sec:highengresultsprevious} we present results for the low-energy spectra of electrons and photons, the energy absorbed by the gas above threshold, and the energy lost to continuum photons, for several DM benchmark models, and give a simple ``approximate $f(z)$'' method of encapsulating the results of the high-energy code. We discuss the effect of a more careful treatment in Section \ref{sec:highsystematicseffective}, and present our results for the modified constraints in Section \ref{sec:newfconstraints}.

\subsection{Notes on $f(z)$}
\label{sec:fz}

We first review the different formalisms used in the literature
to compute the fraction of annihilation energy that is absorbed by the gas, $f(z)$ (see Eq. \ref{enrateselfDM}), which can then be partitioned between the absorption channels discussed in Section \ref{sec:fractions}.

In Eq. \ref{enrateselfDM}, the use of the fraction
$f(z)$ is sometimes referred to in the literature as the ``on-the-spot'' approximation, i.e. the assumption that energy produced by an annihilation  {\it at a given
redshift} is absorbed at that same redshift. This approximation can be accurate for light DM, where electrons are injected below $\sim$ GeV energies, but for higher-energy electrons and photons, there can be a significant delay between the redshift of injection and the redshifts where the bulk of the energy is absorbed. The current use of Eq. \ref{enrateselfDM} goes beyond this approximation: following the computations of \cite{Slatyer:2009yq,Hutsi:2011vx}, it is possible to adopt an $f(z)$ curve that for each redshift describes the
amount of energy \emph{absorbed} at that redshift, including contributions from particles injected at all previous
epochs, and returns it as a fraction of the energy \emph{injection} rate at the redshift of interest.
This formalism (in which $f(z)$ can therefore be 
bigger than unity) unites the compactness of the
on-the-spot formalism with the correctness of a thorough propagation
of the energy cascade. When using $f(z)$, we will adopt this approach, thus moving beyond the on-the-spot approximation.

The second point deserving clarification is the meaning of ``fraction absorbed'' by the gas. Typically, the shower of ``primary'' electrons and photons produced by DM annihilation
has a spectrum extending from low energies up to the
DM mass scale (few GeV -- few TeV). These primaries are propagated down to energies (typically tens to hundreds of keV) below which any electron or photon will be absorbed by atomic processes on timescales much smaller than a Hubble time (see e.g. \cite{Slatyer:2009yq, Slatyer:2012yq}). During the cooling cascade, photons may be produced at energies and redshifts where the universe is transparent or nearly so; these photons carry away energy which is never degraded to the atomic scale or absorbed by the gas.

The $f(z)$ fraction therefore refers to the fraction of energy which is degraded
down to the keV scale, as opposed to escaping as higher-energy photons. The subsequent partition of this energy into the various absorption channels is computed 
separately, as described in Section \ref{sec:fractions}. 

It is clear that an ideal treatment should be able to follow the 
cascade from the DM mass scale down to the atomic processes,
and indeed a recent study \cite{Evoli:2012sc} has presented CMB constraints based on 
such a treatment of the energy cascade. However, that study assumed that the entire cascade occurred at the redshift of injection, effectively imposing an on-the-spot approximation, and neglecting the expansion (and increasing transparency) of the universe during the cascade. We briefly discuss some possible effects of this assumption in Appendix \ref{app:evolicompare}.\footnote{Another difference between this work and \cite{Evoli:2012sc} is our corrected treatment of ICS, see Section \ref{sec:continuum} and Appendix \ref{app:threshold}.} It would \emph{not} be straightforward to extend the approach of this study (similar to our Section \ref{sec:fractions}) to the case without the on-the-spot approximation, as that would require tracking the expansion of the universe (and e.g. corresponding changes in the background gas and photon densities) during the cascade; the improved treatment of ICS we present in this work would also be numerically very difficult to reproduce with a similar Monte Carlo approach (as the IC scatterings at low energy are almost elastic and very abundant). These difficulties motivate the ``split'' approach described in this work, where a detailed Monte Carlo code is used to study the low-energy regime in which these issues are unimportant, while a code designed for the high-energy regime is used to follow the initially injected particles down to the threshold scale.
 
As well as the issue of the on-the-spot approximation, such a treatment is also model-dependent, requiring a separate run for each case under study; an advantage of the ``split'' technique is that only the high-energy part of the cascade depends on the particular DM model, and as we will see, the output of the high-energy cascade can be fairly well parametrized by the single function $f(z)$. However, there are possible systematics introduced by the use of the ``split'' technique, which we will discuss in Section \ref{sec:highsystematicseffective}.

\subsection{Numerical treatment of high-energy processes}
\label{sec:highcalc}

For this work, we adapt the code detailed in \cite{Slatyer:2009yq} to treat the cascade from the annihilation energy down to the scale where redshifting can be safely neglected. Unless noted otherwise, the numerical treatment and the cross sections employed are the same as in that work. Earlier versions of this calculation \cite{Slatyer:2009yq, Slatyer:2012yq} were designed to explore the partition between energy lost to redshifting and free-streaming X-rays and gamma rays, and energy ``deposited'' to the intergalactic medium. This second category included all secondary photons which had cooled below an energy threshold such that their further cooling time was very rapid relative to a Hubble time (by a factor of 1000 in \cite{Slatyer:2009yq}, but the exact threshold does not matter significantly as the photoionization cross section is a very steep function of energy; changing the threshold does not significantly alter the redshift at which the secondary photon is recorded as ``deposited''), and similarly all secondary electrons cooled below the threshold at which ICS is no longer a significant cause of energy loss (when the electron becomes non-relativistic). Since these low-energy electrons and photons were not tracked in detail, it is important to note that secondary photons produced below 10.2 eV in energy could be counted in the ``deposited'' energy budget, although they would contribute to CMB spectral distortion rather than ionization and heating of the gas. For dark matter annihilation studies we are interested in the fraction of energy that can actually affect the recombination history, so we need to accurately calculate and subtract the contribution to the energy fractions of these low energy particles, that do not interact further with the gas.

\subsubsection{Inverse Compton scattering by non-relativistic electrons}
\label{sec:continuum}

In the present study we have implemented an improved treatment of ICS of non-relativistic (or mildly relativistic) electrons and the competing processes of ionization and heating of the gas, to allow the study of these very-low-energy secondary photons. Non-relativistic and mildly relativistic electrons with energies in the few keV - several MeV range play a key role; in this energy range, ionization and heating do not yet clearly dominate the energy losses (they become dominant at lower energies), and scattering of these $\mathcal{O}(1)-\gamma$ electrons on CMB photons only slightly boosts the photon energies. While each individual scattering removes a negligible fraction of the electron's energy, the electrons must lose a large fraction of their initial kinetic energy to enter the regime where ionization/heating dominate, and so most of this energy goes into producing a continuum of slightly-boosted CMB photons (or equivalently, to distorting the black-body spectrum of the CMB, producing a high-frequency tail) via a very large number of scatterings.

As discussed in more detail in Appendix \ref{app:threshold}, this point has \emph{not} been well appreciated in the literature. Previous works \cite{Valdes:2009cq,Evoli:2012sc} have claimed that ICS is sub-dominant below a few hundred keV at $z \sim 600$: this claim was based on an error in the calculation. Treating this population correctly is especially important for few-GeV dark matter candidates, where the dominant relevant annihilation products are few-GeV electrons ($\gamma \sim 10^3-10^4$); ICS of such electrons during the epoch of recombination produces $\sim 0.1-10$ MeV photons, which bracket the energy range where Compton scattering of photons on free electrons is an efficient cooling process (faster than a Hubble time, at recombination). This process produces copious non-relativistic secondary electrons.

For heavier DM candidates, the dominant initial cooling process is pair production, followed by ICS of the resulting electrons. At high energies, where the ICS enters the Klein-Nishina regime, this process produces secondary photons with comparable energy to the primaries; as the primaries cool the gap in energy becomes wider and wider. This generates a broad-spectrum population of electrons, and those below a GeV or so in energy (but above a few MeV) will produce ICS photons with energies above 10.2 eV but below the ``Compton bump'', where photoionization is the dominant energy-loss process.

\subsubsection{Interface with the low-energy calculation}

In addition to the improved treatment of ICS at low energies, allowing us to track the resulting population of ``continuum'' photons, we carefully track all electrons and photons down to a fixed energy threshold, which we typically choose to be 3 keV (except for convergence tests). We discuss and justify this choice in Appendix \ref{app:threshold}. Because a single ionization event can remove a significant fraction of the electron's energy, and free electrons can be produced by Compton scattering by photons and photoionization, this gives rise to a spectrum of sub-threshold electrons, rather than simply a spike at the threshold. We separately track the spectrum of sub-threshold photons removed from the code as ``deposited'', at each timestep (which allows these photons to be separated into those below 10.2 eV and those which will photoionize the gas to produce free electrons below the threshold cutoff). In addition to the photon cooling processes described in \cite{Slatyer:2009yq}, we track the secondary electrons produced by photoionization by above-threshold photons (which were previously simply tagged as ``deposited''). For electrons above the threshold, we count as ``deposited'' energy lost to collisional heating of the gas,  energy absorbed as excitation or ionization (but the secondary electrons from all these processes are tracked until they fall below the threshold), and energy lost by ICS into photons of lower energy than we explicitly track (below 0.1 eV). 

At each timestep\footnote{Timesteps are logarithmic in redshift and are given by $d \ln z = 10^{-3}$, as in  \cite{Slatyer:2009yq}.}, therefore, we compute:
\begin{enumerate}
\item A spectrum of ``deposited'' photons below the threshold (where by ``deposited'' we mean that their energy would entirely be counted as ``deposited'' in the previous version of the code \cite{Slatyer:2009yq}, see Section \ref{sec:fz}), divided into: 
\subitem (a) Photons above 10.2 eV which will ionize or excite the gas.
\subitem (b) Photons below 10.2 eV which will free stream.
\item A spectrum of ``deposited'' electrons below the threshold.
\item A measurement of the energy tagged as ``deposited'' by scatterings of electrons and photons above the threshold, composed of ionization, excitation, heating and (to a small degree) otherwise-unaccounted-for ICS losses.
\item A measurement of the energy \emph{removed} from the background CMB by ICS. The scatterings of mildly non-relativistic electrons produce a spectrum of scattered photons that is very similar to the CMB, but which appears in the spectrum of ``deposited'' photons in $(1)$; unless the spectrum of CMB photons that was present \emph{before} the scatterings is subtracted, this is a double-counting.
\end{enumerate}
This is in addition to the spectrum of \emph{non}-deposited photons at higher energies (where the universe is not very opaque), which are propagated forward to the next timestep \cite{Slatyer:2009yq}.

\subsection{Results and computation of the energy absorption}
\label{sec:highengresultsprevious}

The sum of the energy contained in the first three components, minus the energy lost from the CMB, yields the total ``deposited energy'' output in previous studies; if we denoted the $f(z)$ curve derived e.g. in  \cite{Slatyer:2009yq} as $f_\mathrm{previous}(z)$, then 
\begin{equation} f_\mathrm{previous}(z) = \frac{ (1a) + (1b) + (2) + (3) - (4)}{\text{energy injected at that timestep}}. \label{eq:originalf} \end{equation} 

As a first step, a new effective $f(z)$ curve can be computed from these results, that correctly takes into account the ICS losses to continuum photons. Summing up the other components (from photons between 10.2 eV and the threshold, below-threshold electrons, and above-threshold deposition) gives the total power absorbed by the gas and injected into the regime relevant to the low-energy calculation. This quantity can be multiplied by the energy fractions of Section \ref{sec:fractions}, evaluated for electrons at the threshold energy, to estimate the partition into different channels. In terms of the components above, the new corrected $f(z)$ is given by,

\begin{equation} f_\mathrm{approx}(z) = \frac{(1a) + (2) + (3)}{\text{energy injected at that timestep}}. \label{eq:newf} \end{equation}

The energy from DM annihilation lost to sub-10.2 eV ``continuum''  photons is given by $(1b) - (4)$ (strictly, the spectrum of photons scattered out of the CMB should be subtracted from the total spectrum of ``deposited'' photons (1), but for all redshifts of interest, the vast bulk of the energy in the CMB spectrum is below 10.2 eV, so the component $(1a)$ is unaffected).

This approach is only approximate. A more accurate calculation would take the following effects into account:
\begin{itemize}
\item The spectrum of below-threshold photons needs to be converted into an electron spectrum for the code of Section \ref{sec:fractions} to be directly applicable; this occurs via photoionization, so there is a direct contribution to the ionization level from this process that is not accounted for in the energy fractions $\chi_x$.
\item The spectrum of below-threshold electrons is not a delta function at the threshold energy; a non-negligible fraction of the power is in lower-energy electrons, which tend to lose a higher fraction of their power into heating rather than excitation and ionization.
\item The above-threshold deposition comes from electrons cooling down to the threshold energy, whereas the $\chi_x$ fractions track the cooling of the electron until all its energy has been absorbed by the gas. Consequently, the $\chi_x$ fractions may not precisely describe the partition of the above-threshold deposition into ionization, excitation and heating.
\end{itemize}

We will discuss these effects further in Section \ref{sec:highsystematicseffective}, and show that they have a fairly small impact on the constraints, relative to using $f_\mathrm{approx}(z)$ (including our best estimates for all three tends to strengthen the constraints by $\sim 10-15\%$).

\begin{figure*}
\includegraphics[width=.3\textwidth]{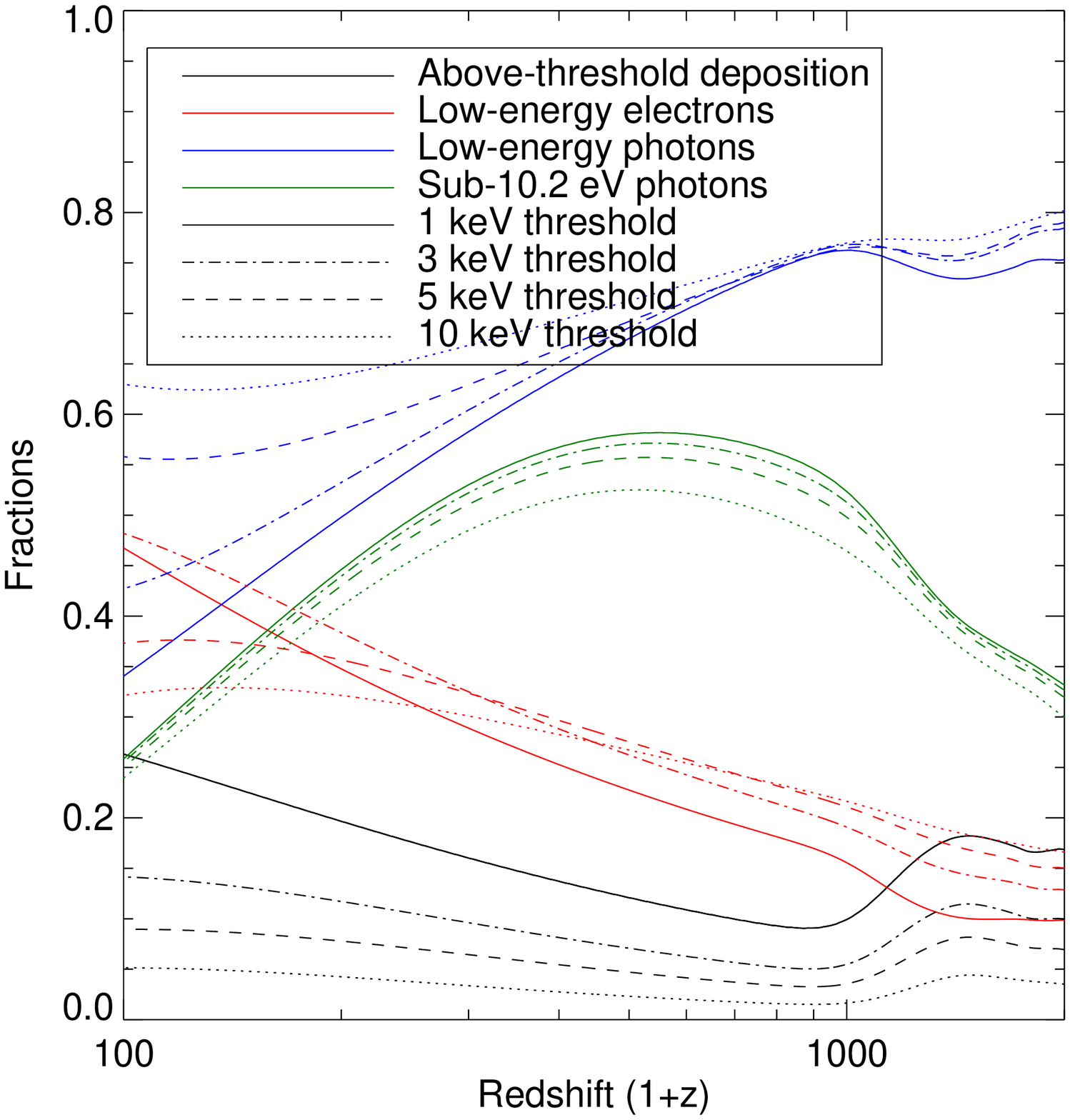}
\includegraphics[width=.3\textwidth]{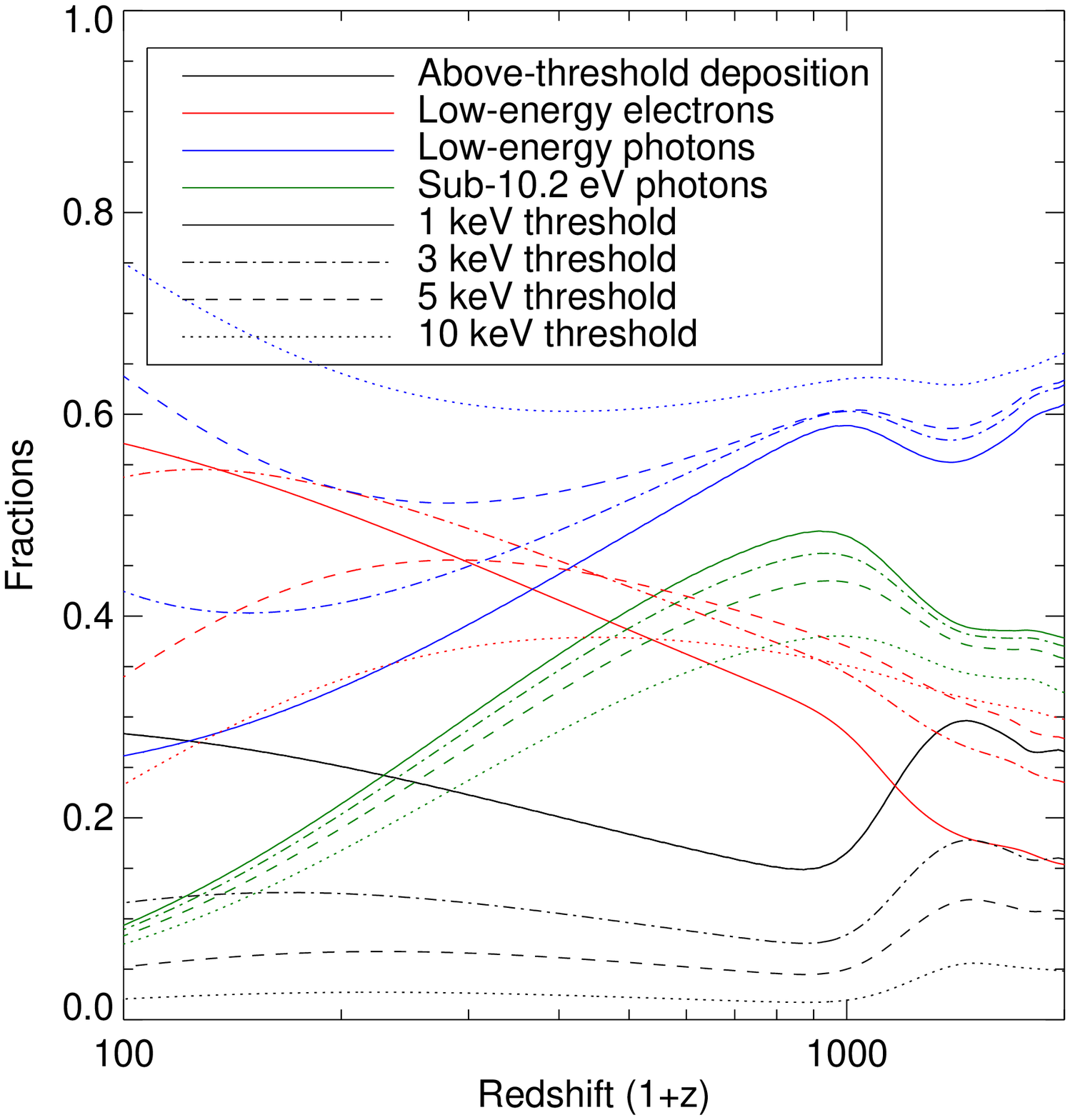} \\
\includegraphics[width=.3\textwidth]{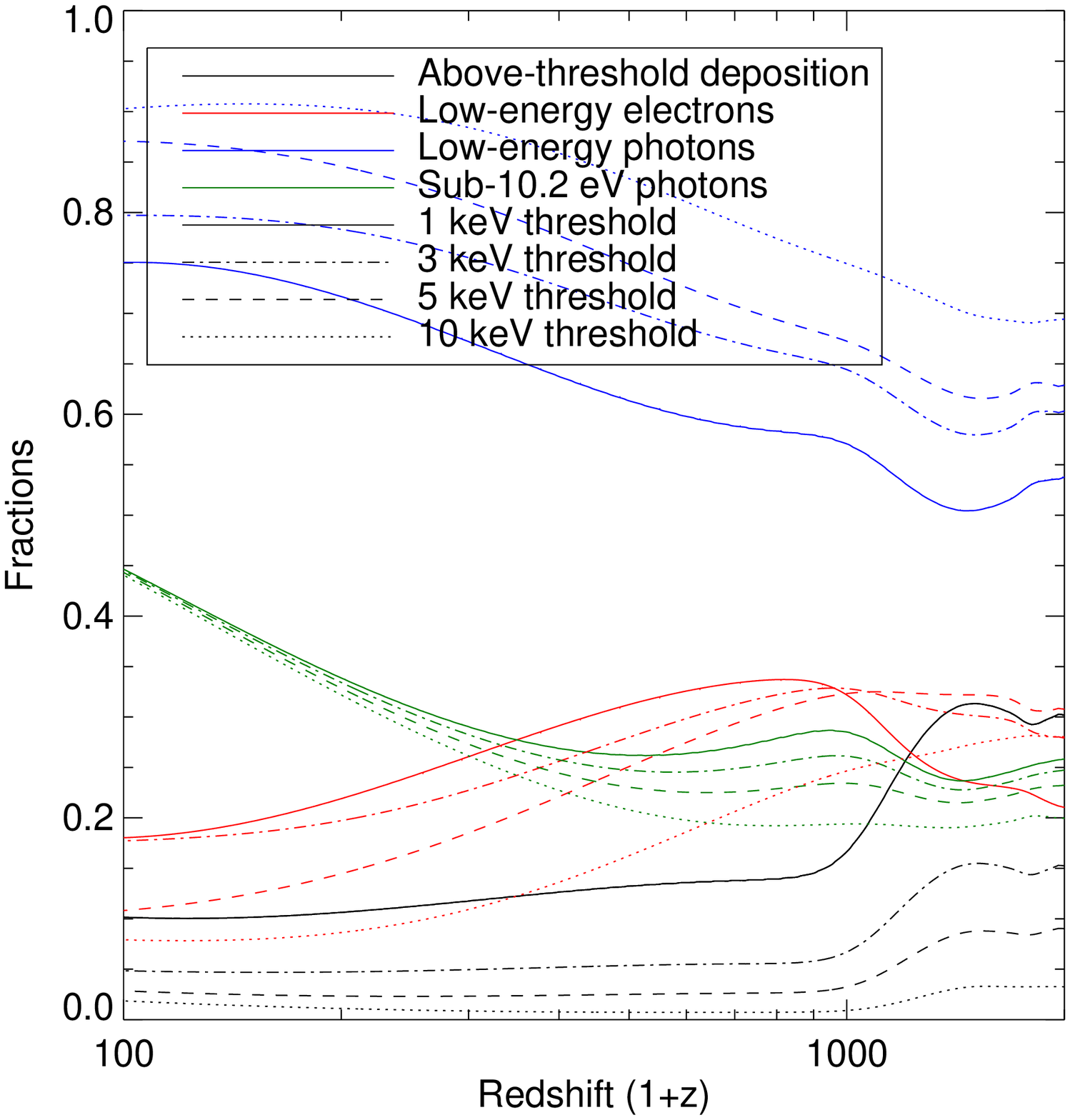}
\includegraphics[width=.3\textwidth]{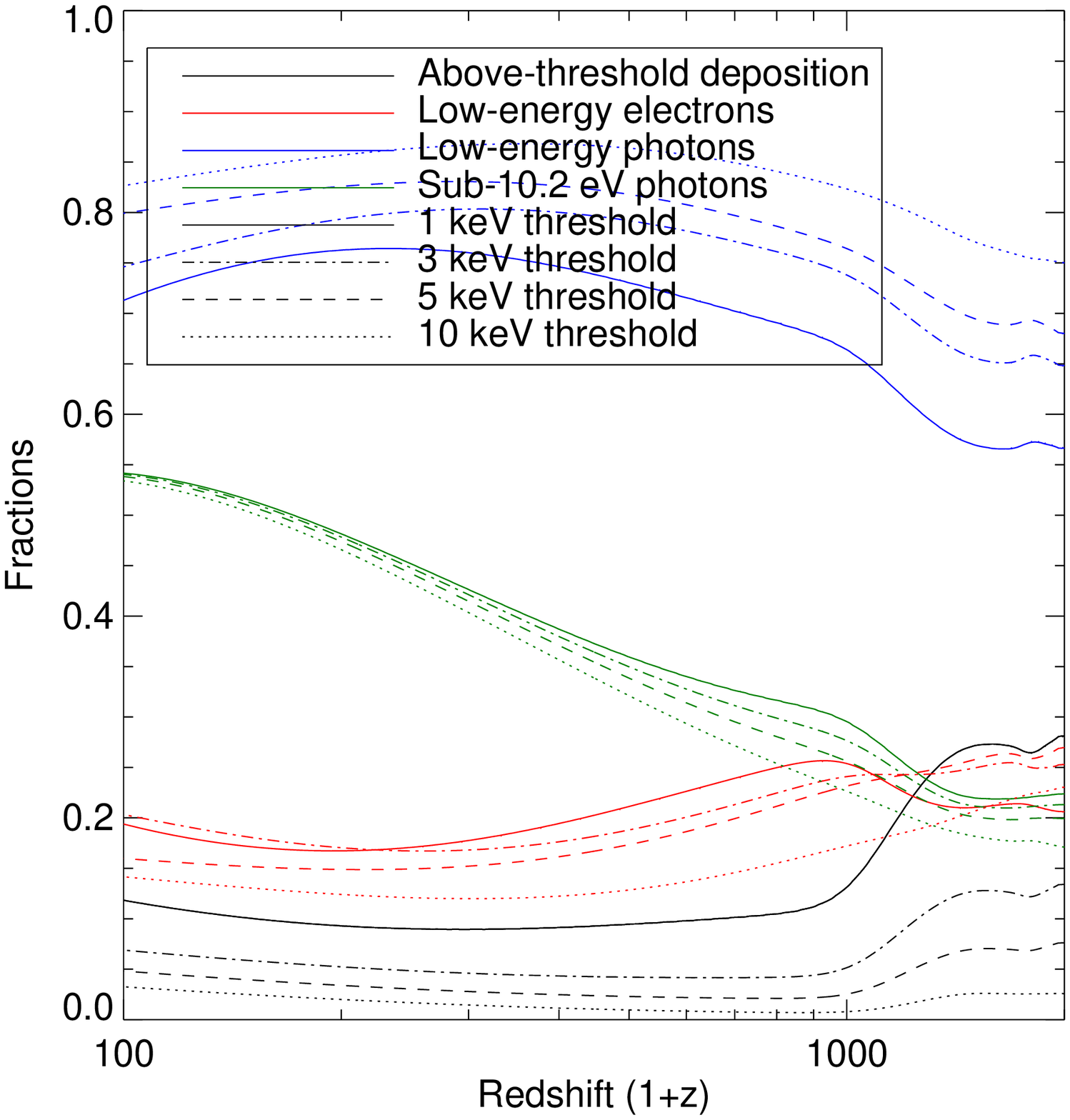} \\
\includegraphics[width=.3\textwidth]{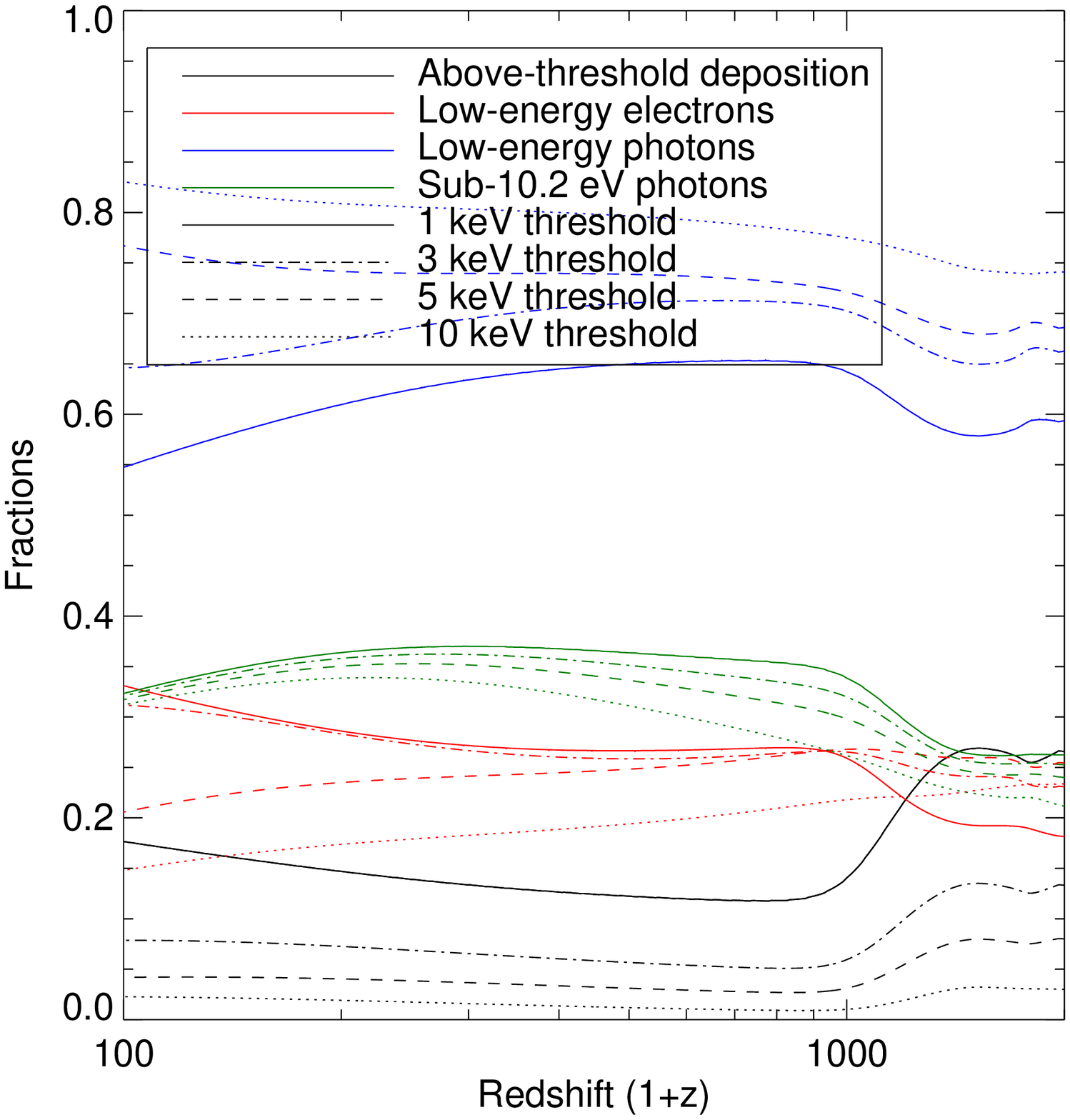}
\includegraphics[width=.3\textwidth]{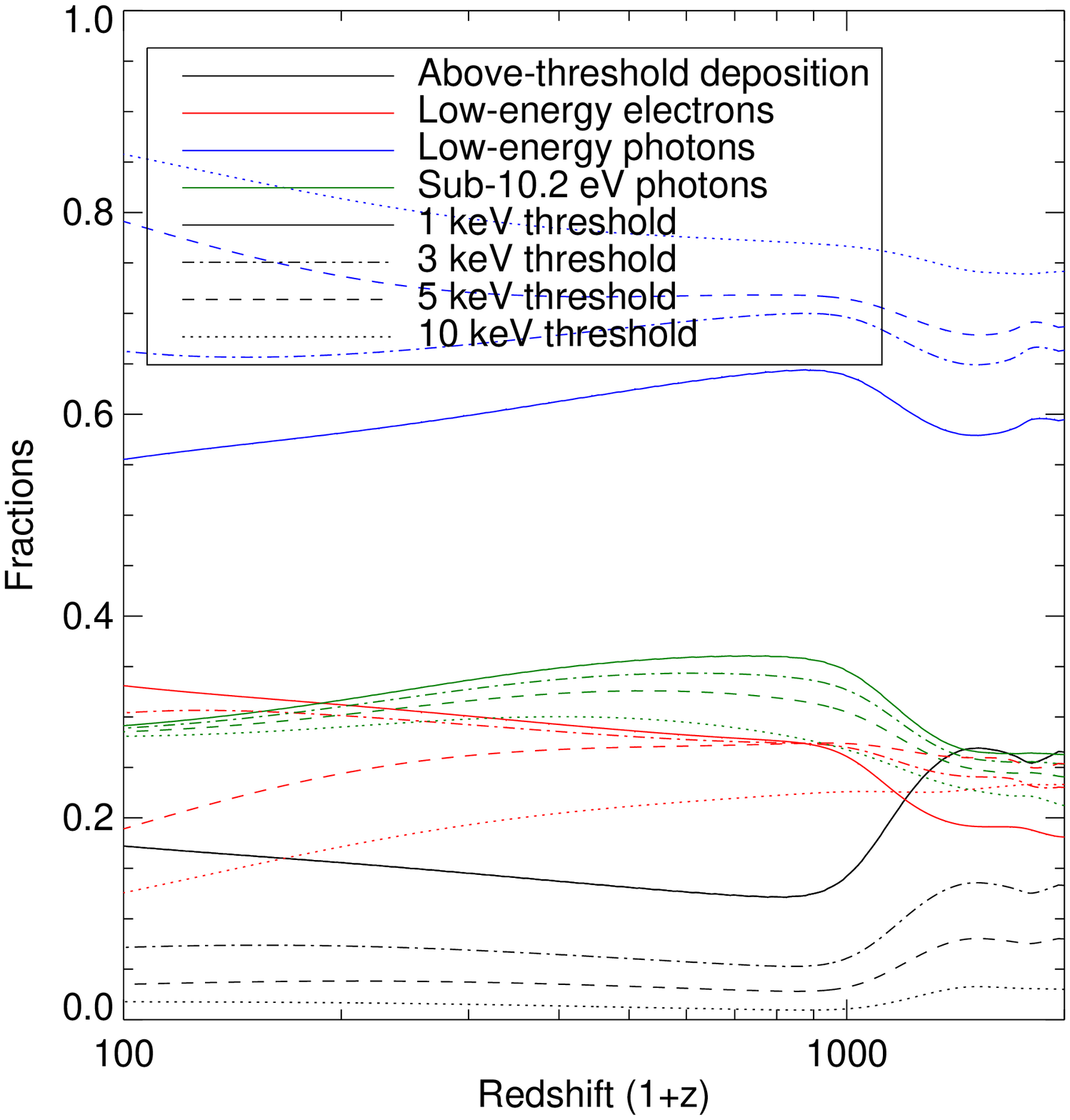} \\
\includegraphics[width=.3\textwidth]{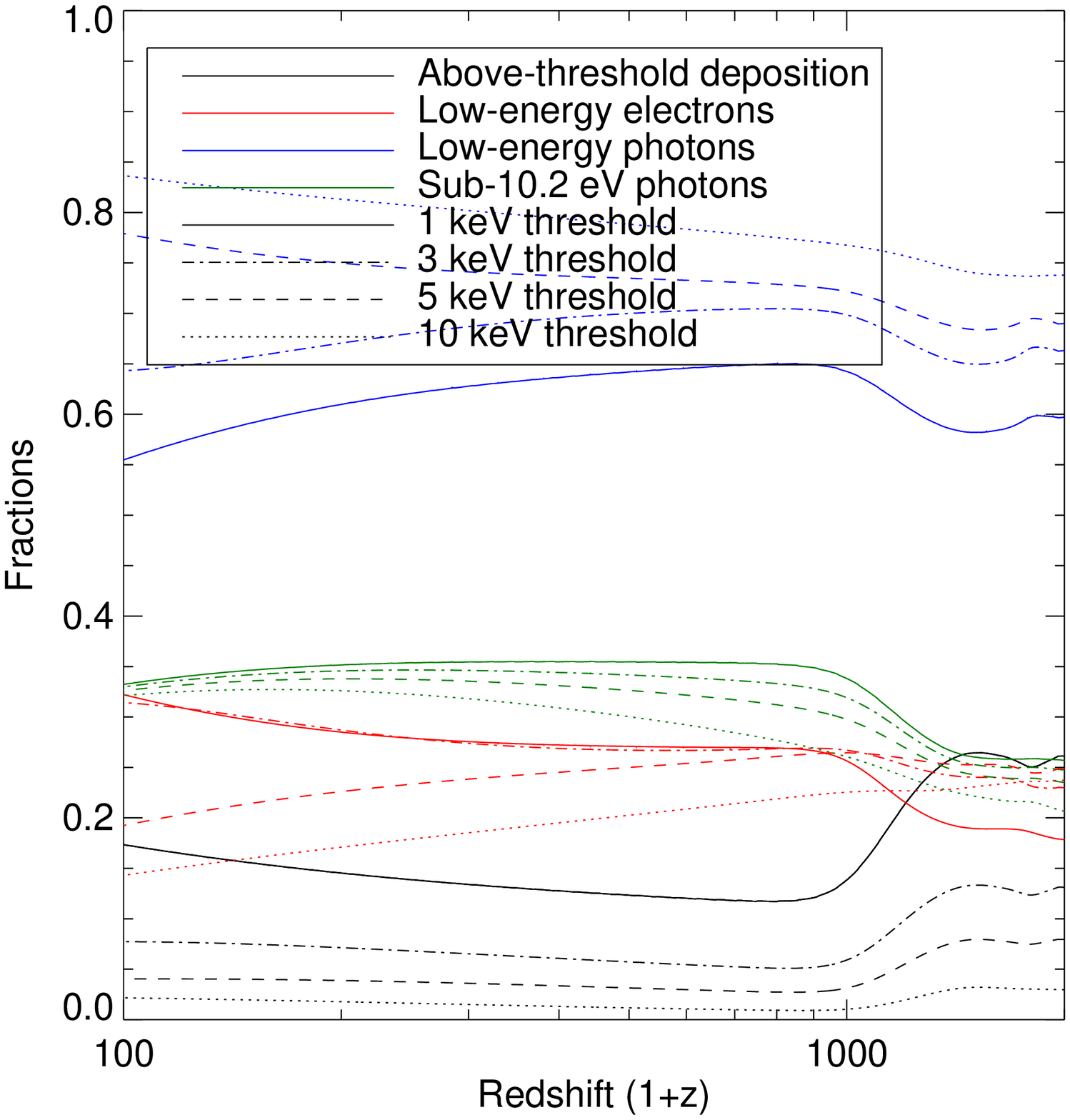}
\includegraphics[width=.3\textwidth]{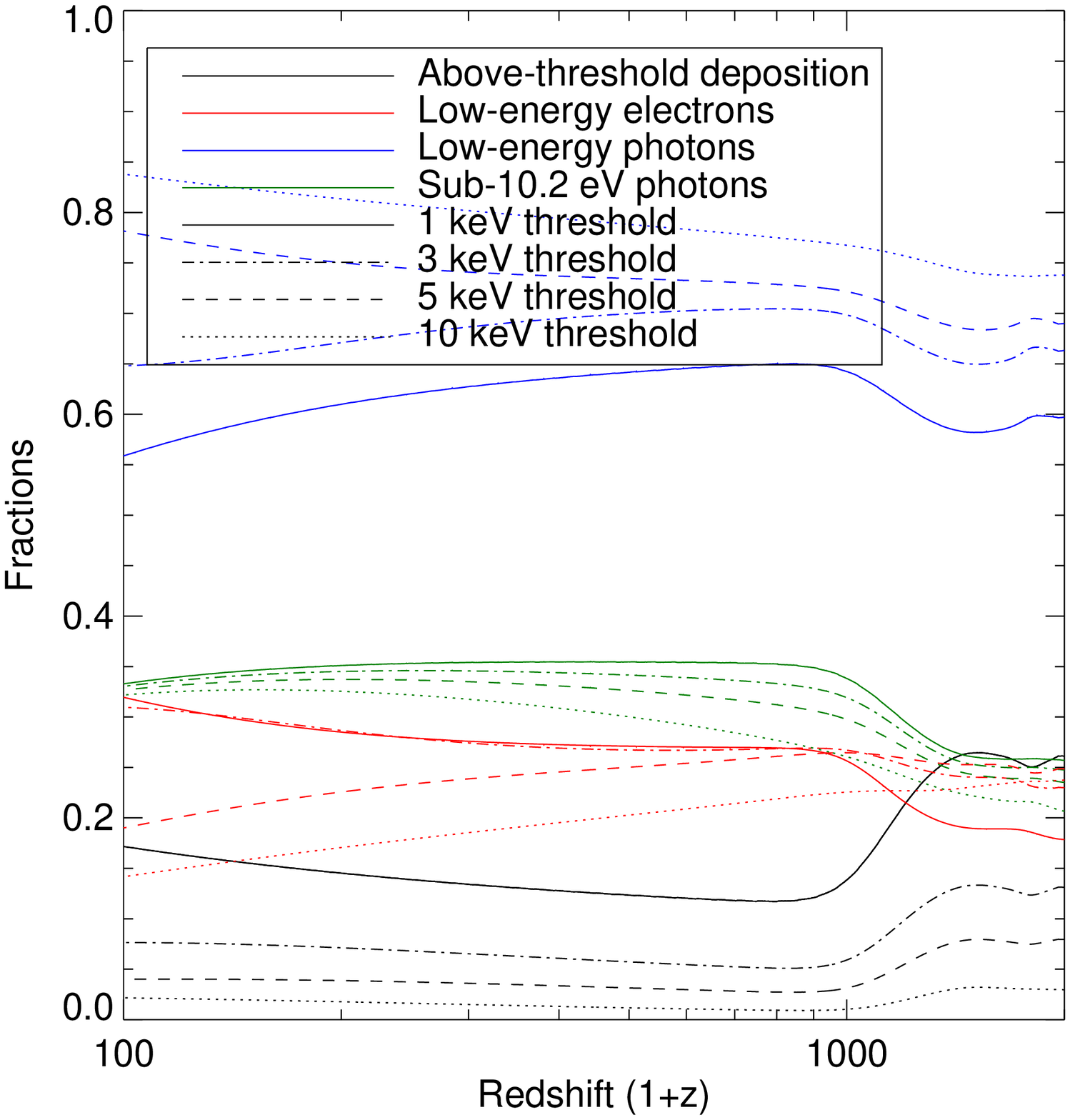} \\
\caption{\label{fig:fractions}
Fraction of ``deposited'' energy attributed to (\emph{red}) sub-threshold electrons, (\emph{blue}) sub-threshold ``deposited'' photons, (\emph{green}) sub-10.2-eV photons (subset of ``deposited'' photons), (\emph{black}) energy ``deposited'' by above-threshold electrons and photons (not tracked in detail). Solid lines use a 1 keV threshold; dot-dashed, dashed and dotted lines employ thresholds at 3 keV, 5 keV and 10 keV respectively. The left-hand panels show the results for DM annihilating directly to electrons, the right-hand panels for DM annihilating directly to muons; the rows correspond to DM masses of 1 GeV, 10 GeV, 100 GeV and 1 TeV DM.}
\end{figure*}

\begin{figure*}
\includegraphics[width=.3\textwidth]{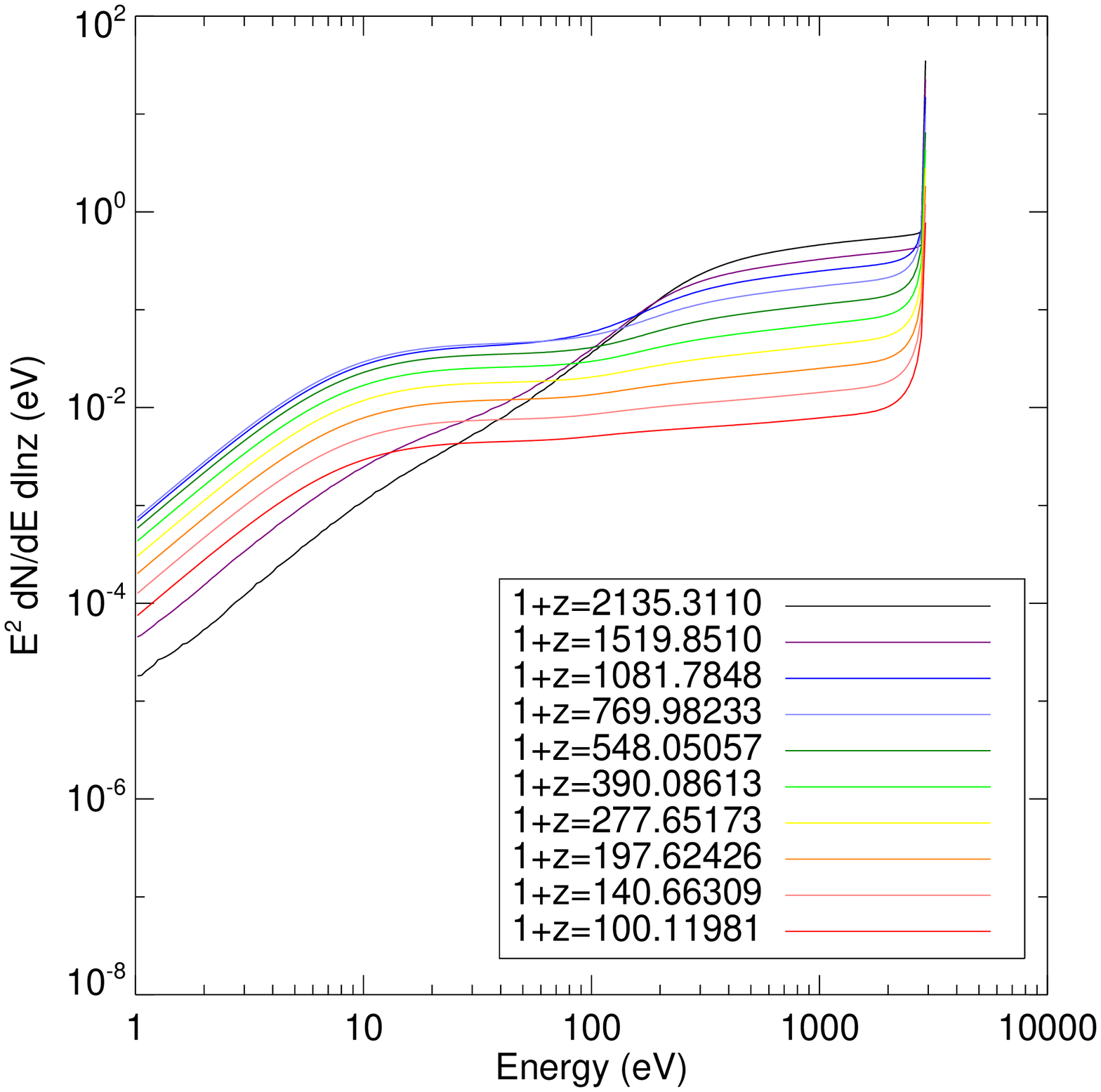}
\includegraphics[width=.3\textwidth]{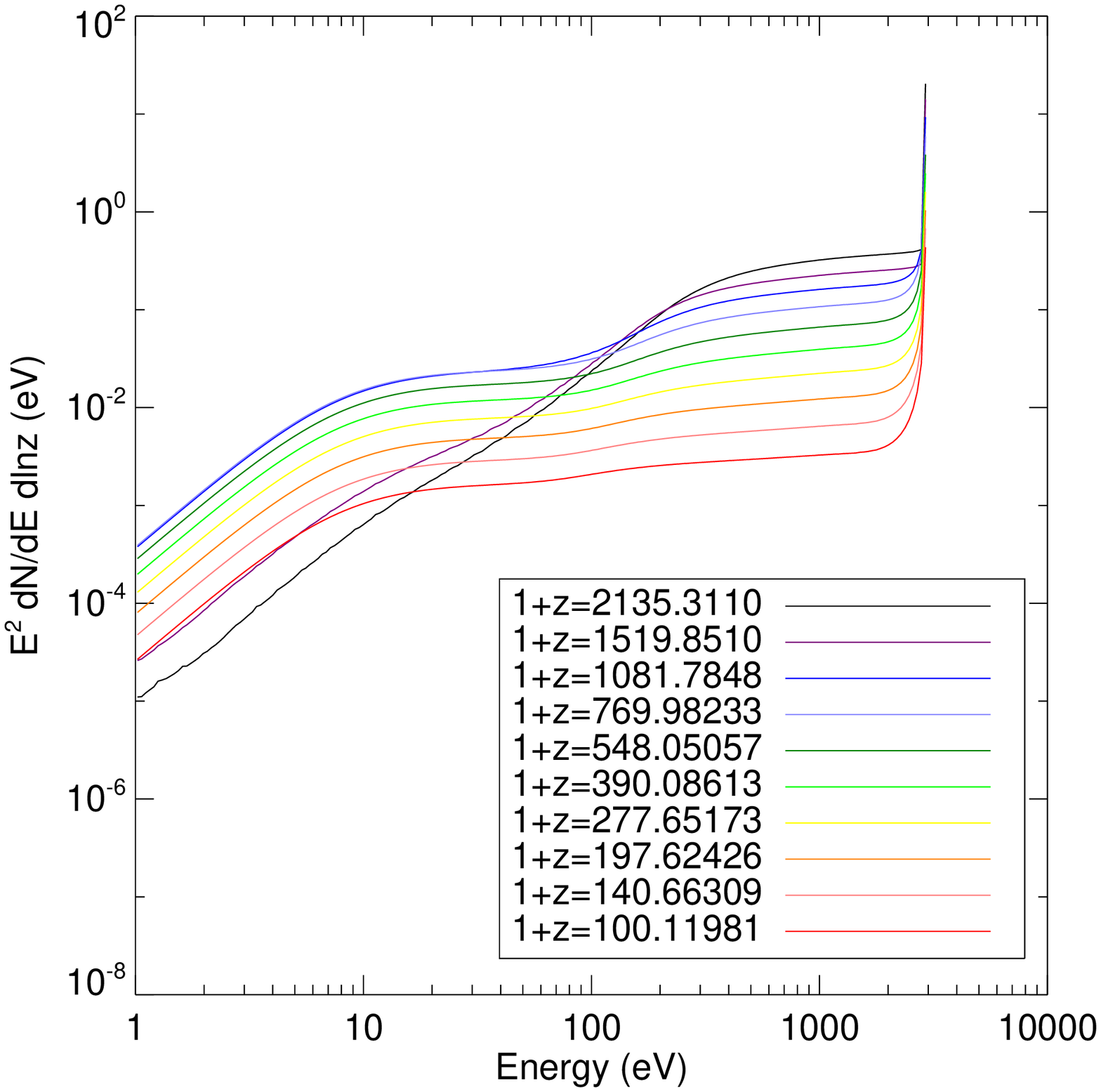} \\
\includegraphics[width=.3\textwidth]{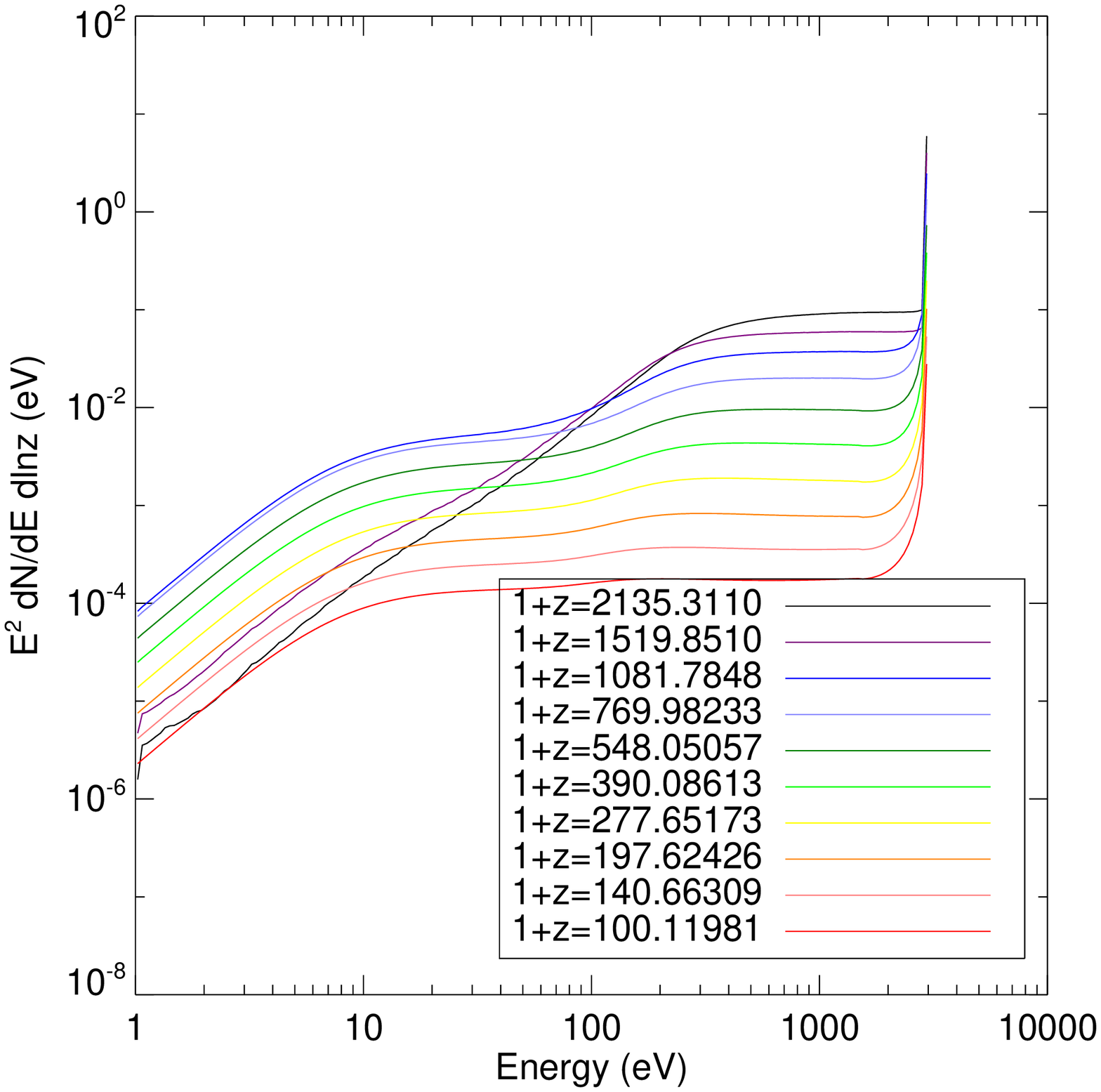} 
\includegraphics[width=.3\textwidth]{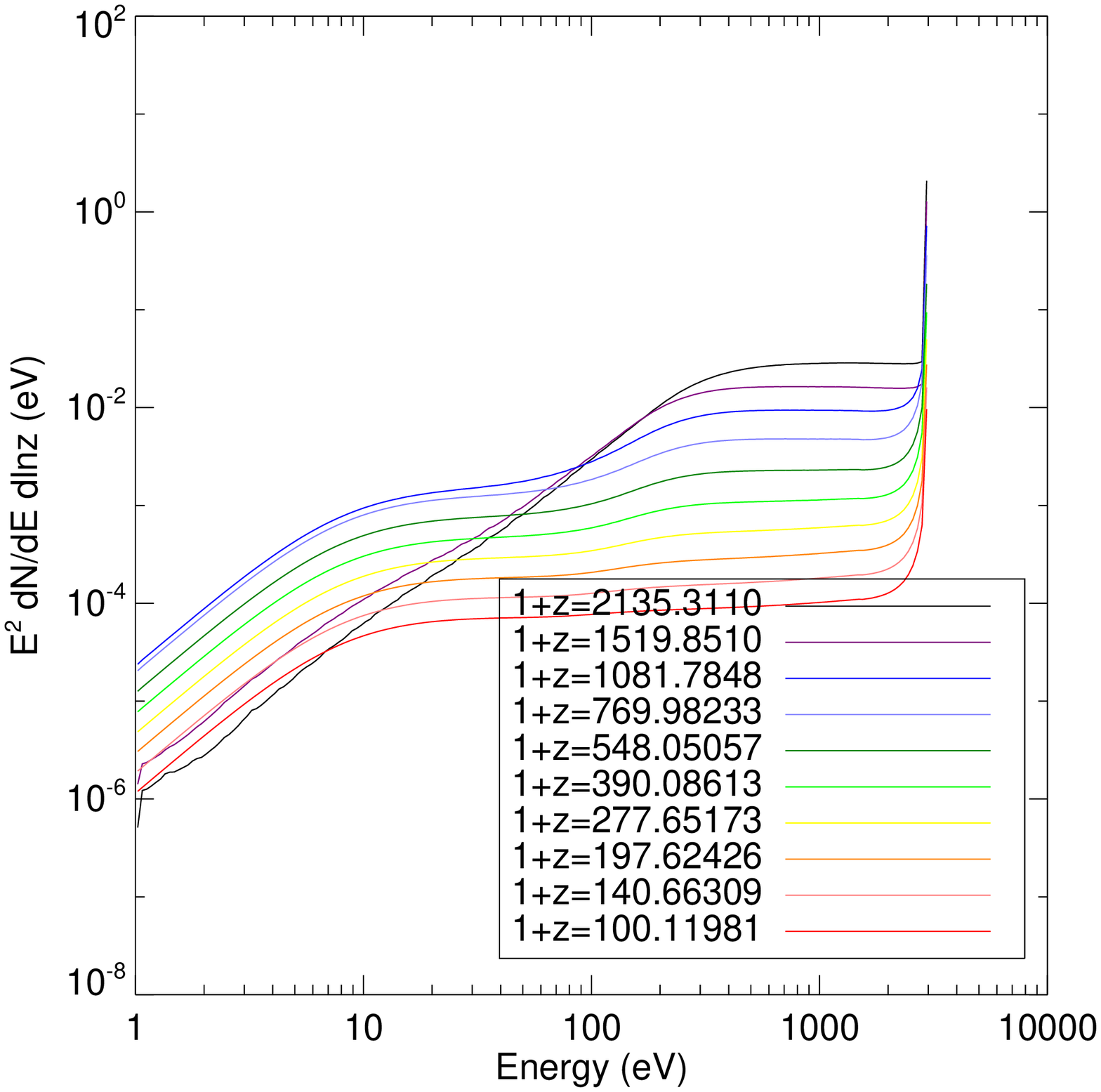}\\
\includegraphics[width=.3\textwidth]{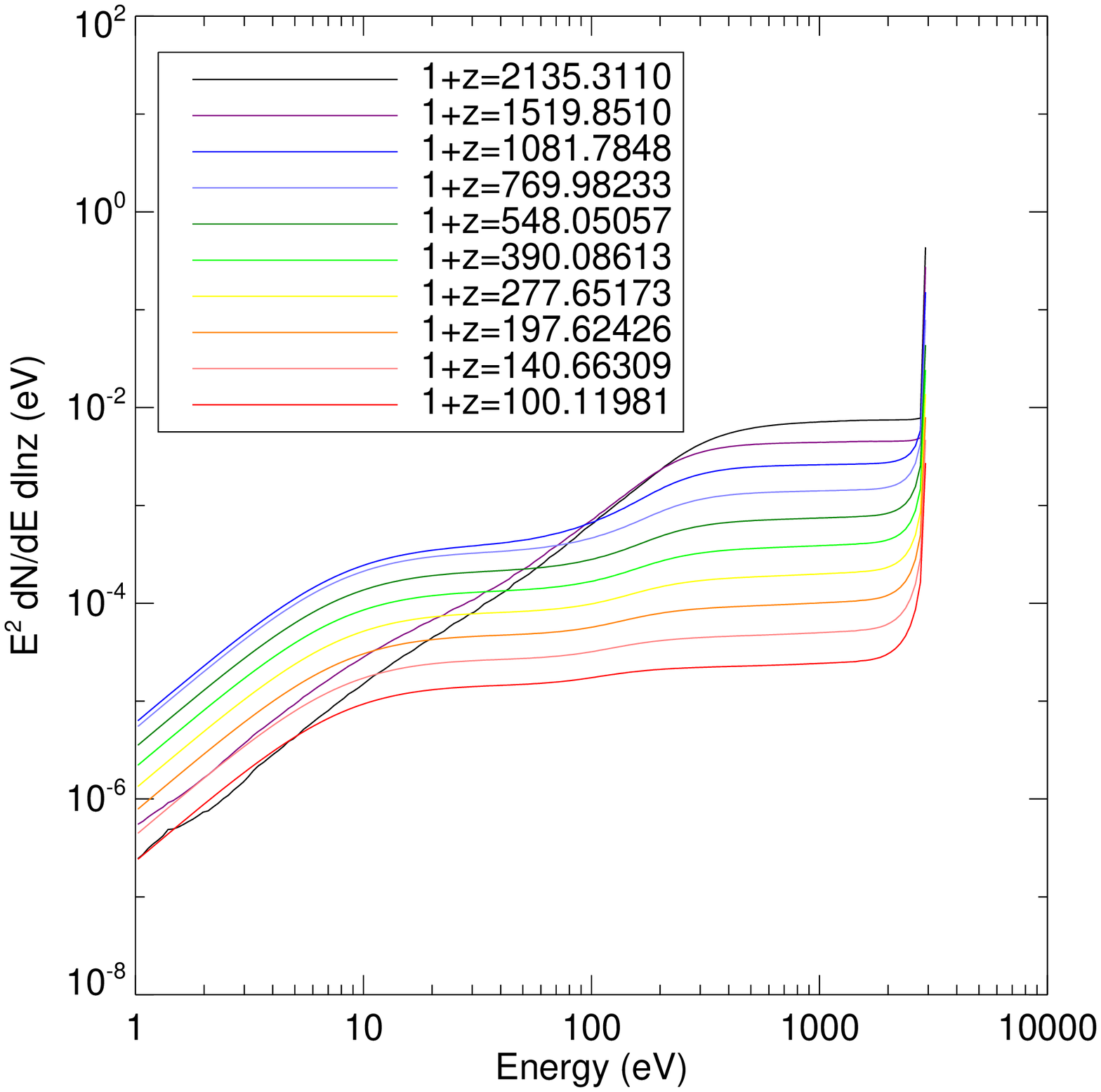}
\includegraphics[width=.3\textwidth]{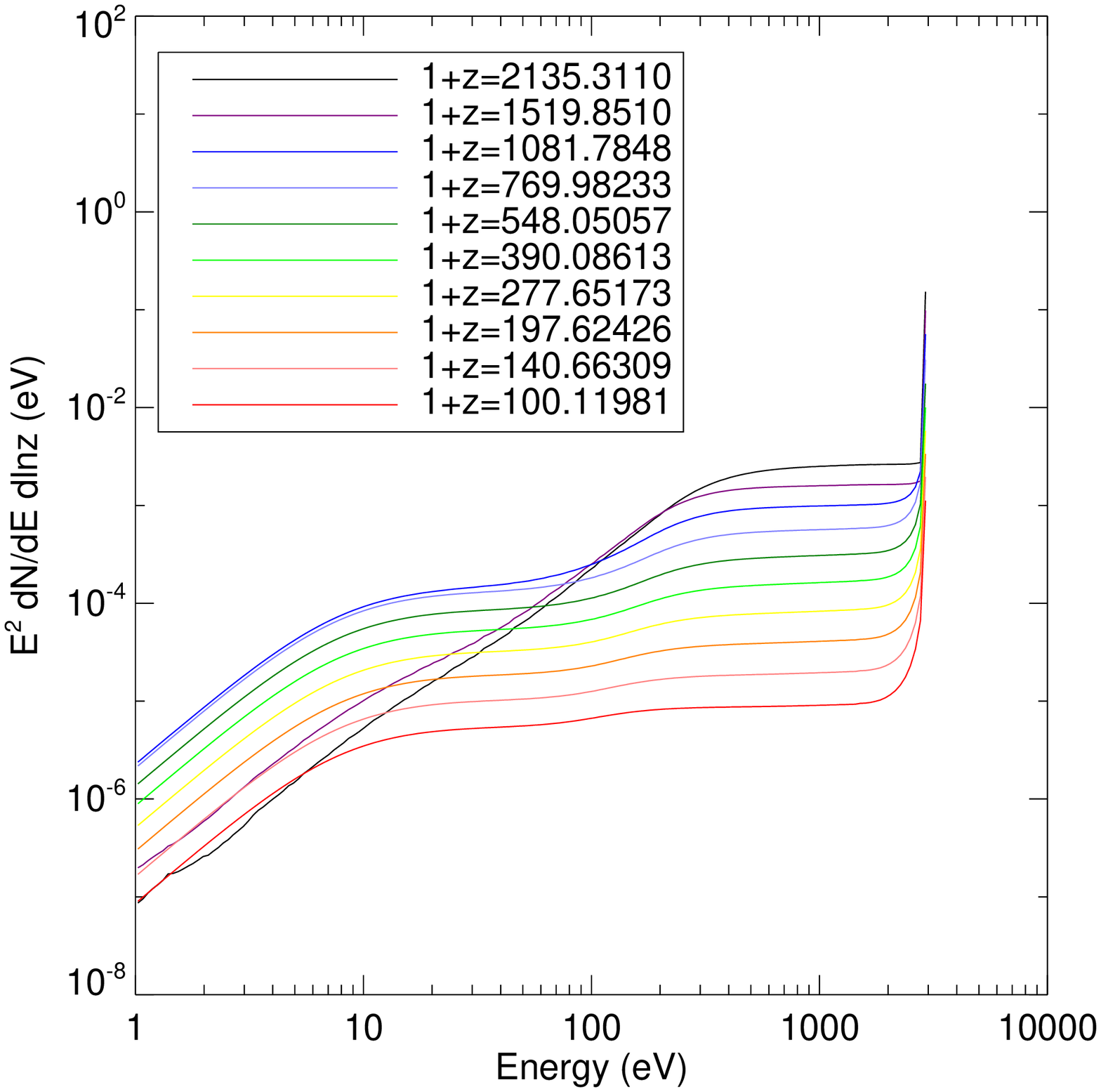} \\
\includegraphics[width=.3\textwidth]{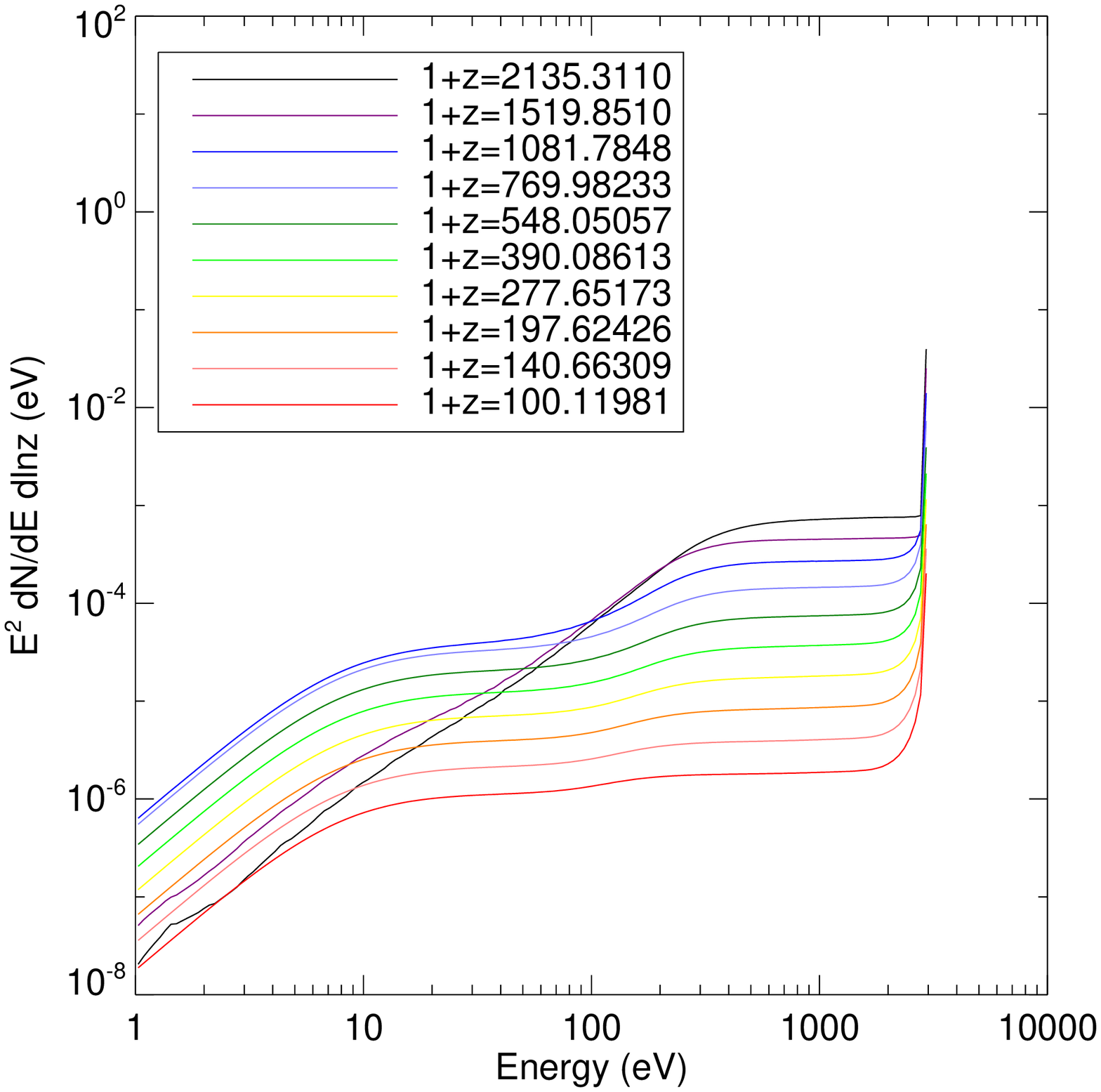}
\includegraphics[width=.3\textwidth]{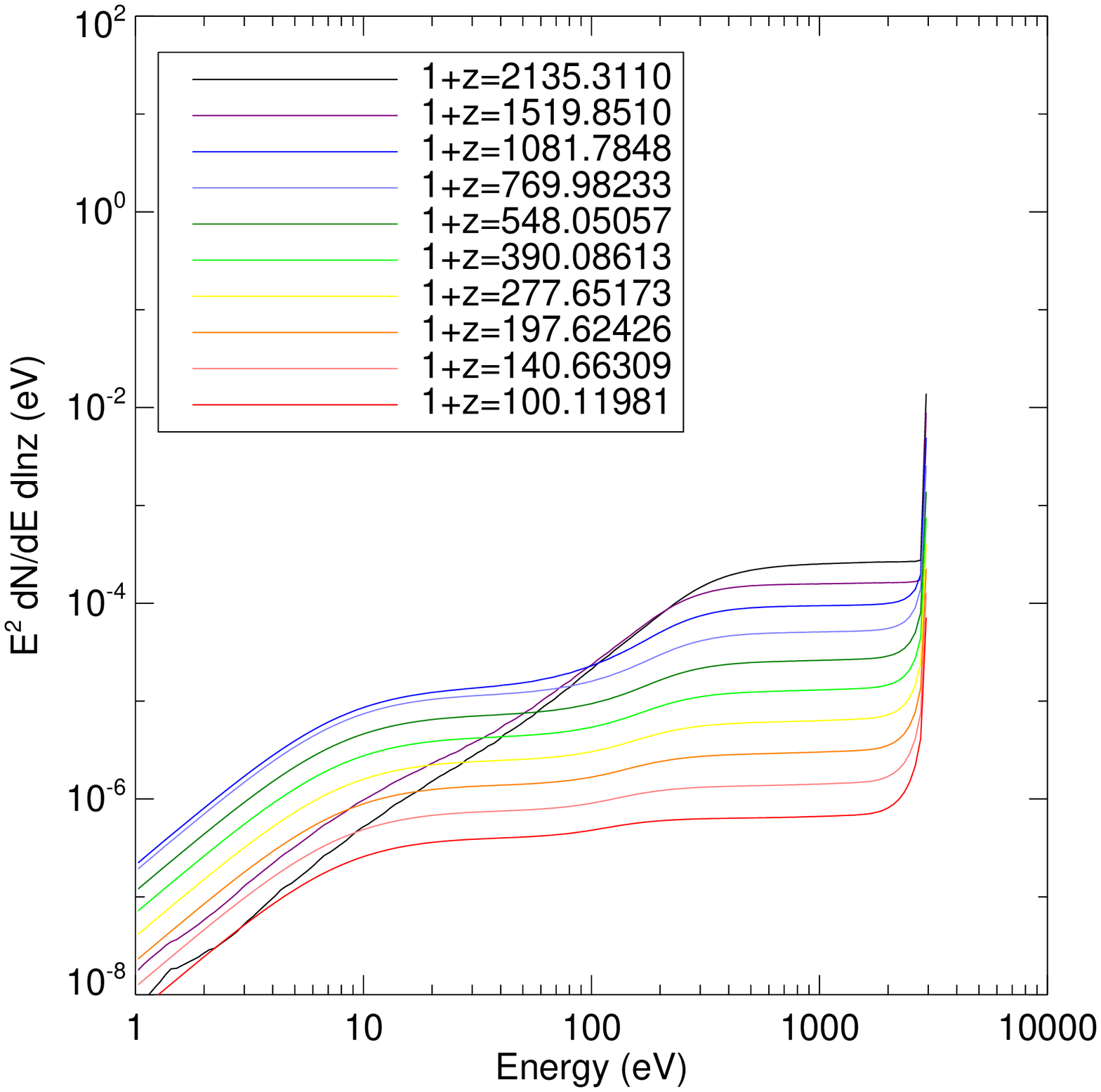}
\caption{\label{fig:elecspec}
Spectra of ``deposited'' electrons, below the 3-keV threshold, in $E^2 dN/dE d \ln z$, at a range of sample redshifts from $z\sim 60$ to $z \sim 2000$. The left-hand panels show the results for DM annihilating directly to electrons, the right-hand panels for DM annihilating directly to muons; the rows correspond to 1 GeV, 10 GeV, 100 GeV and 1 TeV DM. The normalization is per baryon, and assumes an annihilation cross section of $\langle \sigma v \rangle = 3 \times 10^{-26}$ cm$^3$/s.}
\end{figure*}

\begin{figure*}
\includegraphics[width=.3\textwidth]{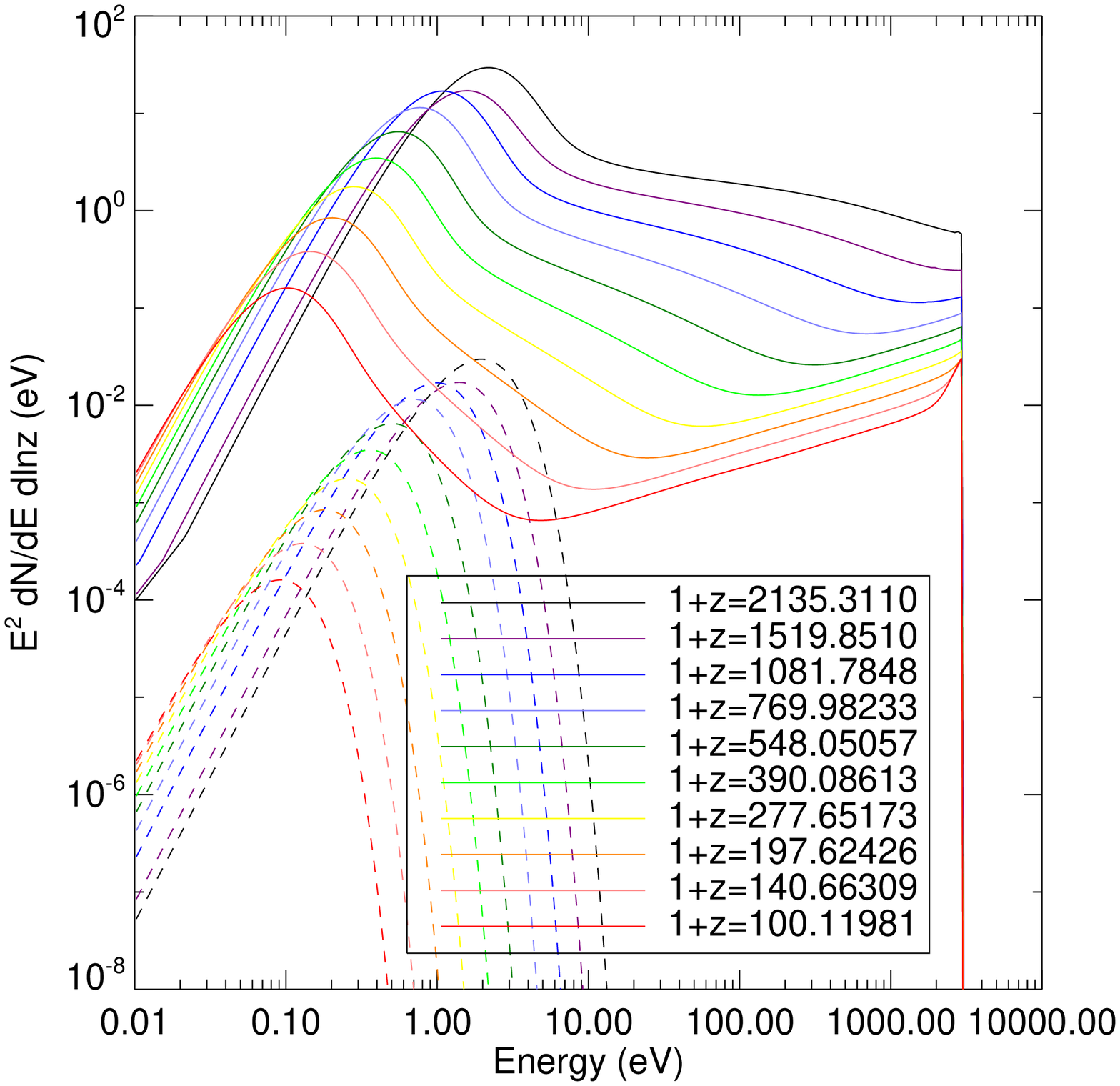}
\includegraphics[width=.3\textwidth]{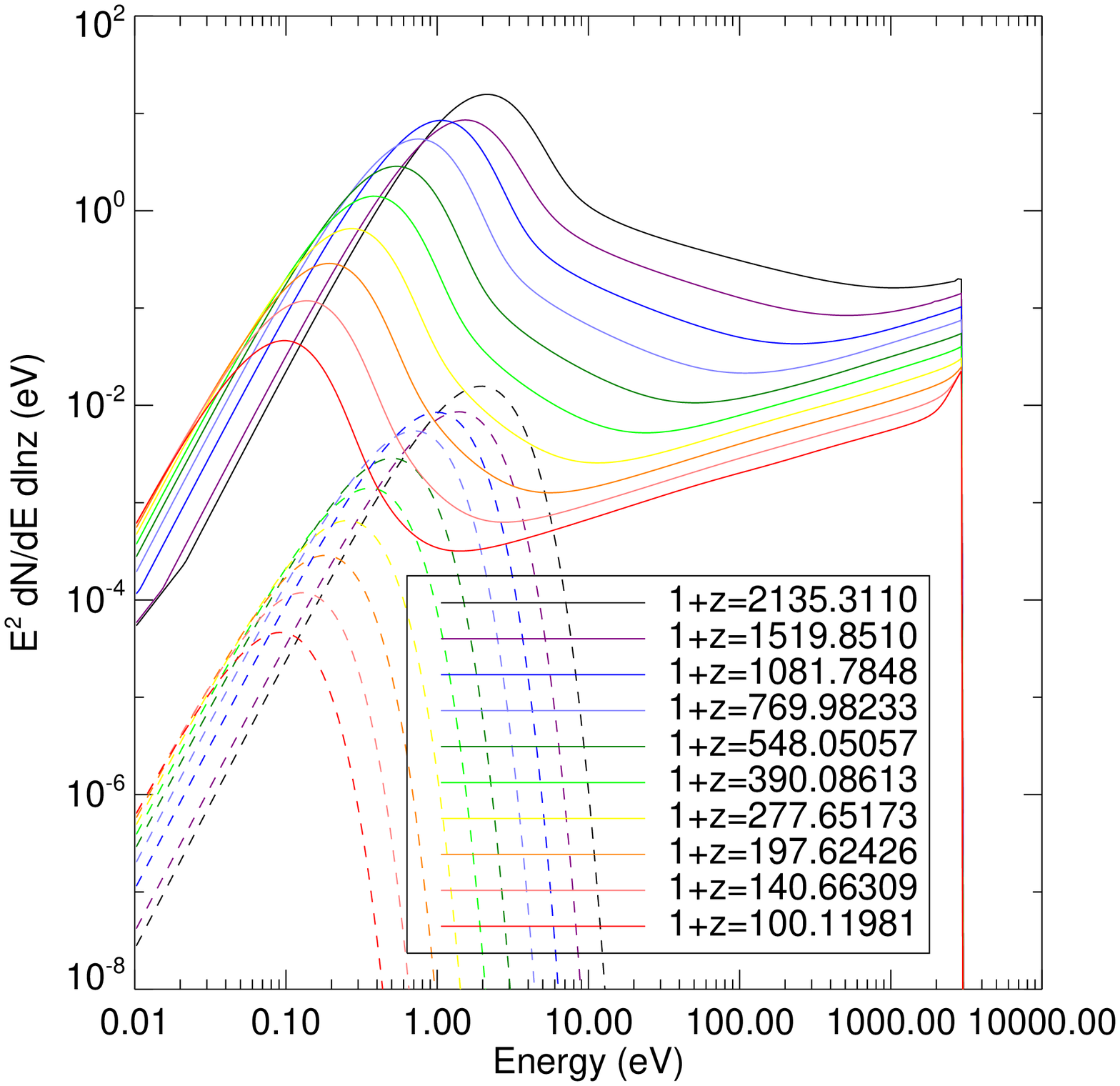} \\
\includegraphics[width=.3\textwidth]{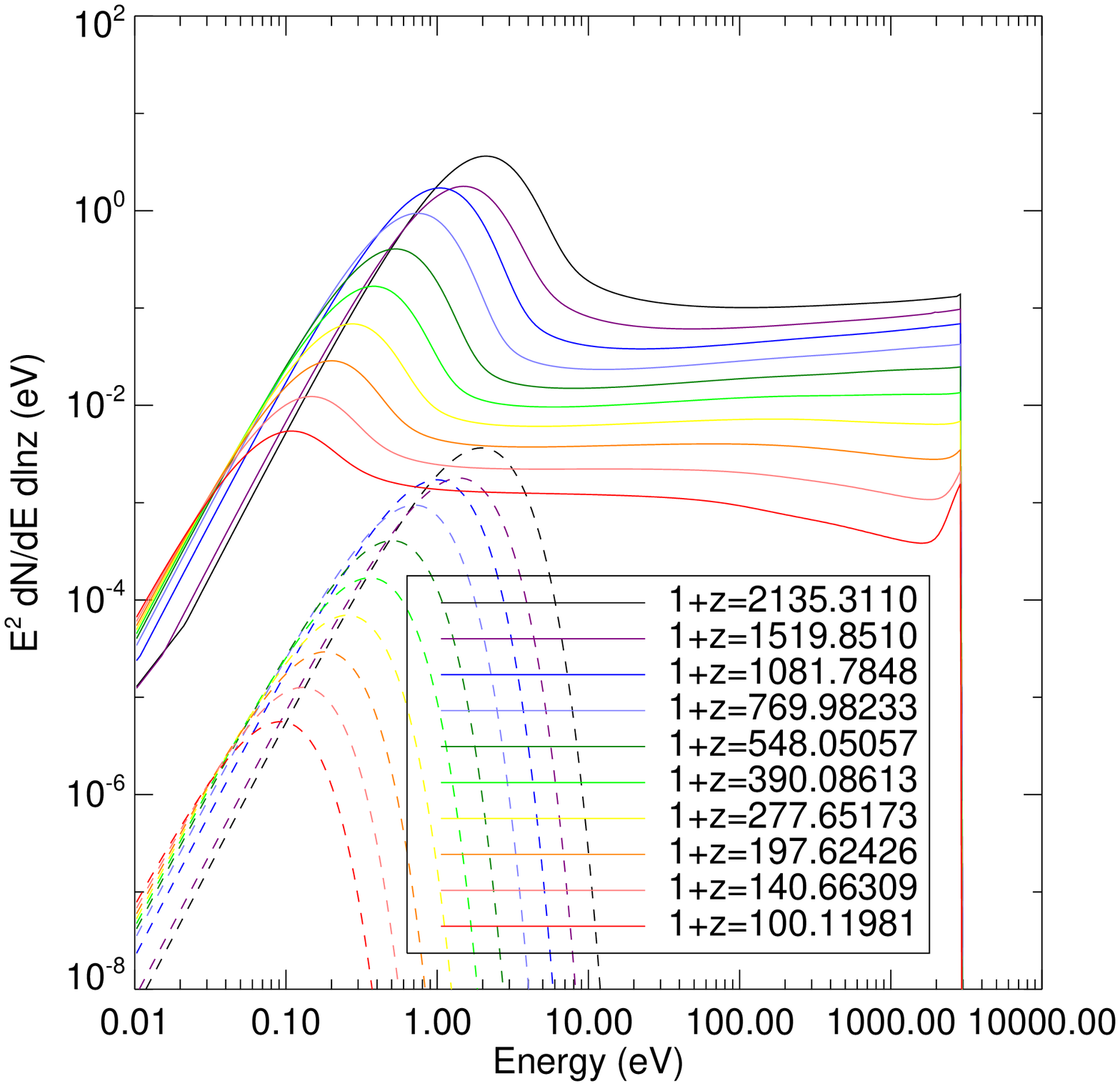} 
\includegraphics[width=.3\textwidth]{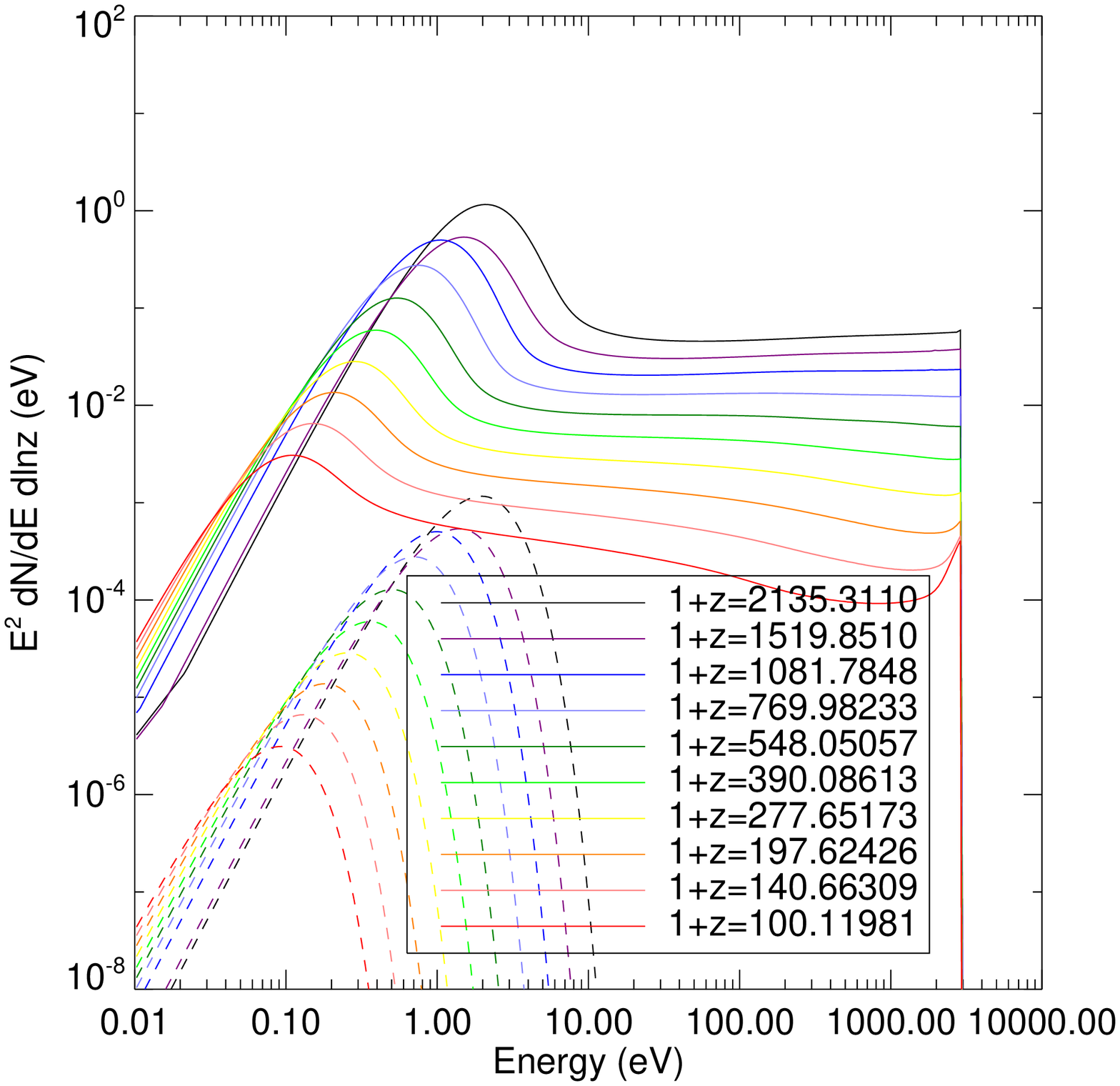}\\
\includegraphics[width=.3\textwidth]{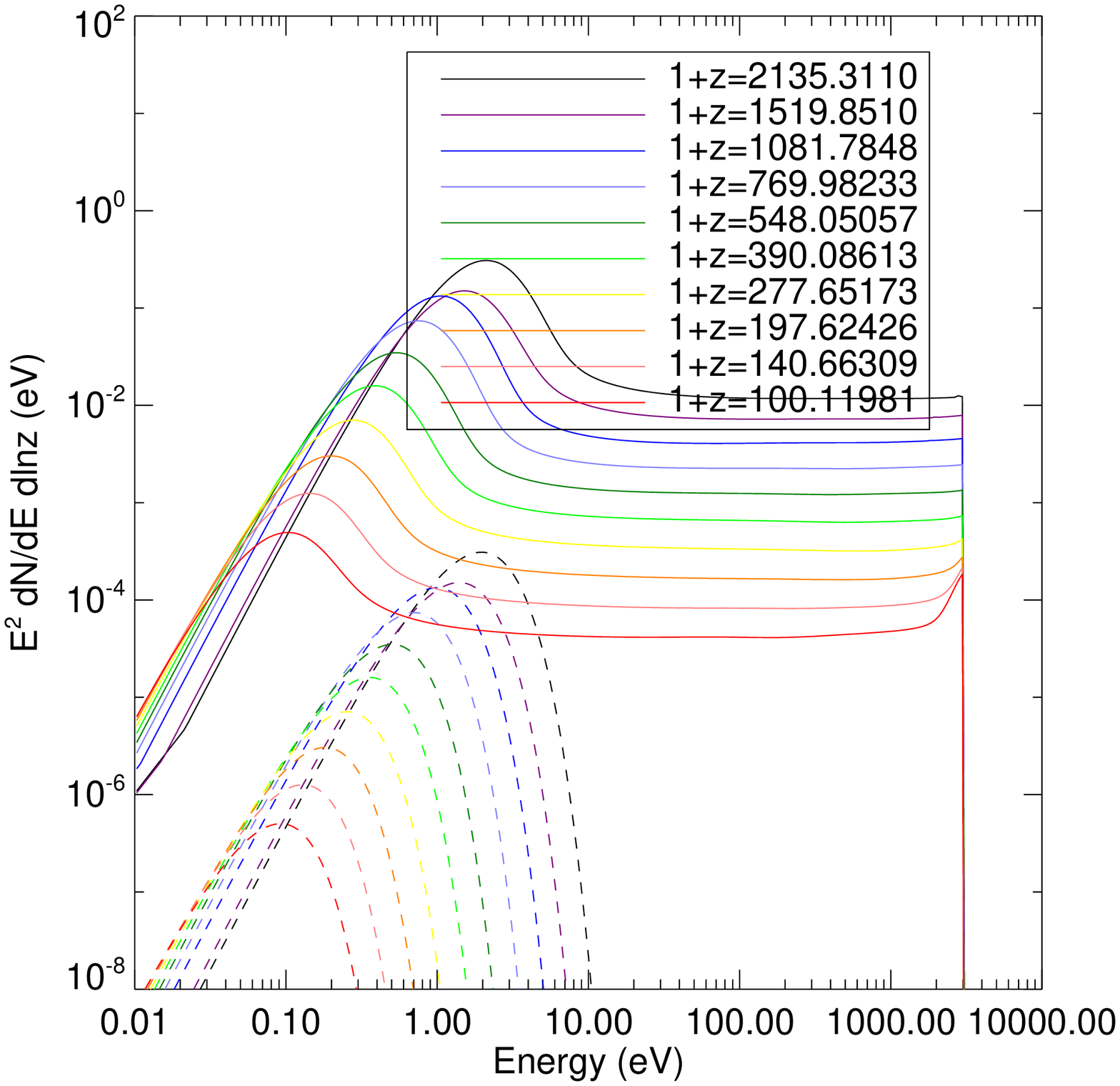}
\includegraphics[width=.3\textwidth]{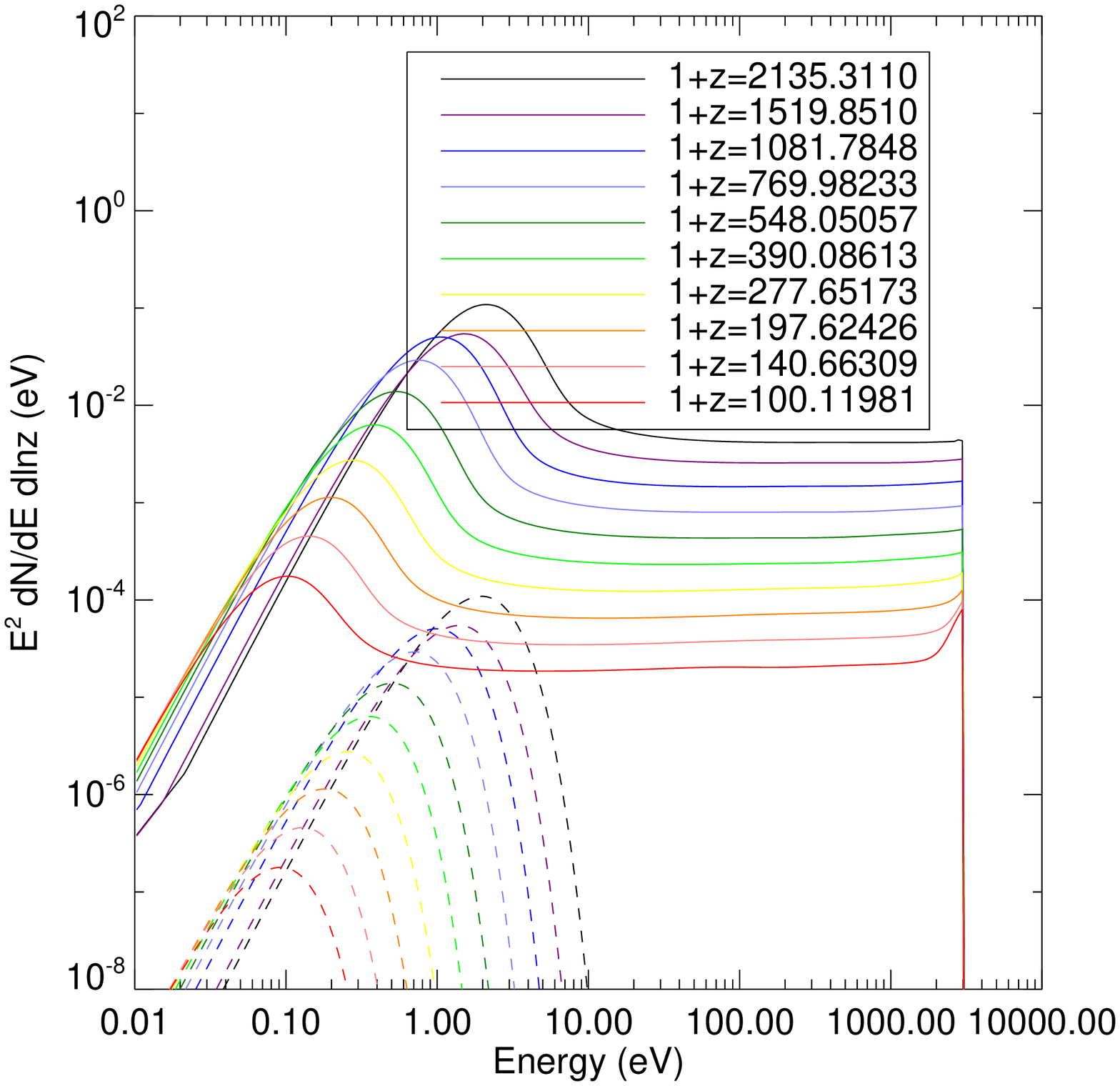} \\
\includegraphics[width=.3\textwidth]{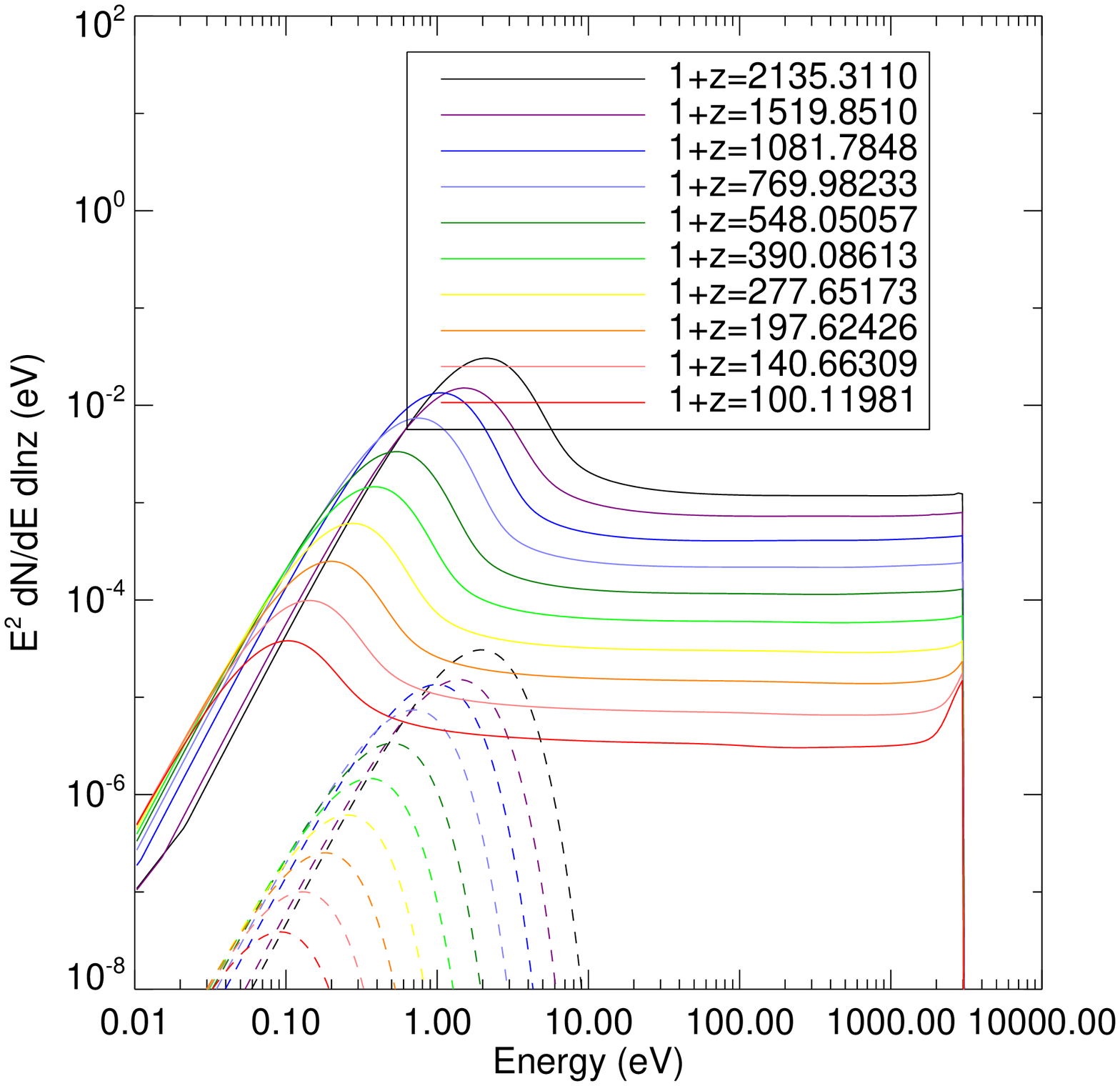}
\includegraphics[width=.3\textwidth]{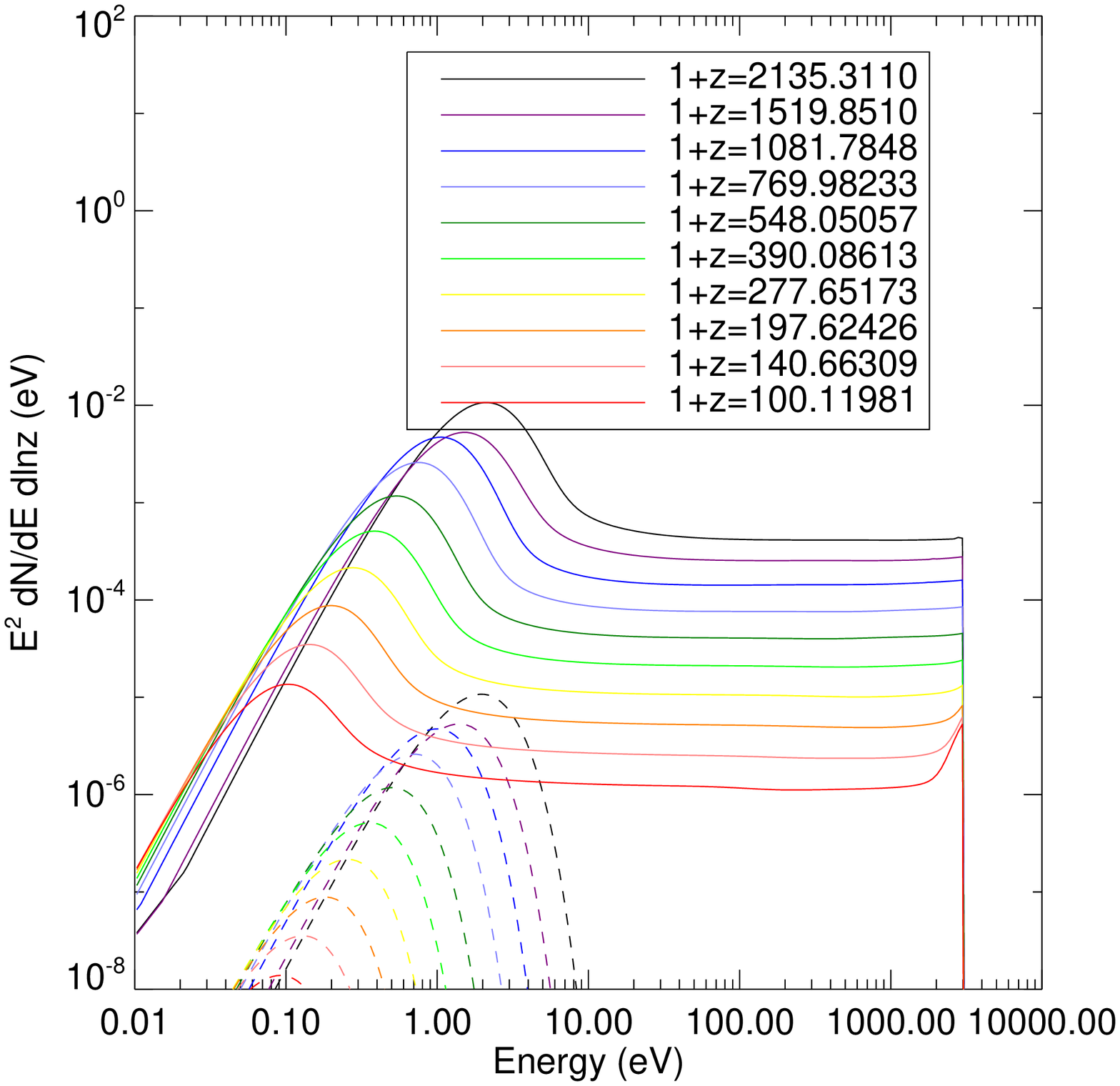}
\caption{\label{fig:photspec}
Spectra of ``deposited'' photons, below the 3-keV threshold, in $E^2 dN/dE d \ln z$, at a range of sample redshifts from $z\sim 60$ to $z \sim 2000$.The left-hand panels show the results for DM annihilating directly to electrons, the right-hand panels for DM annihilating directly to muons; the rows correspond to 1 GeV, 10 GeV, 100 GeV and 1 TeV DM. The dashed lines show the spectrum of the photons \emph{removed} from the CMB by scattering in the corresponding timestep, and should be subtracted from the solid lines to obtain the net change in the photon spectrum. The normalization is per baryon, and assumes an annihilation cross section of $\langle \sigma v \rangle = 3 \times 10^{-26}$ cm$^3$/s.}
\end{figure*}
We display the results of the high energy code for eight sample DM models (1 GeV, 10 GeV, 100 GeV and 1 TeV DM annihilating to $e^+ e^-$ or $\mu^+ \mu^-$) in Figures \ref{fig:fractions}-\ref{fig:photspec}. In Figure \ref{fig:fractions} we plot the fraction of the total ``deposited energy'' in each of the components described above, as a function of redshift. In Figures \ref{fig:elecspec} and \ref{fig:photspec} we plot the sub-threshold electron and photon spectra for a range of sample redshifts. 

\subsection{Improving on $f(z)$}
\label{sec:highsystematicseffective}

As mentioned briefly above, it is possible to improve on the parameterization of the high-energy results as a single function $f(z)$ which multiplies the energy fractions $\chi_x$ at threshold, to determine the final partition of the injected energy between ionization, Lyman-$\alpha$ photons and heating. This parameterization is simple, but may be inaccurate. We defer a detailed exploration of such improvements to future work, and here simply try to estimate their potential effect on the constraints.

The potential differences come from (1) new contributions to ionization, apart from the cooling of sub-threshold electrons, and (2) integrating the energy fractions over the sub-threshold electron spectrum rather than simply taking their values at threshold. Neither of these can be determined simply from a ``total absorbed energy'' $f(z)$ function and the energy fractions at threshold.

However, the constraints on annihilation are driven almost entirely by changes to the ionization history; we have shown in Section \ref{sec:lymanalpha} that the effects of Lyman-$\alpha$ photons are small. Consequently, we can write down an $f_\mathrm{effective}$ curve which describes the total amount of energy absorbed as ionization at a given redshift, divided by $\chi_i \times$ the amount of energy injected at that redshift (with $\chi_i$ being evaluated at threshold). In the limit where $\chi_i$ (at threshold) correctly describes the fraction of absorbed energy going into ionization, this reduces to the previous expression for $f(z)$, and when this $f_\mathrm{effective}$ curve is multiplied by $\chi_i$ and the total injected power, one recovers the correct total power into ionization. While multiplying the $f_\mathrm{effective}$ curve by the other $\chi_x$ fractions will not give exactly the correct power into those channels, the resulting error in the constraints should be small.

We will show that the true constraint can be bracketed by using our best-estimate $f_\mathrm{effective}$ curve and the $f_\mathrm{approx}$ curve given above, and the difference between the two is reasonably small, at the level of $10-15\%$.

\begin{figure*}
\includegraphics[width=.3\textwidth]{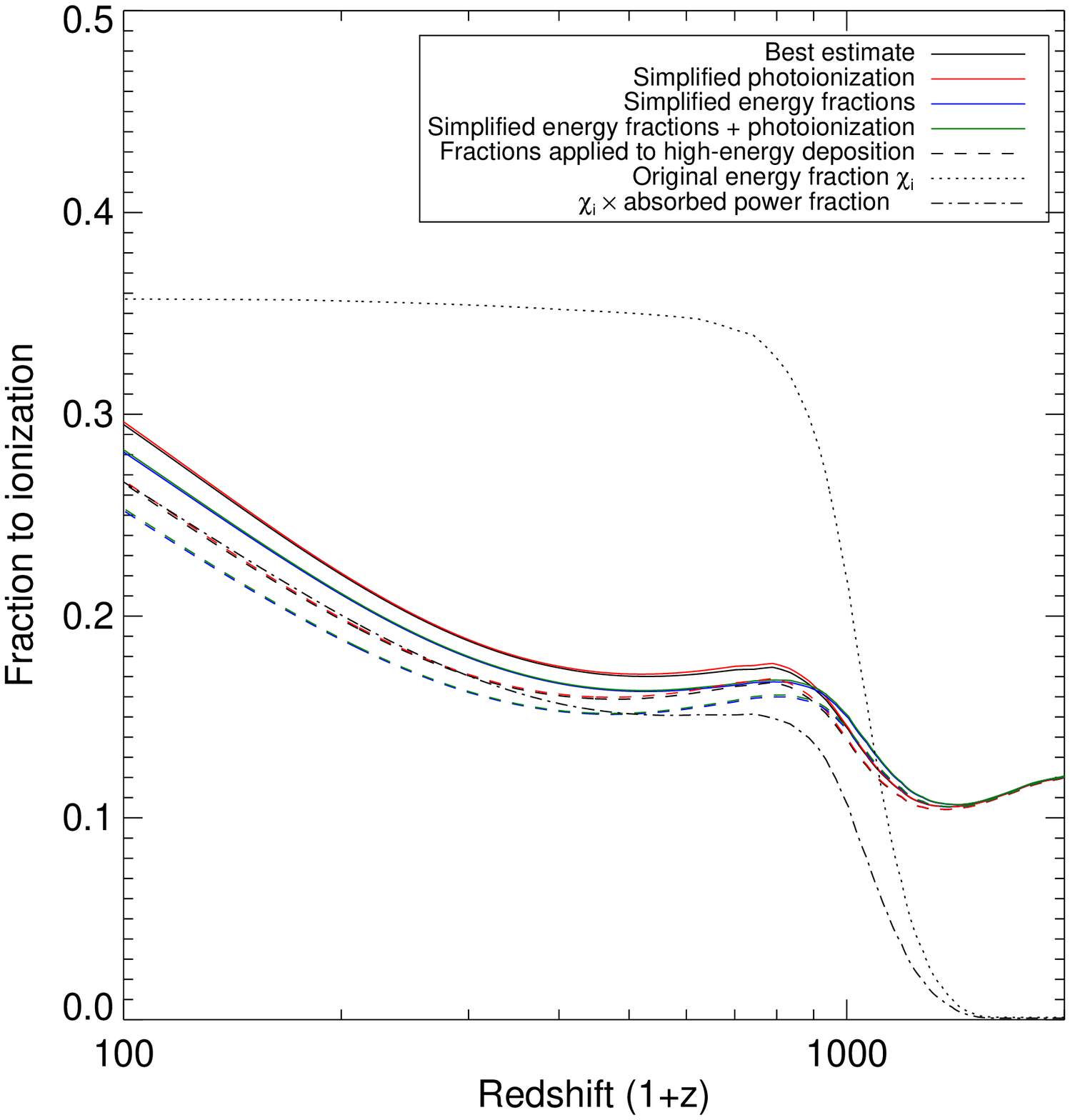}
\includegraphics[width=.3\textwidth]{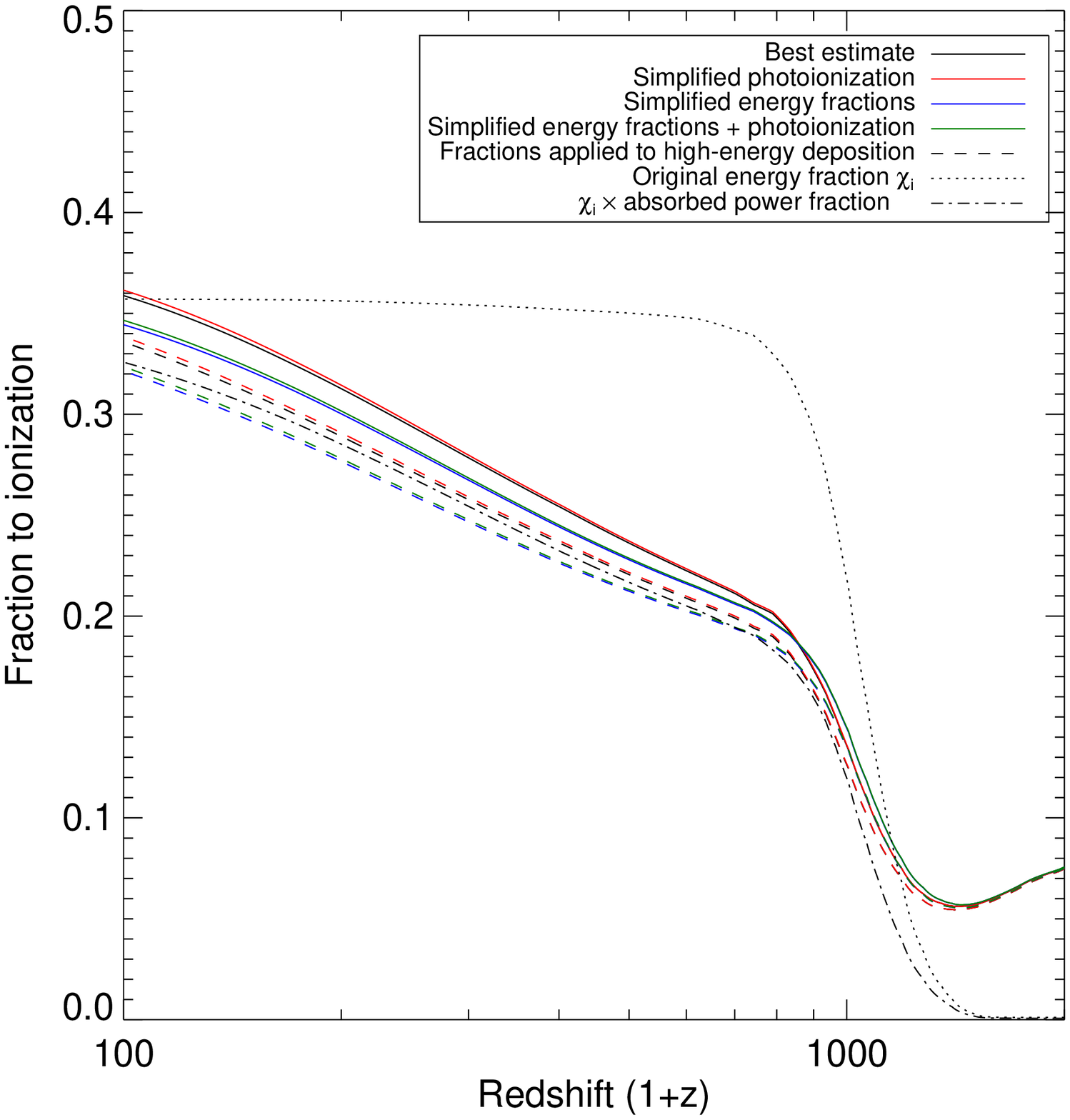} \\
\includegraphics[width=.3\textwidth]{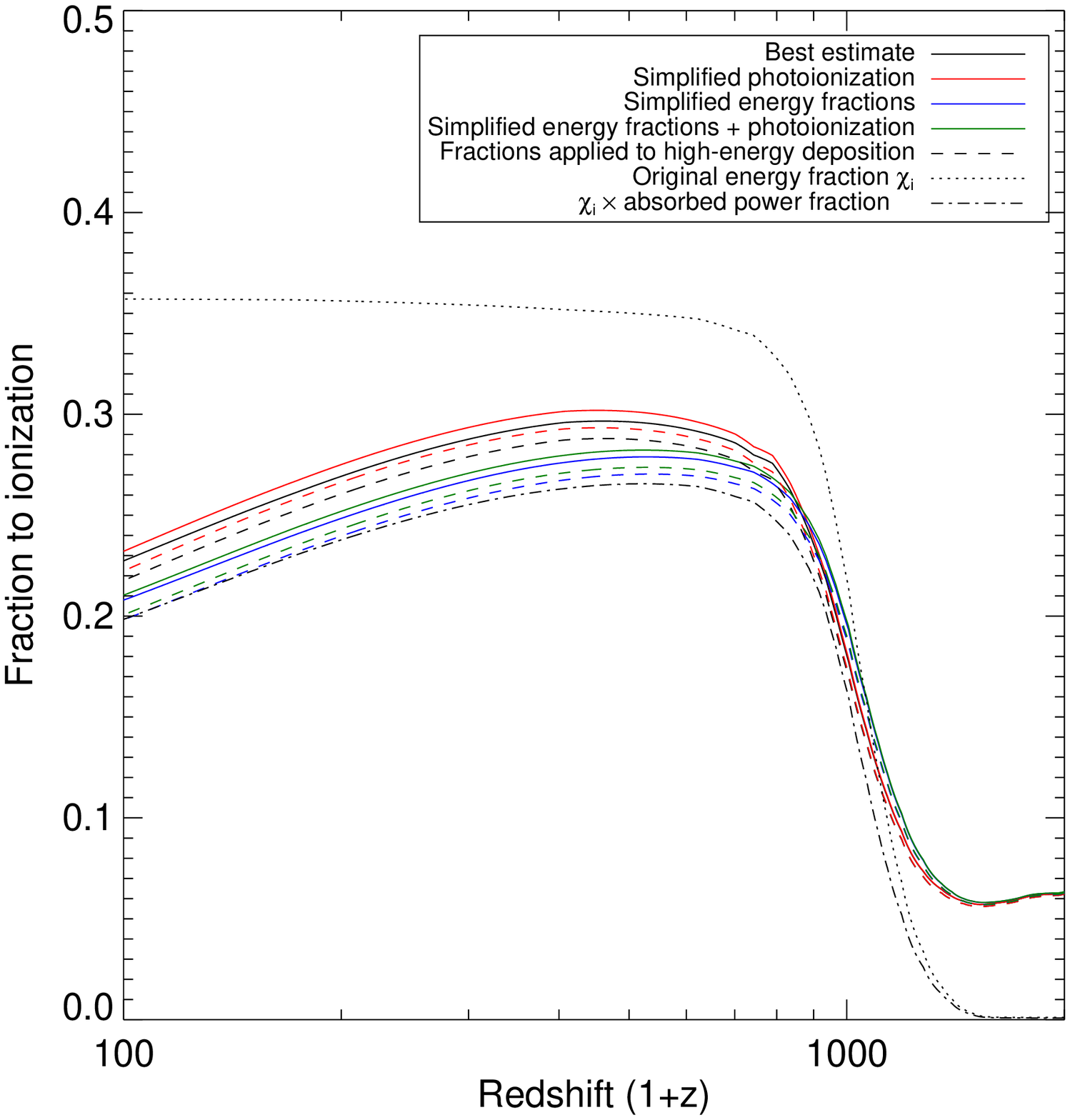} 
\includegraphics[width=.3\textwidth]{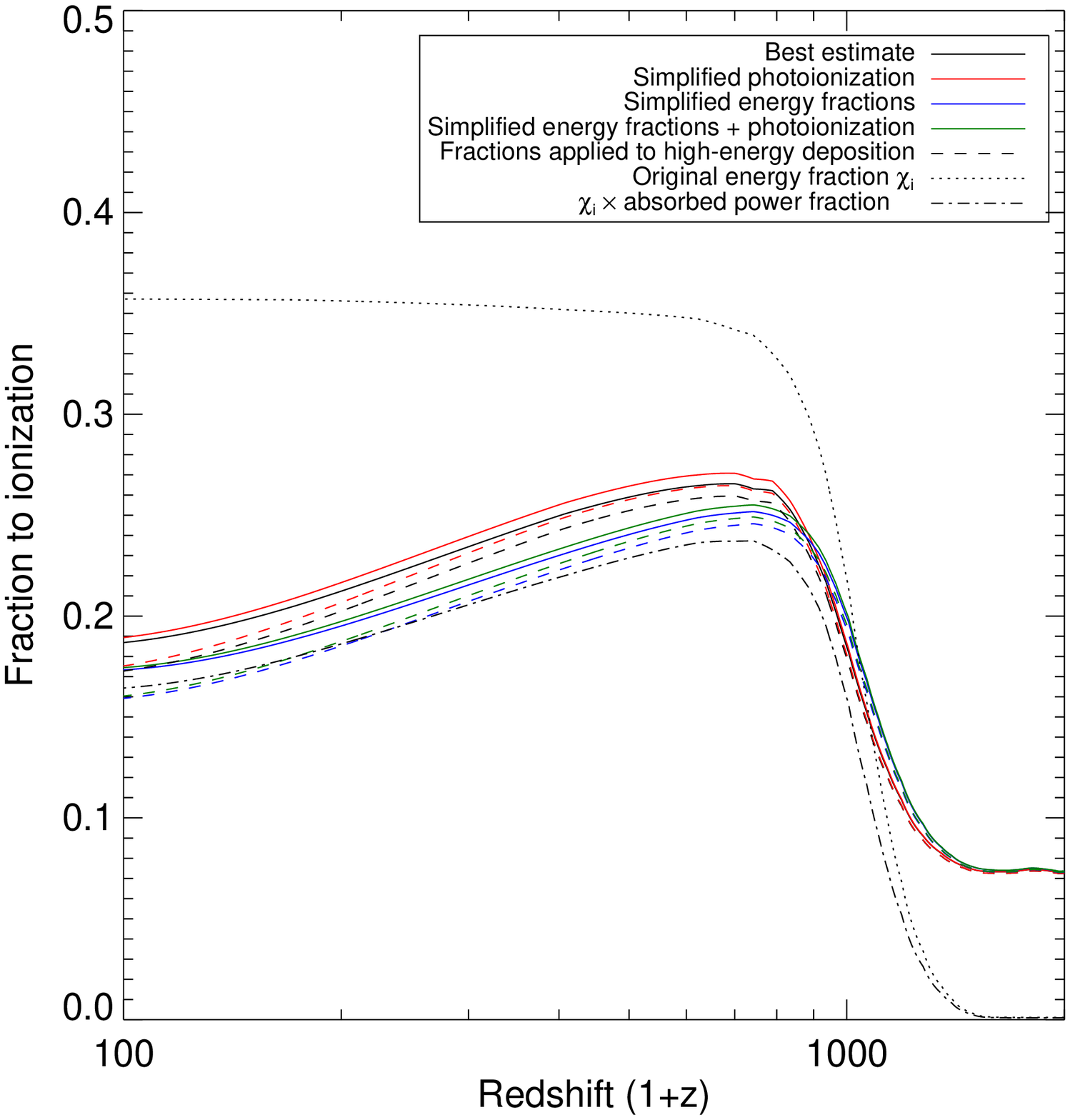}\\
\includegraphics[width=.3\textwidth]{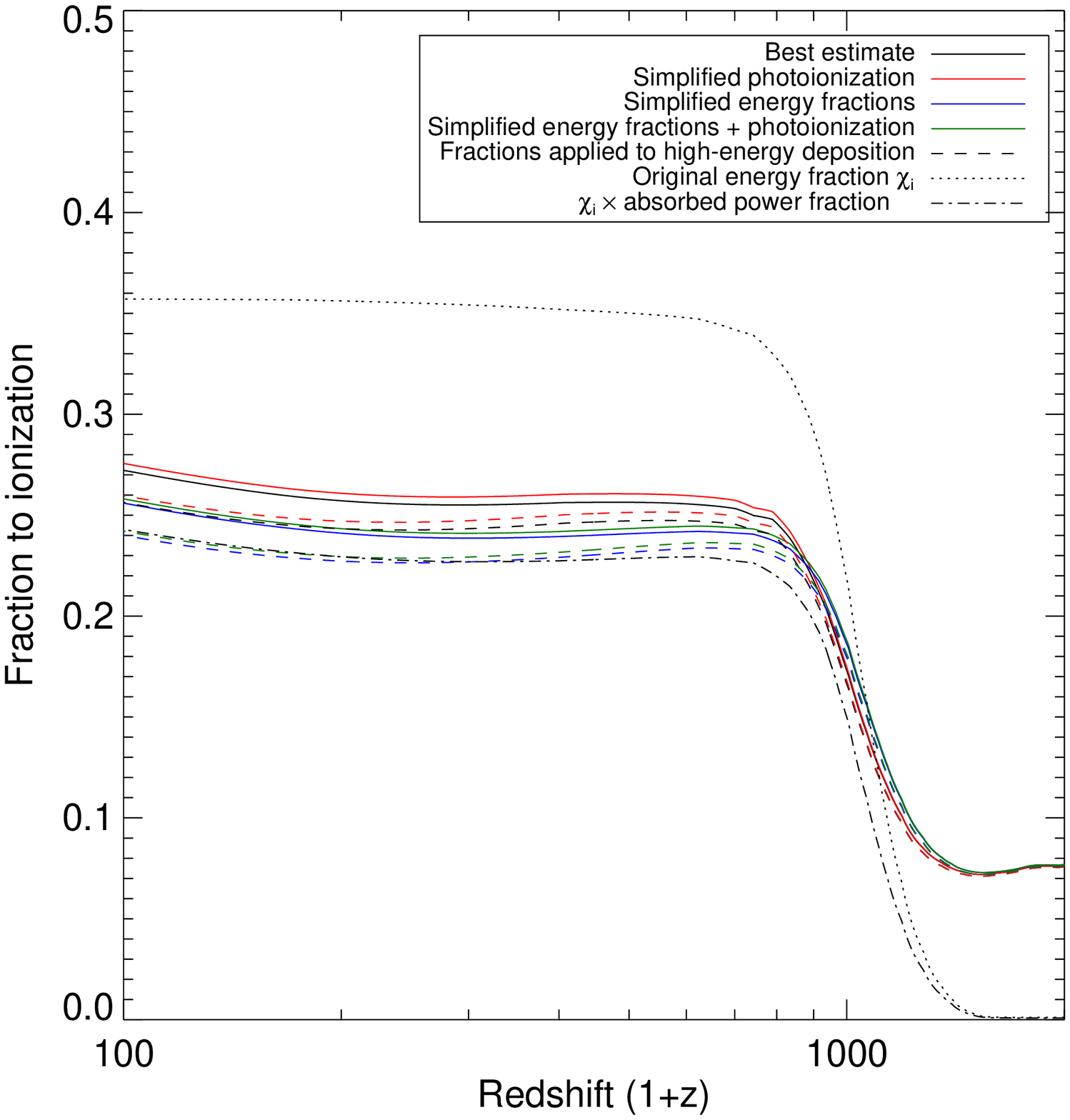} 
\includegraphics[width=.3\textwidth]{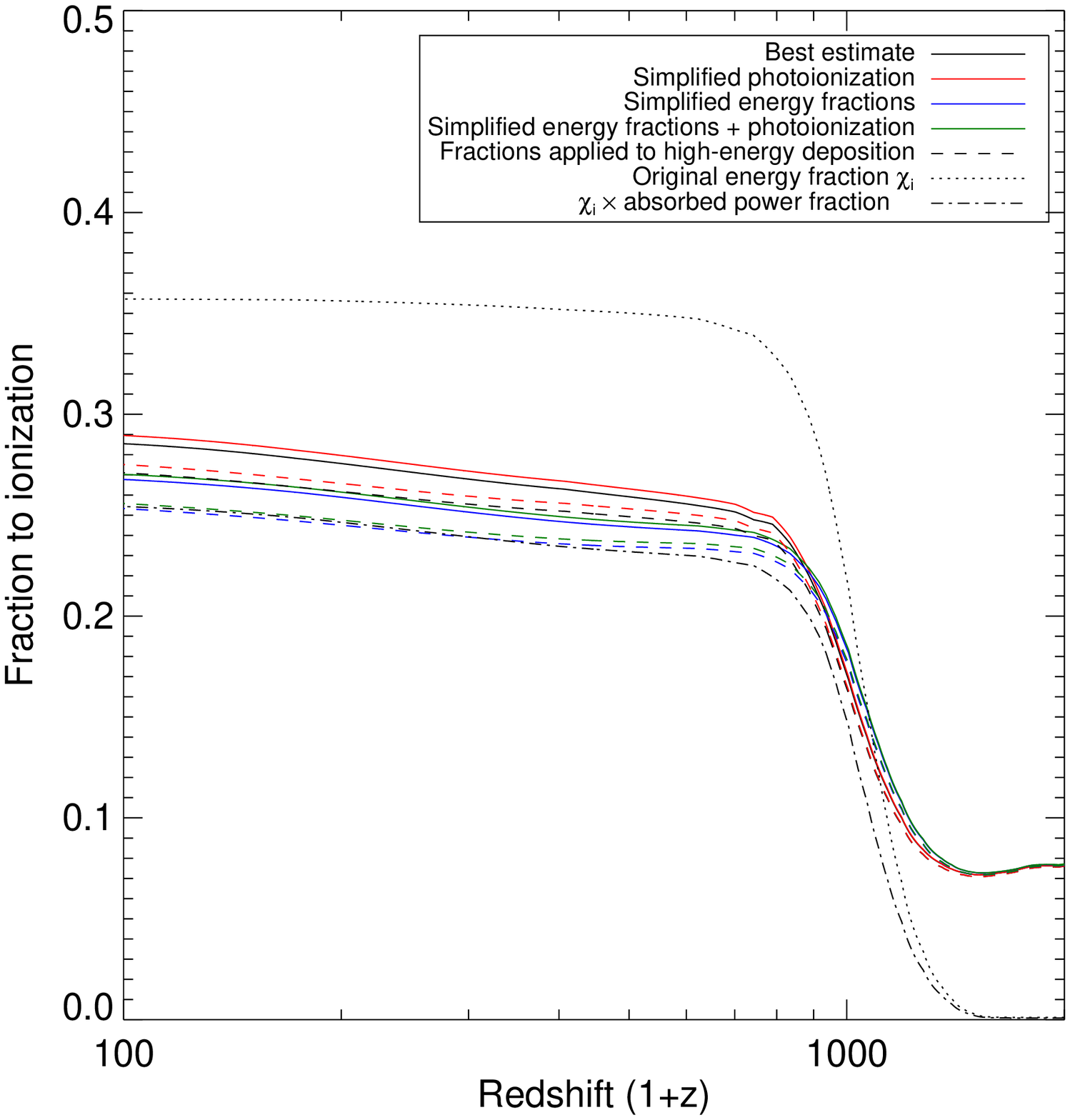}\\
\includegraphics[width=.3\textwidth]{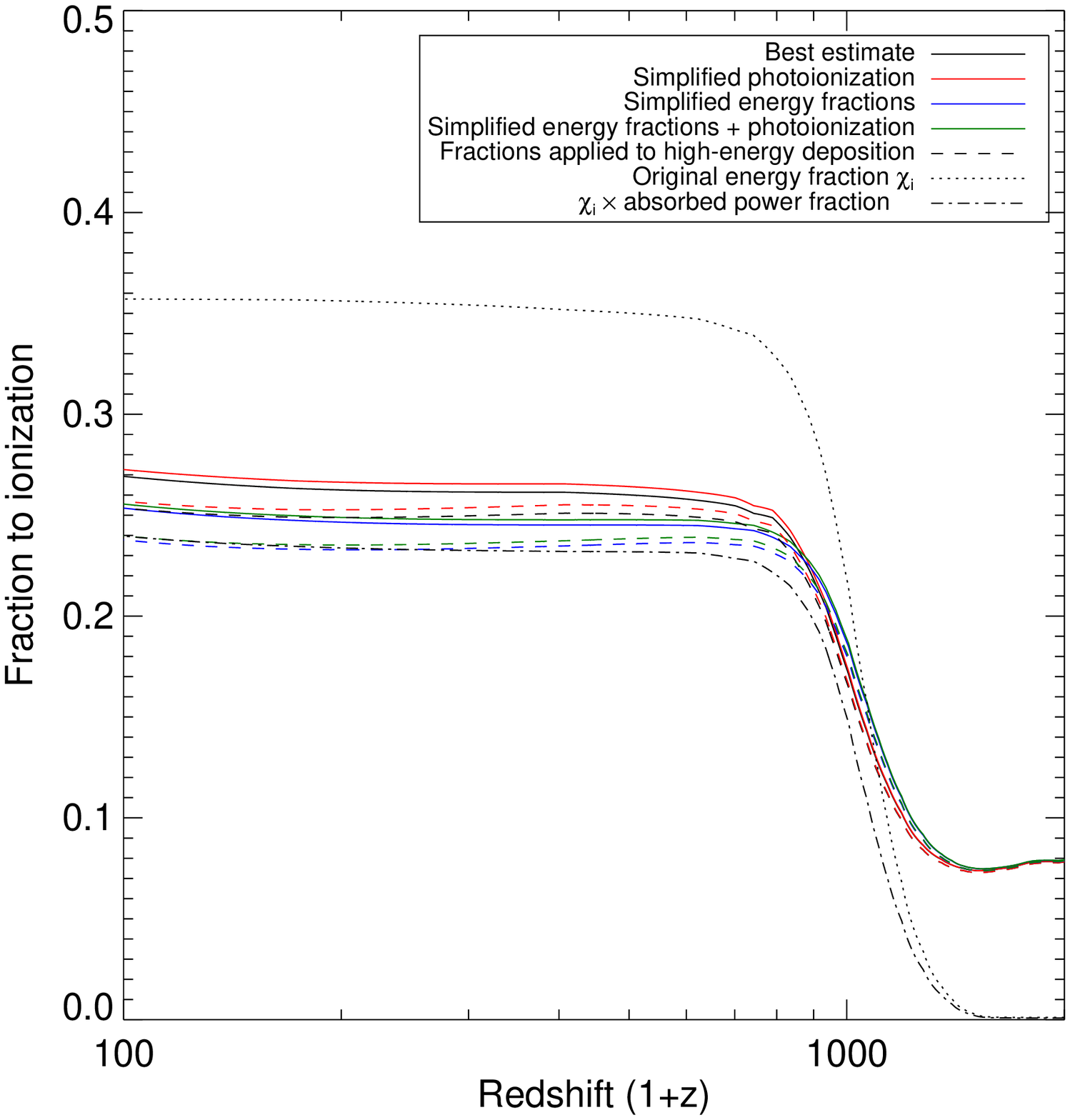} 
\includegraphics[width=.3\textwidth]{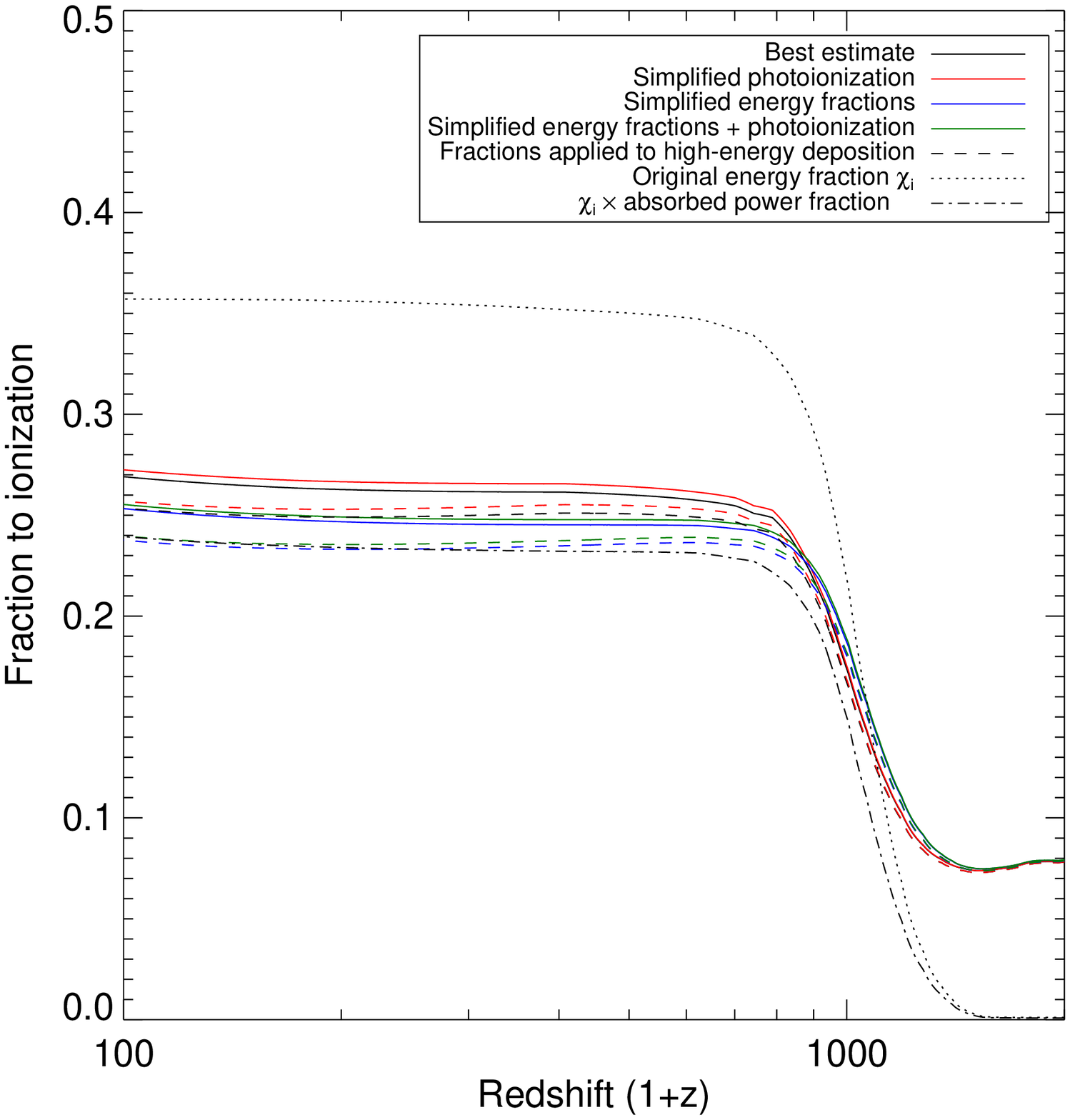}
\caption{\label{fig:highEuncertainties}
Fraction of power ``absorbed by the gas'' proceeding into ionization, in various prescriptions (see text, Section \ref{sec:elecspec}), defined here as $f_\mathrm{effective} \chi_i /f_\mathrm{previous}$. The dotted line shows $\chi_i$ (evaluated at the 3 keV threshold) as a function of $z$.  The dot-dashed line shows  $f_\mathrm{approx} \chi_i /f_\mathrm{previous}$ (see Equation \ref{eq:newf}). For details of the other lines, see the text; black+blue vs red+green lines differ in the treatment of photoionization, black+red vs blue+green lines differ in the treatment of energy absorption by sub-threshold electrons, and dashed vs solid lines differ in the treatment of the high-energy deposition. The left-hand panels show the results for DM annihilating directly to electrons, the right-hand panels for DM annihilating directly to muons; the rows correspond to 1 GeV, 10 GeV, 100 GeV and 1 TeV DM.}
\end{figure*}

\subsubsection{Mapping the low-energy photon spectrum to an electron spectrum}
\label{sec:photoion}

To study the effects of direct photoionization and the importance of the spectrum of below-threshold particles (as opposed to just the total power contained in them), we need to map the below-threshold photon spectrum (e.g. Figure \ref{fig:photspec}) to an electron spectrum. The simplest approach is to assume that each photon above 13.6 eV ionizes hydrogen, depositing 13.6 eV of energy to ionization, with the remaining energy going to the resulting free electron. 

Of course, there are also ionizations on helium, but (consistent with the low-energy code) we assume that these are followed by prompt recombination, producing a line photon, which in turn can ionize hydrogen. In any case, each photon above 13.6 eV should produce 13.6 eV of ionization in addition to any free electrons it generates; we track this contribution separately, before applying the energy fractions to the resulting free electron spectra. Neglecting this direct photoionization contribution would lead to artificially weak constraints.

One might worry that neglecting ionization on helium will change the spectrum of free electrons produced by photoionization, which in turn will modify the fraction of power into ionization, when we properly account for the energy dependence of the fraction $\chi_i$. We have updated the code described above to separately track photoionizations on hydrogen, He and He$^+$, and the subsequent ionizations by line photons from He recombination. Figure \ref{fig:highEuncertainties} shows the differences in the fraction of power proceeding to ionizations, for this approach and the simplified approach where we simply assume every photon ionizes hydrogen, for two methods of treating the energy dependence of the fractions $\chi_x$ (see Section \ref{sec:elecspec}). 

Specifically, we plot $f_\mathrm{effective} \chi_i /f_\mathrm{previous}$, with $\chi_i$ evaluated at threshold. This corresponds to the total power into ionization as a fraction of the total power degraded down to below the few-keV scale (the interpretation of $f_\mathrm{previous}(z)$ is discussed in Section \ref{sec:fz}). If $f_\mathrm{effective}(z)$ were equal to $f_\mathrm{previous}(z)$ (i.e. the losses to continuum photons from ICS could be neglected, as could the differences between $f_\mathrm{approx}$ and $f_\mathrm{effective}$), this quantity would simply be $\chi_i(z)$; the differences between the newly calculated curves, and $\chi_i(z)$, indicate the errors inherent in these approximations. Since the losses to continuum photons are significant, $f_\mathrm{effective}$ is significantly lower than $f_\mathrm{previous}$, except at high redshifts, where direct photoionization dominates ionization from sub-threshold electrons.

Black and blue lines show the results for our best estimate including He ionization, red and green for the simplified approach. For all our example models, the difference between the two is negligible (a few percent or less). 

Note that this direct photoionization contribution is not negligible even when the neutral fraction is very small, because there are no competing cooling processes to be taken into account (in contrast to electrons, which can always lose energy by Coulomb scattering on the electron bath). Consequently, the fraction of absorbed power proceeding into ionization remains non-zero even at quite high redshifts (this can lead to an $f_\mathrm{effective}$ much larger than one, since we normalize to the $\chi_i$ fraction). This behavior is however unimportant to the constraints, as when the universe is almost completely ionized a small change in the ionization fraction is not measurable.

\subsubsection{Energy absorption from the low-energy electron spectrum}
\label{sec:elecspec}

From Figure \ref{fig:elecspec}, we see that the electron spectra are nearly independent of the DM mass and redshift, except for normalization factors, with the bulk of the power being in 10 eV -- 3 keV electrons, and the spectrum in this range being fairly flat in $E^2 dN/dE$ except for the peak at the threshold energy of 3 keV. The photon spectra change with redshift, in shape rather than simply normalization, but the largest effects occur at low (sub-10 eV) energies; at least for DM masses greater than 1 GeV, they are roughly flat in $E^2 dN/dE$ down to $\sim 1-10$ eV energies (depending on the redshift).

The sharp rise in the electron spectra at 3 keV is due to electrons which have been slowly and continuously losing energy, and are not tracked further once their energy falls below 3 keV. If the cooling of the electron population was \emph{purely} continuous (that is, the energy of a given electron was a smooth function of time, and no new electrons were being produced), \emph{all} the electrons would reside at the threshold after being allowed to cool to that point. The presence of continuum below the threshold speaks to the existence of discrete cooling mechanisms and production of electrons below threshold (e.g. via ionization, and inverse Compton scattering to produce photons which subsequently photoionize the gas, yielding free electrons).

The fraction of energy in particles below 100 eV is less than $\sim 20\%$ in all cases. For a conservative constraint, therefore, we might apply the energy fractions described in Section \ref{sec:fractions} to the particles with energies between 100 eV and 3 keV, and assume the energy in sub-100 eV particles is entirely lost; relative to multiplying the total power in sub-3-keV particles by our energy fractions, this cannot weaken the constraints by more than $\sim 20\%$. A more correct estimate of the true constraint would be obtained by taking the fractions $\chi_x$ as a function of energy and integrating over the spectrum of low-energy particles; we leave a more detailed mapping of the energy fractions for future work, but present the results of this procedure using an interpolation from the fractions derived for 14 eV, 30 eV, 60 eV, 100 eV, 300 eV, 1 keV, 3 keV and 10 keV electrons.

Results are shown in Figure \ref{fig:highEuncertainties}. Black and red lines use the ``best estimate'' interpolation approach discussed in this section; green and blue lines neglect electrons below 100 eV and apply the at-threshold energy fractions to electrons above 100 eV. We see that in the redshift range of greatest interest, the relative difference between the two is generally $\sim 5-10 \%$, with the ``best estimate'' approach yielding more ionization.

\subsubsection{Above-threshold deposition}

As described above, the ``above-threshold deposition'' consists of energy lost to collisional heating of the gas,  energy absorbed as excitation or ionization, and energy lost by ICS into photons of lower energy than we explicitly track (below 0.1 eV). Note that the vast bulk of this absorbed energy originates from electrons between the threshold and 10 keV (i.e. only slightly above the threshold, for our baseline choice of 3 keV): this can be seen from Figure \ref{fig:fractions}, where we show the effect of varying the threshold, and see that the above-threshold deposition is at the few-percent level or lower for a 10 keV threshold.

We can separately track the contributions to these different channels, from the high-energy calculation. The results can be compared to the $\chi_x$ fractions of Section \ref{sec:fractions} -- however, relatively more energy tends to be lost into ionization and excitation compared to heating, as these relatively high-energy electrons cool down to the threshold. This is expected and necessary, since the fraction of their energy that goes into below-threshold electrons -- and particularly secondary electrons that can have energies below 100 eV (see the discussion in Section \ref{sec:sourcekev}) -- will, to a greater degree, be deposited via collisional heating. Consequently, multiplying the ``above-threshold deposition'' component by the $\chi_x$ fractions (evaluated at threshold) will tend to underestimate the power deposited in ionization and excitation.
We have performed the cross-check of taking the threshold of the high-energy code to 100 eV, and computing the partition between ionization, excitation, heating and low-energy electrons for electrons with an initial energy of 1 keV (since for such energies, inverse Compton scattering is subdominant), at an ionization fraction of $x_e = 10^{-3}$ corresponding to a redshift of $z \sim 600$, where the signal is expected to peak. (We have also checked the case with $x_e = 10^{-4}$: the results do not change.) We then estimate the partition between ionization, excitation and heating for the low-energy electrons, below 100 eV, using the fractions of Section \ref{sec:fractions} combined with the spectrum calculated by the high-energy code. The overall fraction going into ionization, in this treatment, differs from the detailed result of the low-energy code by less than $10\%$ (this approach gives 0.40 whereas the low-energy code yields 0.37); the bulk of the ionization energy ($\sim 80\%$) comes from particles above the threshold, so this is a meaningful consistency check. This is encouraging, as the high-energy code is not designed for use at such low energies: approximate cross sections are used and only ionizations/excitations from the ground states of H, He, and He$^{+}$ are considered.

In Figure \ref{fig:highEuncertainties} we show the results of taking the fractions from the high-energy code (solid lines), as opposed to simply multiplying the above-threshold deposition component by the $\chi_i$ fraction at threshold (dashed line). We see that using the latter approach generally reduces the power going into ionization by $\sim 5-10\%$ or less.

Overall, we find that our ``best estimate'' curve - using the ionization contribution derived by the high-energy code for the above-threshold deposition, interpolating the $\chi_i$ fraction and integrating over the low-energy electron spectrum, and taking into account photoionization on both H and He - increases the amount of power going into ionization by $\sim 10\%$ over the redshift range of greatest interest, relative to the $f_\mathrm{approx}$ approximation made above (and shown by the dot-dashed line in Figure \ref{fig:highEuncertainties}). We thus expect the constraints to become stronger by roughly the same amount. This curve, denoted $f_\mathrm{best}$, and the $f_\mathrm{approx}$ curve bracket the possible effects of the systematic uncertainties we have studied in this section.

\subsection{Effect on the constraints}
\label{sec:newfconstraints}

\begin{table*}[ht]
\begin{center}
\begin{tabular}{llcccccc}
 DM mass& channel &$\langle\sigma v\rangle$&$\langle\sigma v\rangle$&Ratio&$\langle\sigma v\rangle$&Ratio&Ratio \\
&&$\{f_\mathrm{previous}(z)\}$&$\{f_\mathrm{approx}(z)\}$&$\mathrm{approx./prev.}$&$\{f_\mathrm{best}(z)\}$&$\mathrm{best/prev.}$&$\mathrm{best/approx.}$\\
 $[\rm{GeV}]$ &&[cm$^3$/s]&[cm$^3$/s]&&[cm$^3$/s]&&\\[1mm]
\hline \noalign{\vskip 1mm}
 $    1 $&    electrons  &$ <            3.1\times 10^{-28} $&$ <            7.2\times 10^{-28}  $&$       2.31 $&$ <            6.2\times 10^{-28}  $&$       1.96  $&$       0.85 $\\

 $    2 $&    electrons  &$ <            6.1\times 10^{-28} $&$ <            1.2\times 10^{-27}  $&$       2.01 $&$ <            1.0\times 10^{-27}  $&$       1.68  $&$       0.84 $\\

 $    5 $&    electrons  &$ <            1.6\times 10^{-27} $&$ <            2.3\times 10^{-27}  $&$       1.39 $&$ <            2.2\times 10^{-27}  $&$       1.31  $&$       0.94 $\\

 $   10 $&    electrons  &$ <            3.4\times 10^{-27} $&$ <            4.6\times 10^{-27}  $&$       1.35 $&$ <            4.2\times 10^{-27}  $&$       1.25  $&$       0.93 $\\

 $   20 $&    electrons  &$ <            7.5\times 10^{-27} $&$ <            1.0\times 10^{-26}  $&$       1.39 $&$ <            9.5\times 10^{-27}  $&$       1.27  $&$       0.91 $\\

 $  100 $&    electrons  &$ <            4.3\times 10^{-26} $&$ <            6.4\times 10^{-26}  $&$       1.48 $&$ <            6.0\times 10^{-26}  $&$       1.38  $&$       0.94 $\\

 $ 1000 $&    electrons  &$ <            4.5\times 10^{-25} $&$ <            6.6\times 10^{-25}  $&$       1.48 $&$ <            6.1\times 10^{-25}  $&$       1.37  $&$       0.93 $\\

 $    1 $&        muons  &$ <            8.8\times 10^{-28} $&$ <            1.5\times 10^{-27}  $&$       1.72 $&$ <            1.3\times 10^{-27}  $&$       1.51  $&$       0.87 $\\

 $    2 $&        muons  &$ <            1.8\times 10^{-27} $&$ <            3.6\times 10^{-27}  $&$       2.02 $&$ <            3.2\times 10^{-27}  $&$       1.75  $&$       0.87 $\\

 $    5 $&        muons  &$ <            4.4\times 10^{-27} $&$ <            7.9\times 10^{-27}  $&$       1.77 $&$ <            6.7\times 10^{-27}  $&$       1.51  $&$       0.85 $\\

 $   10 $&        muons  &$ <            9.0\times 10^{-27} $&$ <            1.4\times 10^{-26}  $&$       1.55 $&$ <            1.2\times 10^{-26}  $&$       1.36  $&$       0.88 $\\

 $   20 $&        muons  &$ <            2.0\times 10^{-26} $&$ <            2.8\times 10^{-26}  $&$       1.39 $&$ <            2.4\times 10^{-26}  $&$       1.23  $&$       0.88 $\\

 $  100 $&        muons  &$ <            1.2\times 10^{-25} $&$ <            1.7\times 10^{-25}  $&$       1.41 $&$ <            1.5\times 10^{-25}  $&$       1.30  $&$       0.92 $\\

 $ 1000 $&        muons  &$ <            1.3\times 10^{-24} $&$ <            1.9\times 10^{-24}  $&$       1.48 $&$ <            1.8\times 10^{-24}  $&$       1.37  $&$       0.93 $\\
\hline \noalign{\vskip 1mm}
\end{tabular}
\caption{Constraints at $95\%$ c.l. on $\langle \sigma v\rangle$ from a \emph{Planck}-like experiment. We show results assuming no dark matter annihilation in the mock data and using the previous, approximate or best estimate $f(z)$ functions to analyze it (see Sections \ref{sec:highengresultsprevious} and \ref{sec:highsystematicseffective}). We consider  different dark matter masses and annihilation channels. To make the comparison easier, we also show the ratio between the constraints obtained using approximate (approx.) versus previous (prev.) $f(z)$ in the fifth column, best estimate (best) versus previous in the seventh column, and best estimate versus approximate $f(z)$ in the eighth column.}
\label{tab:resultsfz}
\end{center}
\end{table*}

In this section we study the effect of using the approximate  $f_\mathrm{approx}(z)$ (described in Section \ref{sec:highengresultsprevious}) or the effective ``best estimate'' $f_\mathrm{best}(z)$ (described in Section \ref{sec:highsystematicseffective}) on dark matter annihilation constraints.

In order to assess the impact of the constraints, we follow the same procedure as described in details in Section  \ref{sec:systematics}. We first simulate CMB power spectra assuming no dark matter annihilation for a \emph{Planck}-like experiment. We then analyze the data now assuming a specific DM mass and annihilation channel, and calculate the constraints  using $f_\mathrm{previous}(z)$ (from \cite{Slatyer:2009yq}), $f_\mathrm{approx}(z)$, and $f_\mathrm{best}(z)$, using the Markov Chain Monte Carlo package \texttt{cosmomc}. We calculate constraints for DM masses $m_\chi$ of $1$, $2$, $5$, $10$, $20$, $100$ and $1000$ GeV annihilating into muons or electrons. Contrary to the results of Section \ref{sec:results1}, where the provided constraints  on the model are on the $p_\mathrm{ann}=f\langle \sigma v\rangle/m_\chi$ parameter with constant $f(z)$, here our constraints are directly on the thermally averaged cross section $\langle \sigma v \rangle$, as we test specific models of dark matter corresponding to specific redshift evolution of the $f(z)$ function.  For the analysis, we use the $\chi_x$ fractions calculated as in Section \ref{sec:fractions}, assuming $Y_\mathrm{He}=0.24$, Helium ionization $x_\mathrm{He}=1\times 10^{-10}$, and an initial electron energy of $3$ keV.
Results are presented in Table \ref{tab:resultsfz}.

Replacing the $f_\mathrm{previous}(z)$ curves with the more correct $f_\mathrm{approx}(z)$ (but keeping the same low-energy fractions) leads to weaker constraints by up to a factor $2$. This is  mostly due to the loss to sub-10.2 eV photons, previously included in the ``deposited energy'' budget by previous works. Furthermore, the weakening effect is somewhat stronger for lower masses, for the reasons already described in Section \ref{sec:continuum}: few-GeV dark matter candidates, whose dominant relevant annihilation products are few-GeV electrons, produce $\sim 0.1-10$ MeV photons via ICS, that efficiently produce several non-relativistic secondary electrons via Compton scattering. These electrons then very slowly cool via ICS on CMB photons, producing sub-10.2 eV photons that cannot ionize, excite or heat the gas anymore.
On the other hand, using $f_\mathrm{best}(z)$ instead of $f_\mathrm{approx}(z)$ curves strengthen the constraints by $\sim 10-15\%$, as expected and already anticipated in Section \ref{sec:highsystematicseffective}. 

For thermal relic cross sections ($\langle \sigma v \rangle \sim 2-3 \times 10^{-26}$ cm$^3$/s), the low-mass DM candidates where the weakening is most pronounced are already excluded; for DM masses in the range that will be constrained by \emph{Planck} assuming a thermal relic annihilation cross section, the effect of properly accounting for continuum photons (and the other effects we have discussed) is to weaken the constraints by a factor of $\sim 1.2-1.4$. Thermal relic cross sections are then forecast to be ruled out at $95\%$ confidence for $m_\chi \lesssim 50$ GeV for annihilation to electrons, and $m_\chi \lesssim 20$ GeV for annihilation to muons.
\section{Discussion and conclusions}
\label{sec:conclusions}

CMB data can place  strong constraints on dark matter annihilation parameters. However, a number of different assumptions and parameterizations are usually employed to infer these constraints. In this paper, we have studied whether these approximations are well justified, and whether they can actually bias the recovered results.

In order to calculate constraints, it is necessary to study the way the energy injected by dark matter annihilation (in particles at the GeV/TeV scale) interacts with the surrounding medium, heating, ionizing and exciting it. We have formalized and extended a common approach adopted in the literature, which is to split this calculation into two parts. The first follows the shower of secondary particles produced by DM annihilation from the original energy scale (GeV/TeV) to an energy threshold at  keV-scale. This produces  a continuum spectrum of particles with energies below the threshold, and provides the fraction $f(z)$ of energy that will actually be absorbed by the medium at each redshift $z$. 

The second part of the calculation then follows the propagation of the particles from the keV scale until their energy is fully deposited into the photon-baryon fluid, as heating, ionization, excitation and free-streaming continuum photons (below 10.2 eV). This provides the energy fractions $\chi_x$ that are absorbed in these  different channels.

We have developed and improved codes to follow both parts of the propagation, testing the stability of the results to various assumptions. 

\begin{itemize}
\item For the low-energy calculations, we have shown that changing the Helium abundance or the level of Helium ionization (in a reasonable range) does not meaningfully modify the $\chi_x$ energy fractions; the impact on the inferred constraints on dark matter annihilation is negligible. This demonstrates that the fractions calculated for a fixed amount of Helium ionization can safely be used to calculate dark matter annihilation constraints during cosmological recombination, even when the level of Helium ionization evolves with redshift (by small amounts in the redshift range of interest for these constraints).

\item We have carefully computed the fraction of power going into Lyman-$\alpha$ photons, and found that erroneously neglecting this component would weaken the constraints by $\sim 10\%$. We have also examined the effect of using results from our detailed low-energy code, compared to a simplified parameterization for the ionization/heating/excitation fractions frequently employed in the literature, and found that the improved analysis strengthens the bound on DM annihilation by $\sim 5\%$.

\item We have determined the optimal energy threshold to switch from the high-energy calculation to the low energy one.  The advantage of our high-energy code is that it properly accounts for inverse Compton scattering (ICS), which the low-energy code assumes to be negligible. On the other hand, the advantage of the low-energy code is that it treats the atomic cooling processes much more carefully. Thus, the optimal threshold is the highest energy at which atomic processes dominate over ICS. We have found that this typically occurs at energies around $\sim 3$ keV, for the redshifts where the bulk of the signal originates.

\item A major finding of this work is that at the high redshifts relevant to this signal, inverse Compton scattering is a non-negligible cooling mechanism for electrons with energies greater than a few keV, and therefore it must be properly included down to the threshold. Previous studies have incorrectly neglected this process at energies lower than several \emph{hundred} keV, due to an estimate of the cooling time only valid in the relativistic limit (which does not apply here). Inverse Compton scattering by non-relativistic electrons grants only a very small increase in energy to the scattered photons, and consequently the power lost to this process goes largely into ``continuum'' (sub-10.2 eV) photons, i.e. spectral distortion of the CMB, rather than ionization/excitation/heating.

\item We have improved the high-energy calculation of the $f(z)$ curves by correctly including ICS and following the propagation of the secondaries down to the energy threshold where we switch to the low energy code. This guarantees a very good match between the two ends of the calculations.
We explicitly calculate these improved $f_\mathrm{approx}(z)$ curves for dark matter particles of $1$, $2$, $5$, $10$, $20$, $100$ and $1000$ GeV annihilating into muons or electrons. We find that using these $f_\mathrm{approx}(z)$ curves lead to a weakening of the constraints on the annihilation cross-section $\langle \sigma v\rangle$ by up to a factor $2$ compared to the constraints obtained using previous $f(z)$ curves, because of the power lost into continuum photons (mostly by ICS of non-relativistic electrons). The largest differences are found for light DM, a few GeV or less; for the DM masses and channels at which \emph{Planck} is expected to constrain the thermal relic annihilation cross section, the constraints are weakened by a factor of $\mathcal{O}(1.4-1.5)$ in this approximate treatment.

\item We have then considered other aspects of the high-energy calculation that can be improved to obtain more accurate $f(z)$ curves:
\begin{enumerate}
\item Photon spectrum: the final products of the high-energy code include the energy spectra of below-threshold electrons and photons, whose  further interactions are then followed  by our low-energy code, under the assumption that photons are immediately converted to electrons through photoionization. This process thus produces ionizations, but its contribution to the total ionization energy fractions has been neglected in previous studies.

\item Electron spectrum: the high energy code follows particles until their energy falls below the energy threshold, producing a  continuum spectrum between  $0-3$ keV. However, the low energy calculation is performed assuming a fixed value of the injected electron energy (e.g. $3$ keV), so that the resulting $\chi_x$ fractions are valid for that specific initial energy value.
Thus, one should in principle calculate the $\chi_x$ for several initial electron energies and apply the appropriate fractions to the injected spectrum. For practical reasons, however, it is  convenient to use the same $\chi_x$ fraction calculated for a single energy to the whole spectrum of particles. We have demonstrated that this is a good approximation for particle energies between $100$ eV and $3$ keV, as the $\chi_x$ fractions do not change significantly in this range. On the other hand, $\chi_x$ fractions calculated assuming initial electron energies below $100$ eV become rapidly very different, as most of the energy of these sub-$100$ eV particles contributes to the heating of the medium rather then ionization and excitation. However, we have determined that, at least for the dark matter models we considered, only about $20\%$ of the overall sub-threshold energy lies in particles with energy  below $100$ eV. Even assuming that all this energy is lost, this weakens the constraints by less than $20\%$. To obtain higher accuracy, it is however necessary to consider the proper  $\chi_x$ curves also for the particles below $100$ eV.

\item Above threshold ionization: our low-energy code is designed to properly follow atomic processes that lead to the ionization, excitation and heating of the medium. However, these processes also occur above the energy threshold. Our high-energy code can take these processes into account as well, although in a more approximate way. The energy going into these above-threshold absorption channels is then usually simply added as a contribution to the $f(z)$ curves, and then multiplied by the $\chi_x$ fractions to calculate the amount of energy going into ionization, excitation or heating. 
However, in these above-threshold processes  relatively more energy tends to be lost into ionization and excitation compared to heating than expected from the $\chi_x$ fractions, when we examine the results for the separate channels from the high-energy code. As a consequence, multiplying this component by the  $\chi_x$ fractions (evaluated at threshold) will tend to underestimate the power deposited in ionization and excitation.
\end{enumerate}
Improving these three points as described, we have calculated a ``best estimate'' $f(z)$ curve. We have shown that these  minor improvements strengthen the constraints by less than $15\%$ in all the DM models we considered.
\end{itemize}

In summary, we find that the systematic effect with the greatest impact on  DM annihilation constraints, in previous studies, is the neglect of continuum photons, particularly those produced by inverse Compton scattering of non-relativistic and mildly relativistic electrons.  When this effect is correctly included, the constraints are weakened by up to a factor $2$; note however that the effect is strongest for light dark matter masses, where the thermal relic cross section is already ruled out. The remaining uncertainties, which we have carefully described and corrected for in this paper, change the constraints by less than $20\%$. 

\acknowledgments{
We would like to thank Carmelo Evoli for very fruitful  comments and discussions. SG was supported by Benjamin D. Wandelt's ANR Chaire d'Excellence ANR-10-CEXC-004-01. TRS is supported by the US National Science Foundation under grants PHY-0907744 and AST-0807444. For this work, FI has received support from the EU grant J11E099, ``ERC DARK - Dark Matters".}

\clearpage

\appendix


\section{Energy fractions}
\label{app:engfracs}

In this appendix we present the tabulated results  for the $\chi_x$ energy fractions described in Section \ref{sec:fractions} and \ref{sec:systematics}. We present here results assuming helium abundance $Y_p=0.24$, helium ionization fraction $x_\mathrm{HeII}=1\times10^{-10}$ and an injected electron energy of $3$ keV.

The columns indicate, from left to right, which fraction of the initial electron energy is deposited into 
heat, Ly$\alpha$ excitations, H ionizations, He excitations and photons with $E <$ 10.2 eV, i.e. $\chi_{\rm h}$, $\chi_{\rm \alpha}$, $\chi_{\rm i,H}$, $\chi_{\rm i,He}$ and $\chi_{\rm c}$ respectively, as a function of the ionized fraction $x_e$ listed in the first column.
\begin{table*}\label{table1}
  \begin{center}
    \begin{tabular}{|c|c|c|c|c|c|}\hline
      \hline 
       $x_e$                   &                Gas               & \,\, Excitations \,\,    & \,\,Ionizations \,\,  &\,\, Ionizations \,\,  & \,\,     Excitations\,\, \\
      (ionized fraction) & \,\,\,\,\, Heating \,\,\,\,\, & (Lyman-$\alpha$) & H                            & He                         & ($E<10.2$ eV)\\
      \hline 
      	  0.000100 	& 0.151880	& 0.323526	& 0.350798	& 0.024367	&		0.142202\\
	  0.000300	& 0.174825	& 0.308840	& 0.349058	& 0.023397	&		0.136233\\
	  0.000500	& 0.188520	& 0.302591	& 0.345508	& 0.023737	&		0.133317\\
	  0.001000	& 0.210027	& 0.291280	& 0.341822	& 0.023134	&		0.128324\\
	  0.003000	& 0.258912	& 0.269481	& 0.327298	& 0.023415	&		0.118130\\
	  0.005000	& 0.289871	& 0.256105	& 0.316798	& 0.023029	&		0.111856\\
	  0.010000	& 0.338316	& 0.238304	& 0.301893	& 0.021302	&		0.103617\\
	  0.030000	& 0.458621	& 0.192119	& 0.255925	& 0.018550	&		0.083760\\
	  0.050000	& 0.531628	& 0.165741	& 0.228453	& 0.016601	&		0.072038\\
	  0.100000	& 0.654816	& 0.121705	& 0.175739	& 0.012852	&		0.053388\\
	  0.300000	& 0.849031	& 0.052273	& 0.083885	& 0.007317	&		0.023016\\
	  0.500000	& 0.923644	& 0.026168	& 0.043901	& 0.004914	&		0.011310\\
	  0.800000	& 0.975679	& 0.007178	& 0.013518	& 0.003640	&		0.003117\\
	  0.900000	& 0.987026	& 0.003234	& 0.006406	& 0.003223	&		0.001417\\
	  0.990000	& 0.995299	& 0.000290	& 0.001700	& 0.002924	&		0.000118\\
      \hline
    \end{tabular}
  \end{center}
  \caption{$x_\mathrm{HeII}=10^{-10}$, $E_\mathrm{in}$=3 keV.} 
  \end{table*}

\section{Choice of threshold}
\label{app:threshold}

As discussed in Section \ref{sec:higheng}, inverse Compton scattering (ICS) of mildly relativistic electrons leads to only slight boosts of the CMB photons, and consequently to a (spectrally distorted) population of photons below the 10.2 eV threshold for excitation. ICS is not included in our low-energy code which treats the atomic processes in detail, due to the computational demands of accurately tracking a very large number of nearly elastic scatterings. Consequently, this code can only be reliably used in the energy range where ICS is not the dominant cooling process.

Previous works \cite{Valdes:2009cq,Evoli:2012sc} have stated that ICS is subdominant and can be neglected for electron energies,
\begin{equation} E  < \left( \frac{1+z}{21} \right)^{-1/2} \mathrm{MeV}. \label{eq:hhcrit} \end{equation}
However, this expression is not valid when the inferred cutoff energy is non-relativistic (we have verified this conclusion with the authors of the articles in question): it was originally derived by \cite{Hansen:2003yj}, assuming \emph{relativistic} electrons. At the high redshifts in question here, a naive application of Equation \ref{eq:hhcrit} indicates a non-relativistic cutoff energy, e.g. $\sim 200$ keV for $z \sim 600$, where the effect of DM annihilation on the CMB is thought to peak \cite{Finkbeiner:2011dx}.

 In the non-relativistic regime, following the estimates from \cite{Furlanetto:2009uf}, the cooling times for inverse Compton scattering, collisional heating and ionization are given respectively by,
\begin{align} \tau_\mathrm{ICS} & \approx 10^8 \left( \frac{1+z}{10} \right)^{-4} \mathrm{yr}, \nonumber \\
\tau_\mathrm{heat} & \approx  \frac{5 \times 10^3}{x_e} \left( \frac{E}{\mathrm{keV}} \right)^{3/2}  \left( \frac{1+z}{10} \right)^{-3} \mathrm{yr}, \nonumber \\
\tau_\mathrm{ion} & \approx \frac{5 \times 10^5}{1 - x_e} \left( \frac{E}{\mathrm{keV}} \right)^{3/2}  \left( \frac{1+z}{10} \right)^{-3} \mathrm{yr}.\end{align}

In the regime well after recombination where the ionization fraction $x_e$ is very small ($\lesssim 10^{-2}$), ionization dominates over heating at all energies and redshifts, and the crossover between ICS and ionization occurs at,
\begin{equation}10^8 \left( \frac{1+z}{10} \right)^{-4} = 5 \times 10^5 \left( \frac{E_\mathrm{crossover}}{\mathrm{keV}} \right)^{3/2}  \left( \frac{1+z}{10} \right)^{-3}, \end{equation}
i.e. ICS dominates for energies,
\begin{equation} E > \left( \frac{1+z}{2000} \right)^{-2/3} \mathrm{keV} .\end{equation}
For the redshift range of greatest interest, therefore, ICS always dominates above a few keV. In particular, at $z \sim 600$ where the signal from DM annihilation is thought to be greatest, the crossover energy is $\sim 2$ keV. The main reason for the discrepancy with Equation \ref{eq:hhcrit} is that expression was derived assuming a relativistic limit for the energy loss rate due to inverse Compton scattering, which can badly underestimate the loss rate when extrapolated naively to non-relativistic electrons.

Consequently, it is \emph{not} valid to transition to a code that neglects inverse Compton scattering at hundreds or thousands of keV. This is confirmed in Figure \ref{fig:fractions}, where we see that the energy losses into continuum photons (which originate entirely from ICS) are converged to within a few percent for 1-3 keV thresholds, but can appear (incorrectly) much lower for higher thresholds.

\begin{figure*}
\includegraphics[width=.24\textwidth]{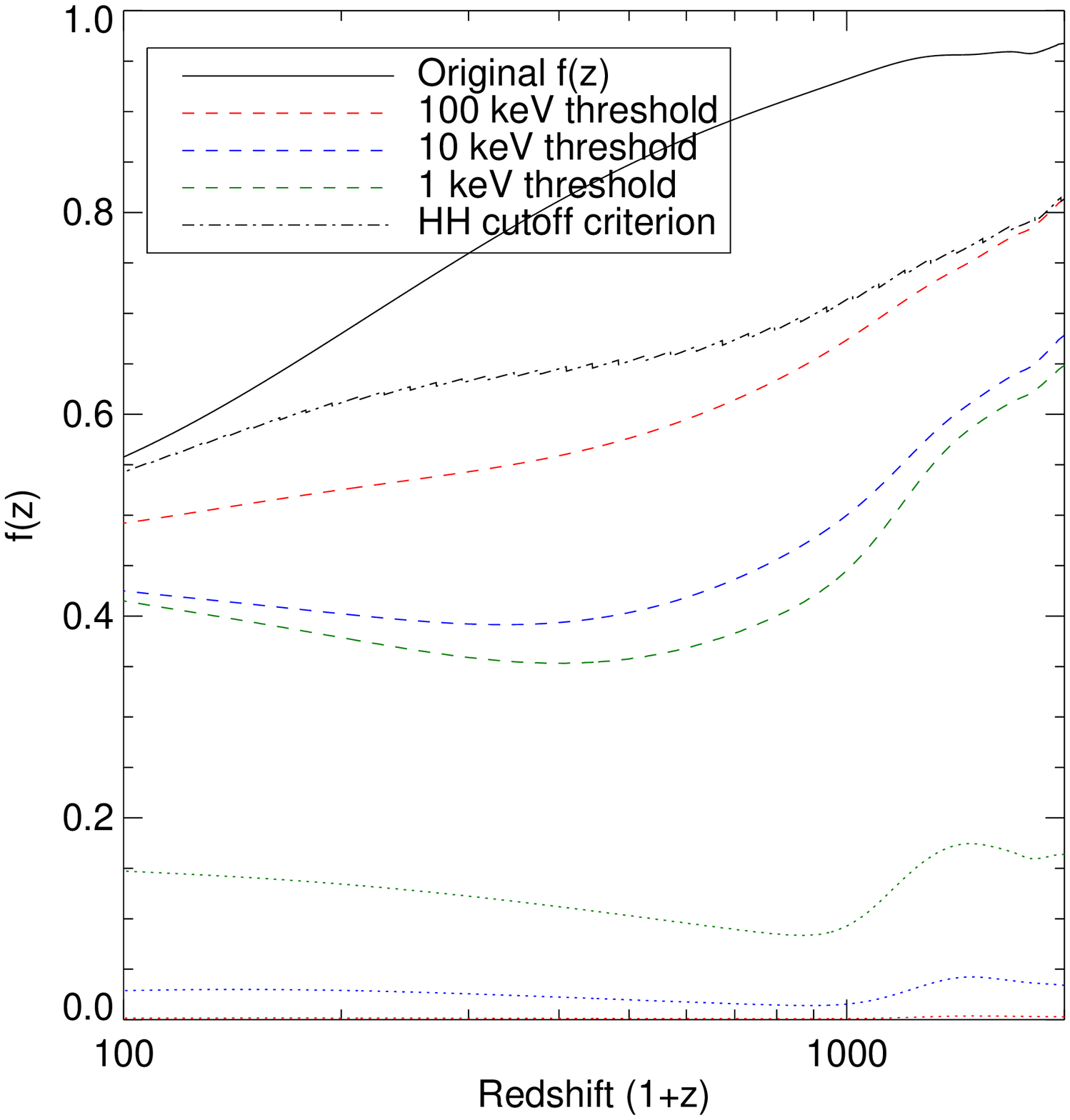}
\includegraphics[width=.24\textwidth]{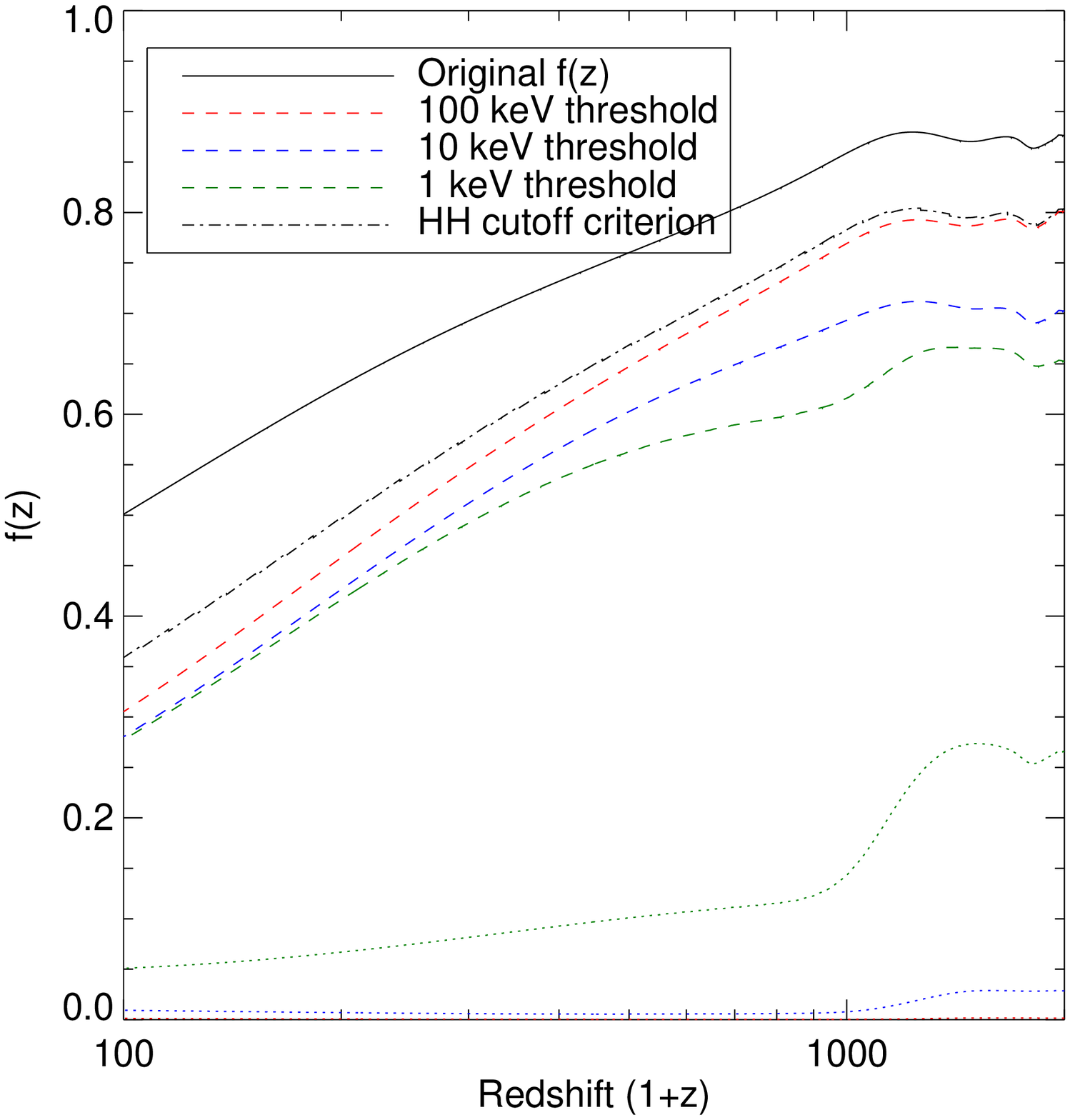}
\includegraphics[width=.24\textwidth]{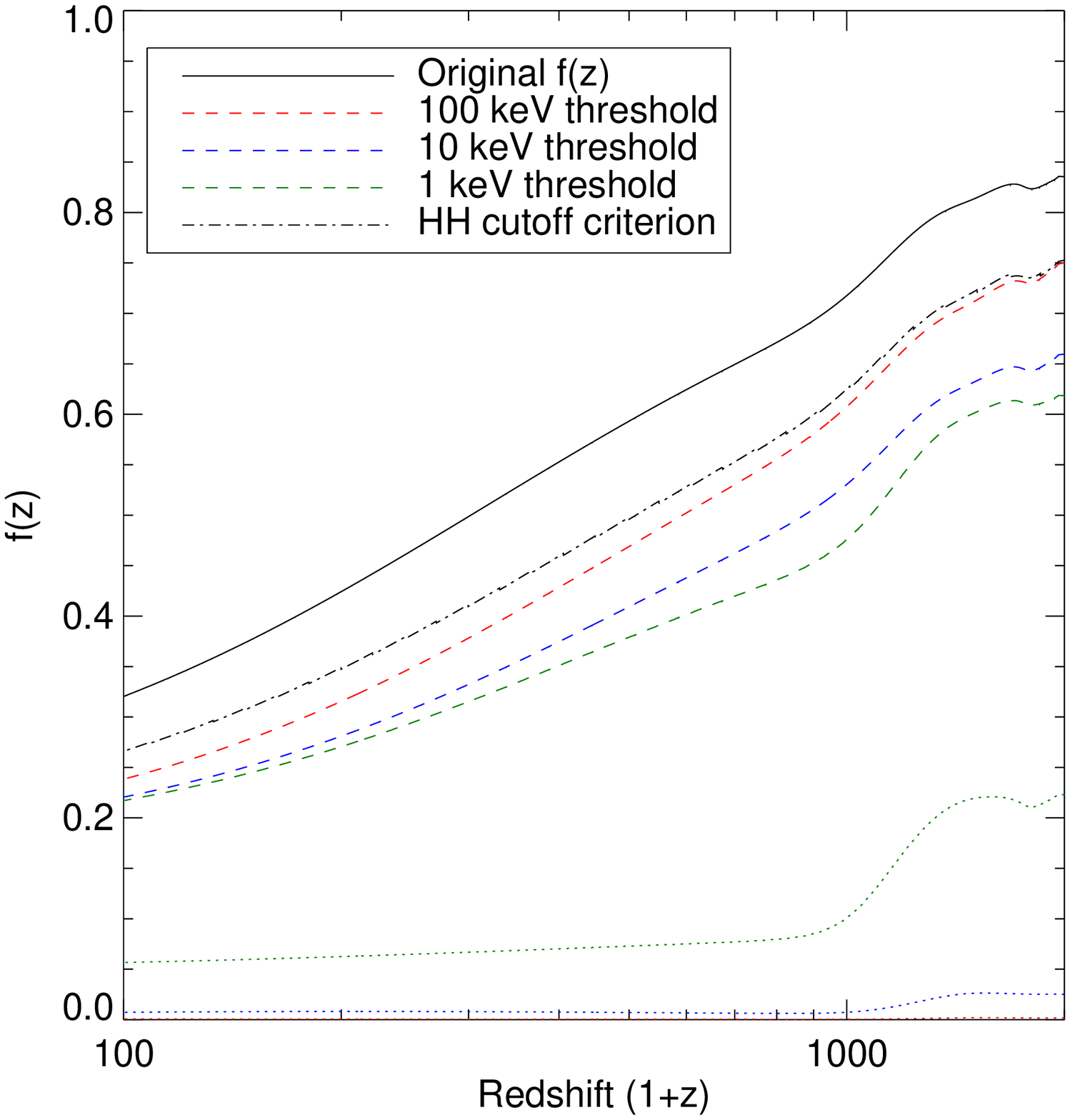}
\includegraphics[width=.24\textwidth]{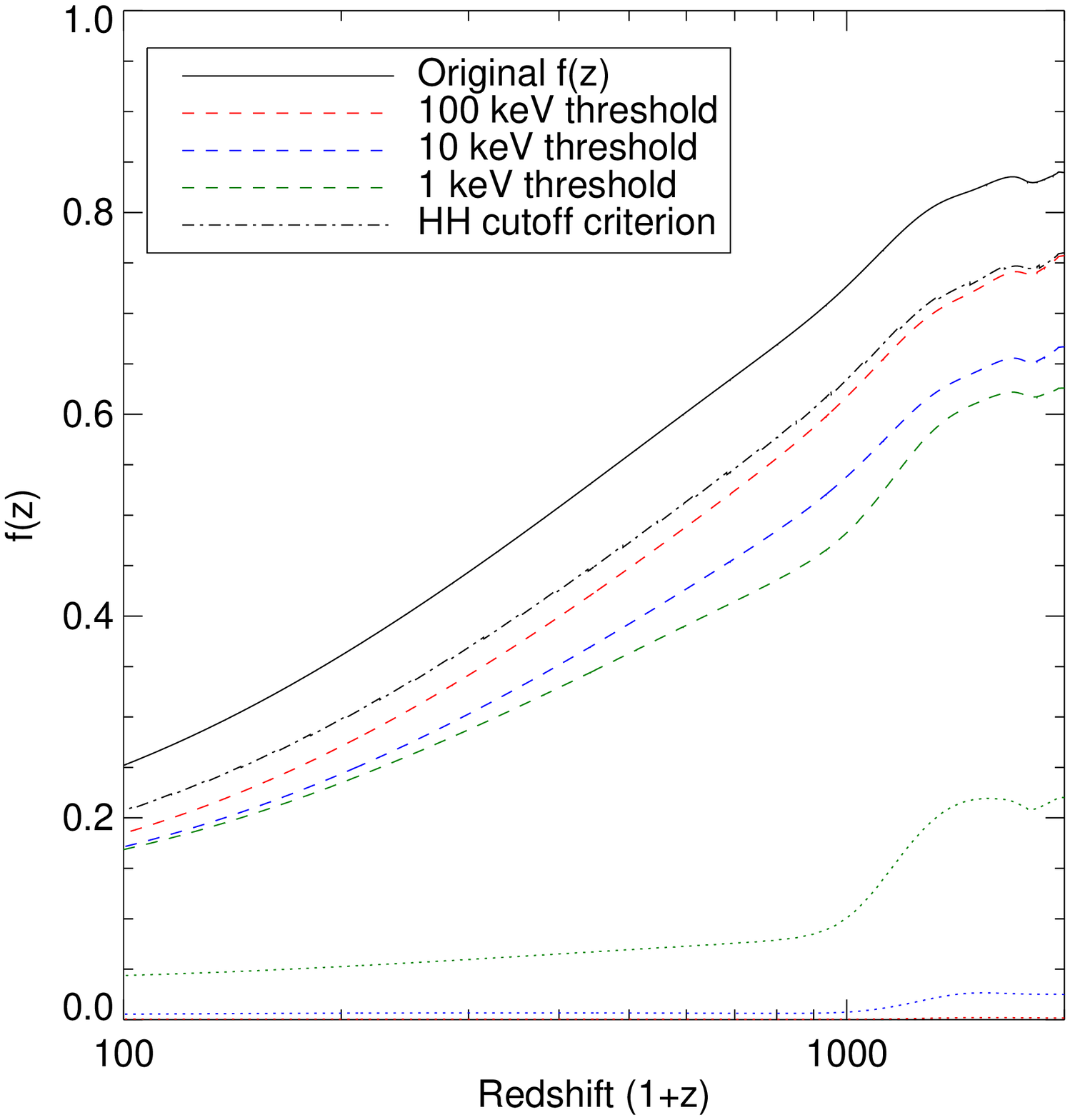} \\
\includegraphics[width=.24\textwidth]{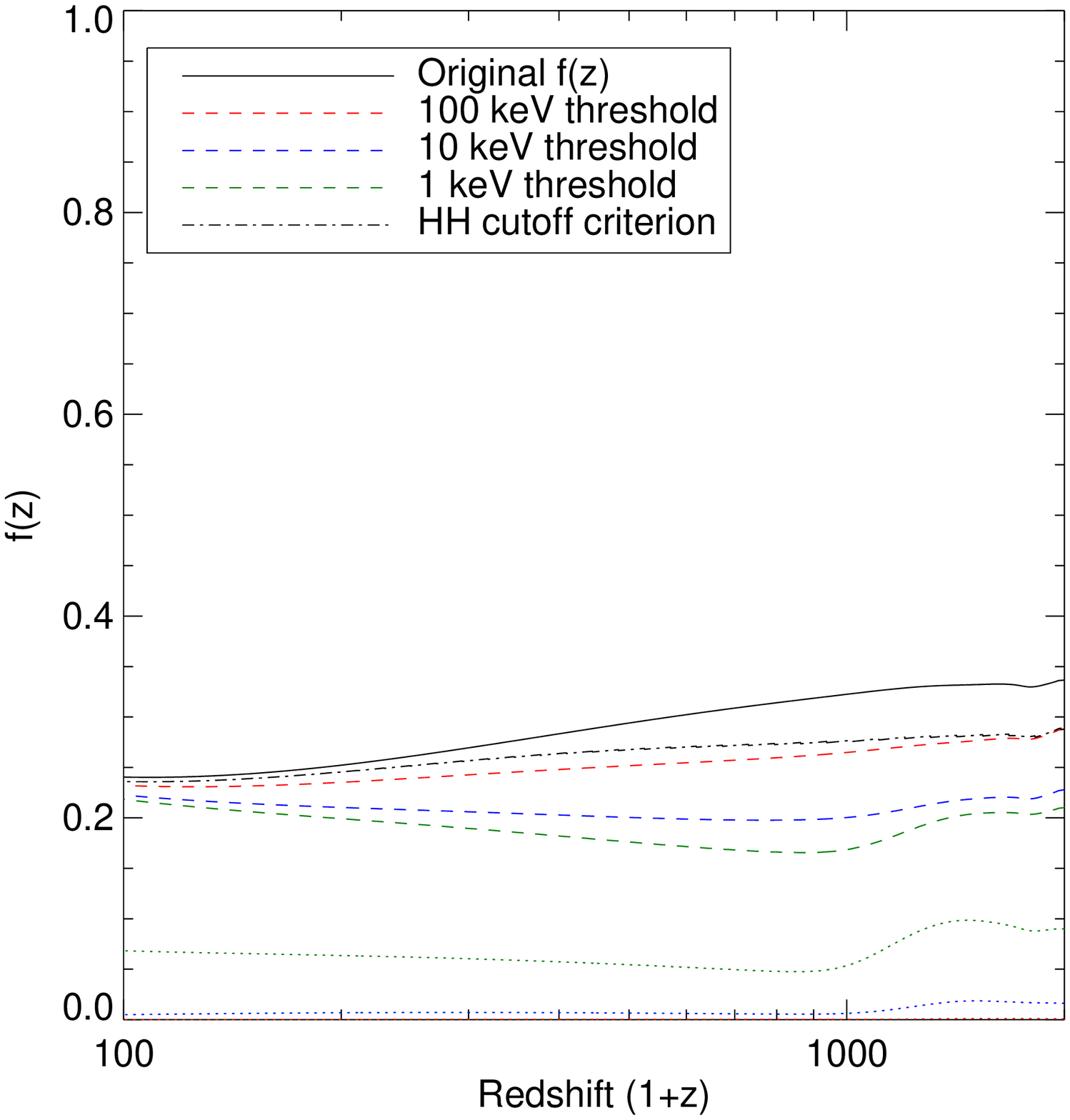}
\includegraphics[width=.24\textwidth]{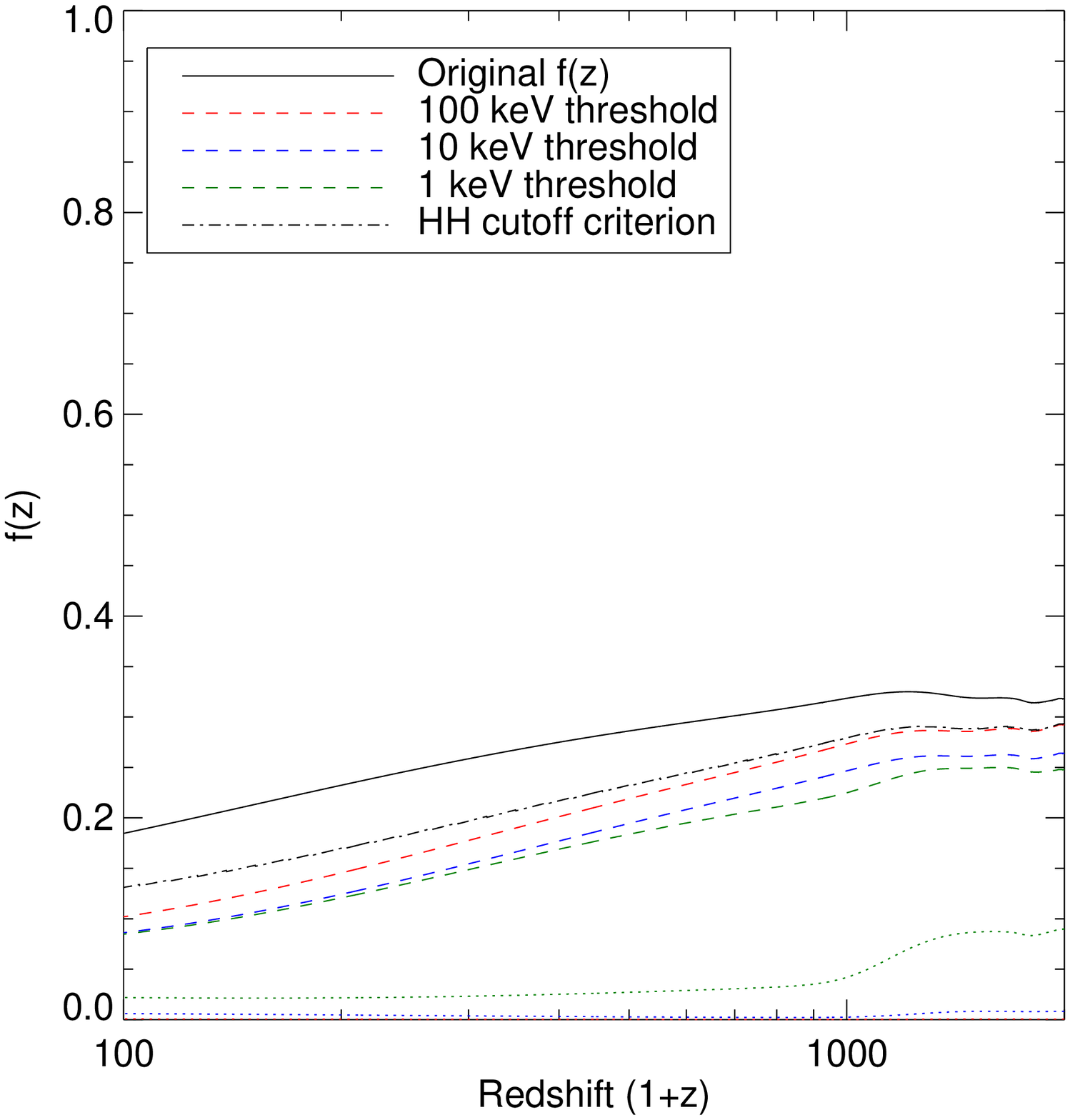}
\includegraphics[width=.24\textwidth]{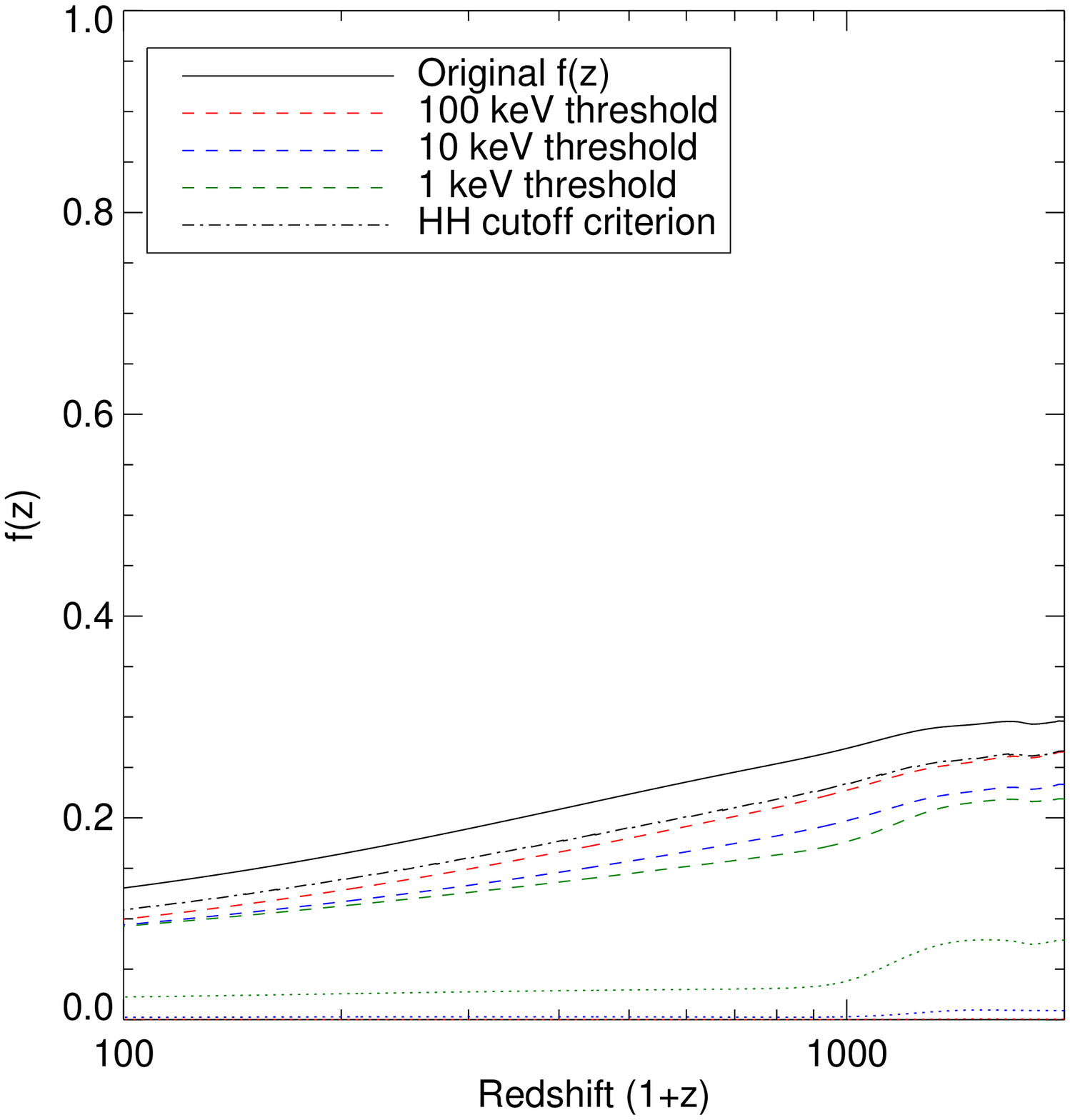}
\includegraphics[width=.24\textwidth]{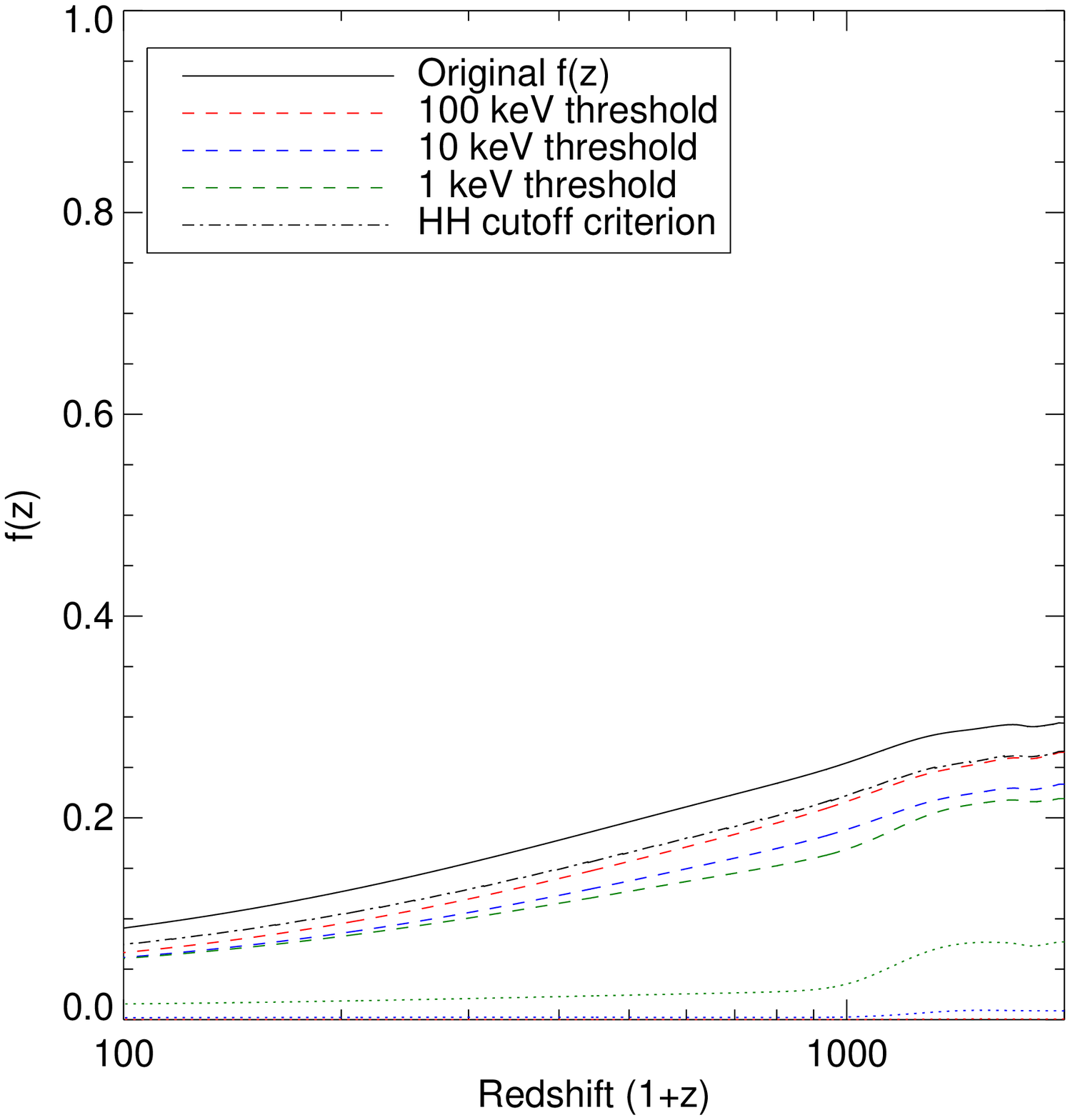}
\caption{\label{fig:varyingpartition}
The absorbed energy fraction $f(z)$ as a function of the threshold down to which the energy losses are followed, after subtraction of the energy transferred into continuum photons (see text), for DM annihilating to (\emph{top}) electrons, or (\emph{bottom}) muons. From left to right the DM mass is 1 GeV, 10 GeV, 100 GeV and 1 TeV. Dotted lines show the fraction of the total energy tagged as ``deposited'' arising from electrons and photons \emph{above} the threshold (this is included in the absorbed energy fraction), with the colors as for the dashed lines showing the absorbed energy fraction: \emph{red} for a 100 keV threshold, \emph{blue} for a 10 keV threshold, \emph{green} for a 1 keV threshold. The dotted lines are negligible for the 10 keV and 100 keV thresholds as in these regimes ICS dominates above threshold and very little energy is tagged as ``deposited'', instead being redistributed to lower-energy electrons and photons.}
\end{figure*}

In Figure \ref{fig:varyingpartition} we show the effect on the $f(z)$ curves, \emph{after} subtraction of continuum (sub-10.2-eV) photons, of taking different threshold energies, and of setting the ICS losses to zero below the cutoff indicated by the Hansen and Haiman criterion (Equation \ref{eq:hhcrit}), as was effectively done in previous work. In this last case we take the threshold (for transition to the detailed low-energy code) to be 1 keV. We have confirmed that lowering the threshold energy below 1 keV, to several hundred eV, has no effect on the deposition curves (as expected from the discussion above).

These results confirm our expectations from the analytic comparison of cooling times. The exact choice of threshold depends on the desired accuracy, and the tradeoff between the improved handling of atomic processes in the low-energy code and its lack of ICS, but in any case it should clearly lie in the 1-10 keV range. We choose 3 keV as our threshold, accepting the percent-level uncertainty from neglect of ICS below this threshold (as estimated from Figure \ref{fig:fractions}).

\section{The effect of the on-the-spot approximation on energy deposition curves}
\label{app:evolicompare}

With decreasing redshift, there is a steep fall in the injected power per baryon from DM annihilation (as $(1+z)^3$). In general, assuming on-the-spot deposition will underestimate the deposited energy at any given redshift, as the reduced fraction of deposited energy from recent annihilations is more than compensated by the residual deposition from earlier annihilations, when the DM density was higher. There is a countervailing effect, in that in the on-the-spot approximation the relevant opacity of the universe is its opacity at the redshift of injection, and it can only grow more transparent at later times; taking the delay in absorption into account might then be expected to reduce the overall energy deposition. However, the first effect generally dominates, giving rise to an $f(z)$ curve that is both higher and flatter with decreasing redshift than would be the case if the on-the-spot approximation held true (since at low redshifts the increasing transparency of the universe, which would naively lead to a steeply falling deposition-fraction curve, is partially compensated by the additional power injected at earlier redshifts).

\begin{figure}
\includegraphics[width=.4\textwidth]{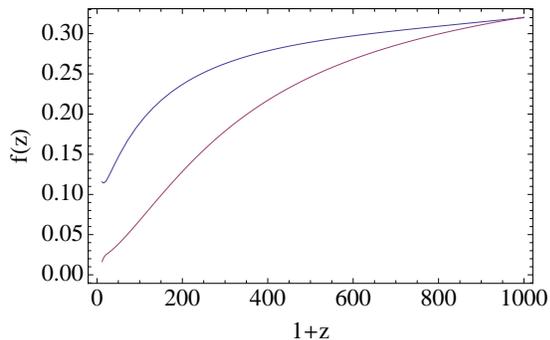}
\caption{\label{fig:onthespotcompare}
For DM annihilating to 10 GeV muons, the $f(z)$ curve (blue) and a curve showing the total fraction of energy ever deposited (purple). The latter curve has been multiplied by 1.18 to match the $f(z)$ curve at $z=1000$, to make the difference in redshift dependence more clear.}
\end{figure}

It is difficult to precisely reproduce the effect of assuming on-the-spot deposition (with a prescription for estimating energy losses to redshifting, as in \cite{Evoli:2012zz}), within our formalism. However, as a proxy, we can look at the total fraction of energy \emph{ever} deposited by a particle injected at a particular redshift; if the on-the-spot approximation held true, this curve would be precisely the same as $f(z)$, and so the degree to which they differ serves as a measure of how badly the approximation is violated. An example is shown for 10 GeV DM annihilating to muons ($\chi \chi \rightarrow \mu^+ \mu^-$) in Figure \ref{fig:onthespotcompare}. Here we show the original $f(z)$ curve defined in Equation \ref{eq:originalf}, as the curves for total energy deposited by particles injected at a particular energy and redshift (taken from \cite{Slatyer:2012yq}) are not yet available for the modified $f(z)$ curves discussed in this work. We expect that the direction of the effect should be similar for the updated calculation.

We see that the $f(z)$ curve falls noticeably less steeply with decreasing redshift, for the example given, and hypothesize that the on-the-spot approximation -- in particular, the failure to account for power deposited by particles injected at earlier redshifts -- may in part be responsible for the discrepancy noted by  \cite{Evoli:2012sc} between the original $f(z)$ curves and the shape of the energy deposition curves in that work's Monte-Carlo-based method (the latter fall much more steeply at redshifts $< 600$). The discrepancy at redshifts close to recombination, on the other hand, is more likely due to the approximate form of the energy fractions used to multiply the original $f(z)$ curve. The implementation of the ICS cutoff in  \cite{Evoli:2012sc} may also contribute to the discrepancy (see Appendix \ref{app:threshold}).

\clearpage

\newcommand\AAP[3]{AAP{\bf ~#1}, #2~ (#3)}
\newcommand\AL[3]{A. Lett.{\bf ~#1}, #2~ (#3)}
\newcommand\AP[3]{Astropart. Phys.{\bf ~#1}, #2~ (#3)}
\newcommand\AJ[3]{Astron. J.{\bf ~#1}, #2~(#3)}
\newcommand\APJ[3]{Astrophys. J.{\bf ~#1}, #2~ (#3)}
\newcommand\ApJ[3]{Astrophys. J.{\bf ~#1}, #2~ (#3)}
\newcommand\APJL[3]{Astrophys. J. Lett. {\bf ~#1}, L#2~(#3)}
\newcommand\APJS[3]{Astrophys. J. Suppl. Ser.{\bf ~#1}, #2~(#3)}
\newcommand\MNRAS[3]{MNRAS{\bf ~#1}, #2~(#3)}
\newcommand\MNRASL[3]{MNRAS Lett.{\bf ~#1}, L#2~(#3)}
\newcommand\NPB[3]{Nucl. Phys. B{\bf ~#1}, #2~(#3)}
\newcommand\PLB[3]{Phys. Lett. B{\bf ~#1}, #2~(#3)}
\newcommand\PRL[3]{Phys. Rev. Lett.{\bf ~#1}, #2~(#3)}
\newcommand\PR[3]{Phys. Rep.{\bf ~#1}, #2~(#3)}
\newcommand\PRD[3]{Phys. Rev. D{\bf ~#1}, #2~(#3)}
\newcommand\SJNP[3]{Sov. J. Nucl. Phys.{\bf ~#1}, #2~(#3)}
\newcommand\ZPC[3]{Z. Phys. C{\bf ~#1}, #2~(#3)}
\newcommand\SCI[3]{Sci.{\bf ~#1}, #2~(#3)}

\end{document}